\documentclass[12pt]{article}
\usepackage[utf8]{inputenc}

\usepackage{amsmath,amsfonts}
\usepackage{mathtools}
\usepackage{bm}
\usepackage{natbib}
\usepackage{color}
\usepackage[table]{xcolor}
\usepackage{graphicx,graphics}
\usepackage{float}
\usepackage{subfig}
\usepackage{setspace}
\usepackage[inline]{enumitem}
\usepackage{xargs, xstring}
\usepackage{stackengine}
\usepackage{rotating}
\usepackage{diagbox}

\usepackage{longtable}
\usepackage{array}
\newcolumntype{L}[1]{>{\raggedright\let\newline\\\arraybackslash\hspace{0pt}}m{#1}}
\newcolumntype{C}[1]{>{\centering\let\newline\\\arraybackslash\hspace{0pt}}m{#1}}
\newcolumntype{R}[1]{>{\raggedleft\let\newline\\\arraybackslash\hspace{0pt}}m{#1}}

\usepackage[toc,page,header]{appendix}
\usepackage{minitoc}

\usepackage[compact]{titlesec}
\titleformat{\section}
{\bfseries\centering}{\thesection.}{1em}{}
\titleformat{\subsection}
{\normalfont\normalsize\bfseries}{\thesubsection}{1em}{}
\titleformat{\subsubsection}
{\normalfont\normalsize\itshape}{\thesubsubsection}{1em}{}
\titleformat{\paragraph}[runin]
{\normalfont\normalsize\itshape}{\theparagraph}{1em}{}

\usepackage{setspace}

\usepackage{tikz}
\usetikzlibrary{positioning}
\usetikzlibrary{arrows.meta, matrix}
\usetikzlibrary{shapes,arrows,chains}
\usetikzlibrary{arrows,positioning,calc,topaths}

\usepackage[nokeyprefix]{refstyle}
\usepackage{varioref}
\usepackage{xr-hyper}
\RequirePackage[citecolor=blue,colorlinks=true,linkcolor=blue]{hyperref}
\usepackage[capitalise,noabbrev]{cleveref}
\crefformat{equation}{(#2#1#3)}
\newcommand{\blind}{1}
\newcommand{\usetitle}{Covariate-informed latent interaction models: Addressing geographic \& taxonomic bias in predicting bird-plant interactions}
\newcommand{\useauthor}{ {\large
Georgia Papadogeorgou$\overset{^1}{,}$
Carolina Bello$\overset{^2}{,}$
Otso Ovaskainen$\overset{^3}{,}$
David B. Dunson$^4$}}
\newcommand{\useinstitution}{ \small
$^1$Department of Statistics, University of Florida, 
$^2$Department of Environmental Systems Science, ETH Z\"urich, Switzerland,
$^3$Department of Biological and Environmental Science, University of Jyv\"askyl\"a;
Organismal and Evolutionary Biology Research Programme, University of Helsinki; and
Centre for Biodiversity Dynamics, Department of Biology, Norwegian University of Science and Technology, 
$^4$Department of Statistical Science, Duke University}
\newcommand{\useauthorshort}{Georgia Papadogeorgou, Carolina Bello, Otso Ovaskainen, David B. Dunson}

\newcommand\numberthis{\addtocounter{equation}{1}\tag{\theequation}}

\newcommand{\birds}{B}
\newcommand{\plants}{P}
\newcommand{\covB}{X}
\newcommand{\covP}{W}
\newcommand{\covsB}{\bm \covB}
\newcommand{\covsP}{\bm \covP}
\newcommand{\latB}{U}
\newcommand{\latP}{V}
\newcommand{\latsB}{\bm \latB}
\newcommand{\latsP}{\bm \latP}
\newcommand{\ssq}{\sigma^2}
\newcommandx{\logit}[1][1={}]{\text{logit}({#1})}
\newcommand{\alldata}{\widetilde{\bm D}}
\newcommand{\allpars}{\bm \theta^*}

\newcommand{\numinteractions}{3,804}
\newcommand{\numbirds}{232}
\newcommand{\numplants}{458}
\newcommand{\numplantsfull}{511}

\allowdisplaybreaks

\begin{document}

%\doparttoc % Tell to minitoc to generate a toc for the parts
%\faketableofcontents % Run a fake tableofcontents command for the partocs

%\part{} % Start the document part
%\parttoc % Insert the document TOC

\def\spacingset#1{\renewcommand{\baselinestretch}%
{#1}\small\normalsize} \spacingset{1}

\if1\blind{
\vspace{-10pt}
  \title{\bf \usetitle \vspace{-5pt}}
  \author{\useauthor \vspace{-5pt}}
  \date{\useinstitution}
%    \date{}
  \maketitle
} \fi

\if0\blind
{ \title{\bf \usetitle} \author{} \date{} \maketitle \vspace{-40pt}
} \fi

\begin{abstract}
%Climate change and r
Reductions in natural habitats urge that we better understand species' interconnection and how biological communities respond to environmental changes. However, ecological studies of species' interactions are limited by their geographic and taxonomic focus which can 
%lead to severe under-representation of certain species and 
distort our understanding of interaction dynamics. 
% We illustrate that ignoring the studies' focus can result in poor performance.
We focus on bird-plant interactions that refer to situations of potential fruit consumption and seed dispersal.
We develop an approach for predicting species' interactions that
%\begin{enumerate*}[label=(\alph*)]
%\item
accounts for errors in the recorded interaction networks,
%\item 
addresses the geographic and taxonomic biases of existing studies,
%\item 
is based on latent factors to increase flexibility and borrow information across species,
%\item 
incorporates covariates in a flexible manner to inform the latent factors, and
%\item 
uses a meta-analysis data set from 85 individual studies.
%\end{enumerate*}
We focus on interactions among \numbirds{} birds and \numplantsfull{} plants in the Atlantic Forest, and identify 5\% of pairs of species with an unrecorded interaction, but posterior probability that the interaction is possible over 80\%. Finally, we develop a permutation-based variable importance procedure for latent factor network models and identify that a bird's body mass and a plant's fruit diameter are important in driving the presence of species interactions, with a multiplicative relationship that exhibits both a thresholding and a matching behavior.
\end{abstract}

\noindent
{\it keywords:}
%adjacency matrix;
Bayesian methods;
% bipartite graph;
ecology;
graph completion;
latent factors;
% missing data;
% species interactions;
% trait matching;
variable importance
% shrinkage priors

\newpage

\spacingset{1.5} % DON'T change the spacing!
\section{Introduction}

\setlength{\abovedisplayskip}{6pt}
\setlength{\belowdisplayskip}{6pt}

% \subsection{Species interactions in the Atlantic Forest}

Animal-plant interactions have played an important role in the generation of Earth's biodiversity \citep{Ehrlich1964butterflies}. Hundreds of species form complex networks of interdependences whose structure has important implications for the stability of ecosystems \citep{Sole2001complexity},  their robustness to species extinctions \citep{Aizen2012specialization, Dunne2002food}, and their resilience in the face of environmental change \citep{Tylianakis2008global}.
Climate change and the reduction in species' natural habitats necessitate that we urgently understand species' interdependence in order to better predict how environmental changes will affect species' equilibrium and co-existence. 

Predicting and understanding species interactions is a long standing question in ecology. However, accessing all the possible interactions in a mutualistic network is a huge task that requires significant experimental effort \citep{Jordano2018sampling}.
Individual studies might focus on recording the interactions of only a given set of species. Even for the species under study, most measured networks are recorded in a specific geographical area where only a subset of species occurs. As a result, recorded networks are substantially incomplete and not comprehensively representative of species interactions, irrespective of the researchers' observational effort. These individual study characteristics lead to over-representation of a subset of species and under-representation of others implying that the resulting measured networks are taxonomically and geographically biased.
Even if measured networks from individual studies are compiled into one overarching network including all recorded interactions, these biases will propagate since cryptic species that occur in uncharted regions, or are not the explicit focus of the individual studies, will remain under-represented.
Even though these biases and their implications are well-recognized \citep{Baldi2003island, Seddon2005taxonomic, Pysek2008geographical, 
%Trimble2012geographical,
Hale2016ecological},
%, Jordano2018sampling},
most models for species interactions do not account for them \citep[e.g.][]{Bartomeus2013understanding}.
%, Gravel2013inferring}.
Some advances are emerging in the literature
\citep{Cirtwill2019quantitative, Weinstein2017comparing, Graham2018towards}, but the models therein do not provide a comprehensive treatment of species' traits and phylogenetic information.
Our goal is to use incomplete networks to understand whether {\it a given bird would eat the fruit of a given plant if given the opportunity}, and to learn which species traits are important in forming these interactions.

% \subsection{The question from a statistical perspective}
From a statistical perspective, a bird-plant interaction network can be conceptualized as a {\it bipartite graph}, where the birds and plants form separate sets of nodes, and an edge connects one node from each set. If a certain animal-plant interaction has been recorded, the corresponding edge necessarily exists. However, absence of a recorded interaction does not mean that the interaction is not possible and the networks are measured {\it with error}.
Modeling the probability of connections on a graph measured with\textit{out} error has received a lot of attention in the statistics literature, and examples stretch across social \citep{Newman2002random, 
%Eckmann2004entropy,
Wu2010evidence}, biological \citep{
%Han2004evidence, Sporns2004organization,
Chen2006detecting, Bullmore2009complex}, and ecological \citep{Croft2004social, Blonder2011time} networks, among others.
Since the literature on network modeling is vast, we focus on approaches for bipartite graphs.
In an early approach, \cite{Skvoretz1999logit} adapted the $p^*$ network models to the bipartite setting.
%Within this string of literature, there have been a number of frequentist and Bayesian approaches performing community detection in bipartite graphs. 
Community detection in bipartite graphs (referred to as co-clustering) was first introduced in \cite{Hartigan1972direct} and it has flourished in the last couple of decades \citep[e.g.,][]{Dhillon2003information, Shan2008bayesian, Wang2011nonparametric, Razaee2019matched}.
% with applications in genetics \citep{Cheng2000Biclustering, Kluger2003spectral, Madeira2004biclustering} and user-movie networks \citep{Ungar1998formal, Hofmann1999latent, Yang2002capturing}, among others.
%
%\cite{Dhillon2001} used the left and right eigenvalues of a scaled adjacency matrix to simultaneously cluster words and documents. \cite{Dhillon2003information} developed an algorithm to cluster the nodes of a bipartite graph such that the optimal clustering maximizes mutual information between the clustered random variables. Relatedly, \cite{Banerjee2007generalized} viewed community detection in bipartite graph as a matrix approximation problem, where the reconstruction minimizes information loss while preserving co-clustering statistics of the original graph.
% From a Bayesian perspective, \cite{Shan2008bayesian} specified a generative model for co-clustering which allows for mixed membership of the rows and columns. %, and proposed a variational Bayes approach for estimation.
%\cite{Wang2011nonparametric} performed simultaneous and nested clustering of the rows and columns of a bipartite graph, and \cite{Razaee2019matched} developed an approach to matching communities of the one set of nodes to those of the other.
%
Our approach is more closely related to network modeling using latent factors \citep{Hoff2002latent, Handcock2007model}
%, Hoff2008}
and its extension to multilinear relationships \citep{Hoff2005bilinear, Hoff2011hierarchical, Hoff2015Multilinear}, where the nodes are embedded in a Euclidean space and the presence of an edge depends on the nodes' relative distance in the latent space. 
Since our observed networks have missing edges, our approach also has ties to modeling noisy observed networks \citep{
%DeChoudhury2010inferring,
Jiang2011network, Wang2012measurement, 
%Chatterjee2015matrix,
Priebe2015statistical, Chang2020estimation}.

Our goals are to complete the bipartite graph of species interactions given the recorded, error-prone networks from individual studies, and  to understand which covariates are most important for driving species interdependence.
We develop a Bayesian approach to modeling the probability that a bird-plant interaction is possible based on a meta-analysis data set from 85 studies on the Atlantic Forest. 
The proposed approach
\begin{enumerate*}[label=\alph*)]
\item models the probability of a link in the bipartite graph,
\item incorporates the missingness mechanism caused by the taxonomic and geographic bias of individual studies, and the possibility that an interaction was not detected,
\item uses covariate information to inform the network model and improve precision, 
\item employs a latent variable approach to link the model components,
%, which aids prediction for bird-plant pairs that did not co-occur,
\item quantifies our uncertainty around the estimated graph, and
\item uses posterior samples in a permutation approach to acquire a variable importance metric.
\end{enumerate*}
To our knowledge, our approach is the first to employ latent network models for noisy networks, to use covariates to inform the latent factors via separate models instead of including them in the network model directly, and to study variable importance in latent factor models.

\section{A multi-study data set of bird--plant interactions in the Atlantic Forest}
\label{sec:data}

The Atlantic Forest is threatened due to overexploitation of its natural resources, and it currently includes only 12\% of its original biome \citep{Ribeiro2009brazilian}. In this biome, plants rely heavily on frugivore animals for their seed dispersal, and reductions in frugivore populations lead to disruptions in the regeneration of ecosystems. To better understand species' interactions and how biological communities respond to environmental changes, we study bird-plant interactions in the Atlantic Forest. We use an extensive data set which includes frugivore-plant interactions from 166 studies for five frugivore classes \citep{bello2017atlantic}.
A recorded interaction represents a setting where a frugivore was
% could have been 
% Caro told me to change "chould have been" to "was" -- I thought that her paper suggested that some interactions might not have led to seed dispersion.
involved in the plant's seed dispersal process, in that it handled a fruit in a manner that may have ended in consumption and subsequent dispersal of the seed. Other types of fruit handling that could not have led to seed dispersal were excluded from the data set, wherever this information was available.
Since we focus on bird-plant interactions (excluding mammals or other classes), we maintain 85 studies that include at least one such interaction.
These 85 studies recorded interactions for \numbirds{} birds and \numplantsfull{} plant species, but only \numplants{} of the plant species were involved in
an interaction with a bird (Supplement \ref{supp_sec:list_species} includes the list of species). The number of {\it unique} recorded bird-plant interactions was \numinteractions{}.

One of the key characteristics of our data is that unobserved interactions might be possible. For an interaction to be recorded there has to exist at least one study for which both species co-occur at the study site, they interact, and the interaction was detected and recorded.
However, individual studies are often limited in terms of the species or geographical area they focus on. 
Species-oriented studies record only a subset of the interactions that are detected: an animal-oriented study focuses on a {\it given} animal's diet whereas a plant-oriented study focuses on learning which animals eat the fruits of a {\it given} plant.
Hence, measured networks from such studies do not represent species comprehensibly and are {\it taxonomically biased}. In contrast, network studies record {\it any} interaction that is observed. However, studies of either type often focus on a small area where not all animal and plant species occur, and are hence {\it geographically biased}. As a result, a complete record of interactions is almost impossible to acquire, even for species of explicit interest.

\begin{figure}[!b]
\vspace{-10pt}
\centering

\subfloat[Geographic bias]{
\includegraphics[width=0.26\textwidth,trim=0 20 0 100, clip]{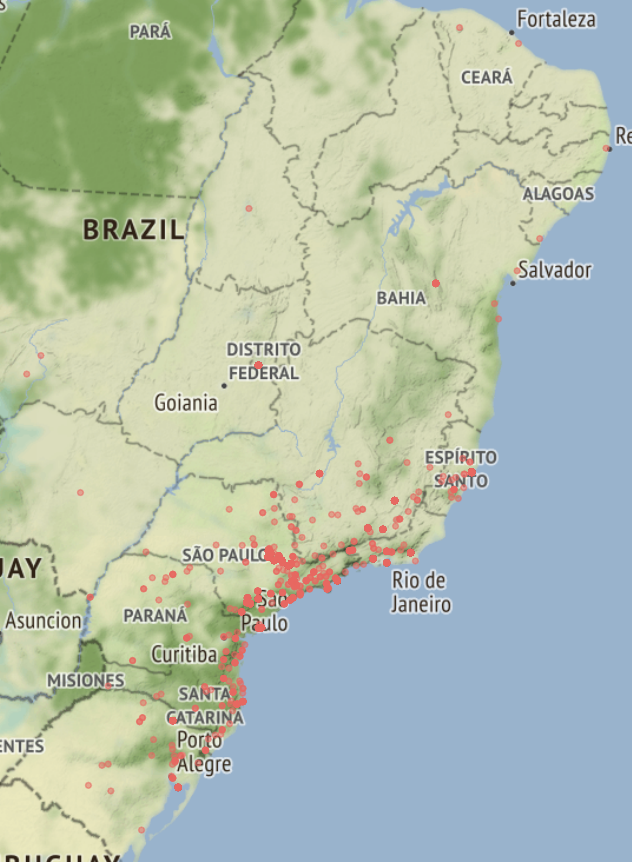}
\label{fig:illust_geo_bias}}
\hspace{0.05\textwidth}
\subfloat[Taxonomic bias]{\includegraphics[width=0.51\textwidth,trim=0 20 10 0, clip]{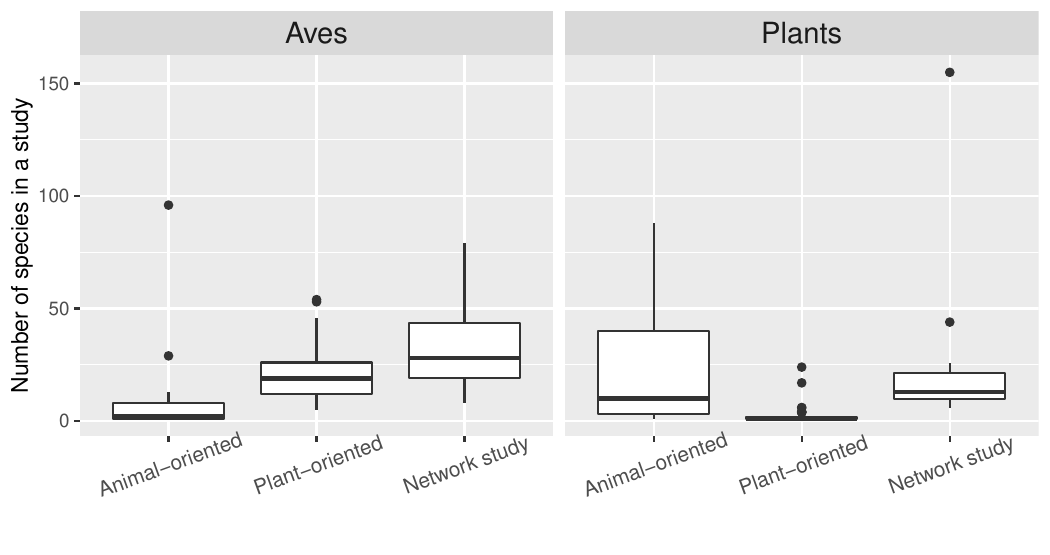}
\label{fig:illust_tax_bias}}

\caption{Geographic and Taxonomic Bias. (a) Locations of recorded interactions with reported coordinate information amounting to 68\% of recorded interactions. (b) Number of unique bird and plant species with recorded interactions within each study, by study type (animal/plant-oriented, network study).}
\label{fig:observed_interactions}

\end{figure}

The taxonomic and geographical biases of the individual studies propagate when compiling the recorded interactions into one combined network.
Since most studies are located in the southeast Atlantic Forest (see \cref{fig:illust_geo_bias}), interactions among species that do not co-occur in this area are less likely to be detected. Out of the eight bioregions of the Atlantic Forest biome, 45\% of interactions were recorded in the Serra do Mar bioregion, and there was no recorded interaction in the S\~ao Francisco bioregion. Therefore, the combined network will over-represent species that occur in the regions that are heavily studied and under-represent species that do not, implying that the combined network is itself geographically biased.
Out of the 85 studies in our data, 19 were animal-oriented, 45 were plant-oriented, and 19 were network studies (the remaining 2 were a combination).
\cref{fig:illust_tax_bias} shows the number of unique species observed in each study by study type. Animal-oriented studies have recorded interactions on a much smaller number of bird species than plant-oriented studies, and the reverse is true for plant species.
These trends in over- and under-representation of certain species will persist in the combined network, as species that were the focus of species-oriented studies will be more heavily represented.
In fact, our data over-represent trees and shrubs and under-represent other types of plants, whereas birds with recorded interactions correspond to only 27.1\% of the birds residing in the Atlantic Forest \citep{bello2017atlantic}. An analysis of the number of recorded interactions in \cite{bello2017atlantic} indicates that new studies continue to discover previously undetected interactions, implying that the recorded interactions are only a subset of those that are possible.

Our data include key bird and plant physical traits such as the diameter and color of the plant's fruit, and the bird's body mass and gape size, which are available with varying amounts of missingness.
These covariates may influence the success of a frugivory interaction, and researchers are interested into understanding this relationship \citep{rossberg2013food, Fenster2015quantifying, Dehling2016morphology, Descombes2019plant}.
In ecological studies, it is often assumed that species that are more genetically related have more similar traits and share more interactions. Phylogenetic trees have been used to represent such correlations across species \citep{Ives2011generalized}, and incorporating species' phylogenetic information can improve our understanding of species interactivity \citep{Benadi2021quantitative}. In some cases, phylogenetic information has agreed with observed correlations in species' traits or interaction profiles \citep{Mariadassou2010uncovering},  %Gilbert2007phylogenetic,
but not in others \citep{Rezende2007nonrandom}.
We acquire phylogenetic information for bird species from \url{https://birdtree.org} \citep{Jetz2012global} and for plant species using the \texttt{V.PhyloMaker} R package \citep{Jin2019phylomaker}.

\section{Learning species interactions addressing geographic and taxonomic bias}
\label{sec:model}

We use $i = 1, 2, \dots, n_\birds$ and $j = 1, 2, \dots, n_\plants$ to represent birds and plants respectively. For every bird $i$, $\covsB_i = (\covB_{i1}, \covB_{i2}, \dots, \covB_{ip_\birds})'$ represents $p_\birds$ measured physical traits, and similarly $\covsP_j =  (\covP_{j1}, \covP_{j2}, \dots, \covP_{jp_\plants})'$ for plant $j$.
Each species has an individual detectability score representing the probability that it would be detected to interact if the interaction occurred. We denote this as $p_i$ for birds and $q_j$ for plants.
Each study $s = 1, 2, \dots, S$ has recorded an interaction for each pair of species, or not. We compile the record of measured interactions across all studies in a three-dimensional array $\bm A$ of dimension $n_\birds \times n_\plants \times S$ where the $ijs$ entry, $A_{ijs}$, is equal to 1 if study $s$ recorded an $(i,j)$ interaction, and equal to 0 otherwise.
We are interested in inferring the $n_\birds \times n_\plants$ matrix $\bm L$, the entries of which represent whether bird $i$ would interact with plant $j$ if given the opportunity ($L_{ij} = 1$), or not ($L_{ij} = 0$). 
% Since whether the species are interactive  or not does not depend on the individual studies, the true interaction matrix $\bm L$ does not depend on $s$.
We assume that there was no human error in recording interactions, and a recorded interaction is truly possible (hence if $A_{ijs} = 1$ for at least one $s$, then $L_{ij} = 1$ necessarily).
In contrast, pairs without any recorded interaction might still be interactive.
Our goal is to infer the value of $L_{ij}$ for pairs $(i,j)$ without a recorded interaction.
In our study, $n_\birds = \numbirds{}$, $n_\plants = \numplantsfull{}$, and $S = 85$. A glossary is included in Supplement \ref{supp_sec:notation}.

\subsection{Study focus and species co-occurrence}
\label{subsec:setup}

To elucidate a model for the probability that two species are interactive, we first investigate the conditions under which a specific pair would be recorded to interact in a given study.
In a measured network, an interaction would be recorded if all of the following held:
\begin{enumerate*}[label=\alph*)]
\item the species interact if given the opportunity,
\item the species are of interest in the particular study,
\item the species co-occur in the study area, and
\item the researchers detected the two species interacting.
\end{enumerate*}
If any of the above does not happen, the given study necessarily would not record the specific interaction.
Violations of b) and c) capture the potential of taxonomic and geographical bias of the given study. To address these biases, one should take the focus and species occurrence for each study into consideration. We let $\bm F$ denote a 3-dimensional binary array of dimension $n_\birds \times n_\plants \times S$ representing the focus of each study. In general, $F_{ijs} = 1$ implies that if an interaction between $i$ and $j$ was detected, it would have been recorded, and $F_{ijs} = 0$ only occurs when study $s$ is animal- or plant-oriented, and the species $i$ or $j$ are not species of interest.
For species occurrence, we let $\bm 
O$ be a similarly defined array where $O_{ijs} = 1$ indicates that species $i$ and $j$ both occur in the geographic area of study $s$, and $O_{ijs} = 0$, otherwise.
The focus of each study is known. In contrast, even if the individual study area is well-defined, lack of perfect knowledge of which species exist in that area keeps us from knowing which interactions are even {\it possible to be observed} \citep{Poisot2015beyond}. In our study, we consider $\bm O$ fixed and known. We discuss its choice in \cref{sec:application} and extensions that assume $\bm O$ unknown where applicable.

\subsection{The covariate-informed latent interaction model}

Our approach is based on linking
%\begin{enumerate*}[label = \alph*)]
%\item
measured networks,
%\item
species true interactivity,
%\item
detectability, and
%\item
trait information
%\end{enumerate*}
using latent factors $\latsB_i =(\latB_{i1}, \latB_{i2}, \dots, \latB_{iH})^T$ for bird $i$ and $\latsP_j = (\latP_{j1}, \latP_{j2}, \dots, \latP_{jH})^T$ for plant $j$. The number of latent factors $H$ can be conceptualized as very large to include all important species' traits, measured or not.
To elucidate a likelihood for the measured networks and trait information, $P \big( \bm A, \{ \covsB \}, \{ \covsP \} \mid \bm F, \bm O \big)$, we make assumptions that are summarized below and discussed in detail in Supplement \ref{supp_sec:likelihood}.

\subsubsection{The likelihood for the measured networks}

We allow for species' (measured or latent) covariates to drive detectability and the interactions that they are able to form. For example, a bird's size might be informative of both as larger birds are more visible and their larger beaks allow them to consume fruits of all sizes. Conditional on the species' detectability scores and their true interaction profiles, we assume that species traits do not inform which interactions are recorded in any other way.

Measured networks might exhibit dependence across species or study sites in the following ways. Studies that are geographically close or focus on similar species will have similar patterns of recorded interactions. An impossible interaction will be unrecorded across all studies. Species that are hard to detect will have a low number of recorded interactions across all measured networks. We assume that these are {\it all} types of dependencies that can manifest across the measured networks, and measured networks are independent across species and studies once we condition on study focus, species occurrence and detectability, and the true underlying interaction matrix. The assumption that observing a possible interaction is conditionally independent across studies has been previously employed within a related context \citep{Weinstein2017comparing}.
We assume that the record of each interaction depends only on the individual species and study characteristics. These assumptions allow us to write the likelihood of the measured networks conditional on (measured and latent) covariates, the true interaction matrix, study focus, species occurrence and detectability, $P \big( \bm A = \bm a \mid \bm L, \{ \latsB \}, \{ \latsP \}, \{ p \}, \{ q \}, \{ \covsB \}, \{ \covsP \}, \bm F , \bm O \big)$,
as 
\(
\prod_{i,j,s} P \left( A_{ijs} = a_{ijs} \mid L_{ij}, F_{ijs}, O_{ijs}, p_i, q_j \right)
\)
(see Supplement \ref{supp_subsec:network_likelihood}).
We specify
\begin{equation}
P(A_{ijs} = 1 \mid L_{ij} = l, F_{ijs} = f, O_{ijs} = o, p_i, q_j) =
\begin{cases*}
0,        & if $lfo = 0$, \quad and \\
p_iq_j,   & if $lfo = 1$,
\end{cases*}
\label{eq:model_probA_givenL}
\end{equation}
which implies that the conditional likelihood for the measured networks simplifies to
\begin{equation}
\prod_{\substack{{i,j,s} \\ F_{ijs}O_{ijs}L_{ij} = 1}}
(p_i q_j)^{a_{ijs}} (1 - p_i q_j)^{1 - a_{ijs}}
\prod_{\substack{{i,j,s} \\ F_{ijs}O_{ijs}L_{ij} = 0}} I \left( a_{ijs} = 0 \right).
\label{eq:network_likelihood}
\end{equation}
Equation \cref{eq:model_probA_givenL} acts as a specification for a ``missingness mechanism'' for the unrecorded interactions. It expresses that an impossible interaction will never be recorded. It also specifies that a study is informative of whether an interaction is possible only if the species co-occur in the area and they are part of the study focus, as a way to account for geographic and taxonomic biases. Even if all of these hold, an interaction can still be unrecorded in the study with probability $1 - p_i q_j$. From \cref{eq:network_likelihood} we see that the measured data are informative about species detectability through how often a given species is recorded to interact versus not among all studies and species of the other type for which such interaction is possible to be observed. Therefore, detectability scores are not informed by measured networks from studies for which the species is not of focus or does not interact with the focal species.

\subsubsection{The latent factors for model specification}

The likelihood discussed above cannot be used directly since it conditions on unmeasured quantities (the true interaction matrix, the unmeasured detectability scores and the latent covariates) along with the measured ones.
% covariates, study focus and species co-occurrence.
Building towards an observed data likelihood, we specify a joint distribution over the unmeasured variables conditional on the measured ones (see Supplements \ref{supp_subsec:latent_likelihood} and \ref{supp_subsec:covariates_likelihood} for the mathematical details).
We assume that the indicators of species' true interactions are independent conditional on species' characteristics.
% For the true interaction matrix, whether an interaction is possible depends on the species' characteristics, conditional on which the indicators of species' true interactions are independent.
Therefore, we ignore the possibility that species co-occurrence and competition might imply that an interaction that occurs in one location might not occur in another.
For the species' detectability, we assume that it is independent across species, and only depends on individual characteristics and not those of other species. To specify the distribution of the latent features conditional on the measured covariates, we combine it with the likelihood of the measured traits, and specify instead the likelihood of the measured traits given the latent features, and the marginal distribution of the latent features. Under these assumptions, we can write the distribution of the latent parameters times the likelihood of measured covariates, 
\( \displaystyle 
p \left( \{ \latsB \}, \{ \latsP \}, \bm L, \{ p \}, \{ q \} \mid \{ \covsB \}, \{ \covsP \}, \bm F, \bm O \right) p\left(\{ \covsB \}, \{ \covsP \} \mid \bm F, \bm O \right) 
\) as
{\small
\begin{equation}
\begin{aligned}
\Big[ 
\prod_j p \left( q_j \mid \latsP_j, \covsP_j \right) 
\prod_j p(\covsP_j \mid \latsP_j)
\Big]
\Big[
\prod_i p \left( p_i \mid \latsB_i, \covsB_i \right) 
p(\covsB_i \mid \latsB_i)
\Big] \\
\times
\Big[ \prod_{i,j} p \left( L_{ij} \mid \latsB_i, \latsP_j, \covsB_i, \covsP_j \right) \Big]
p(\{ \latsB \}) p(\{ \latsP \}),
\label{eq:latent_trait_likelihood}
\end{aligned}
\end{equation}
}
which is combined with \cref{eq:network_likelihood} for the full distribution over our measured and latent variables.
In Supplement \ref{supp_subsec:covariates_likelihood} we discuss how one could simultaneously model $\bm O$, and incorporate geographical covariates as predictors for species co-occurrence in different locations.
% \[
% p(\latsB, \latsP \mid \covsB, \covsP, \bm F) p(\covsB, \covsP \mid \bm F)
% = 
%p(\covsB, \covsP \mid \latsB, \latsP, \bm F) p(\latsB, \latsP \mid \bm F)
%\]

For appropriately chosen link functions $f_m$ and  $g_l$, we assume the {\it trait submodel}:
\begin{equation}
\begin{aligned}
& f_m^{-1}(E(\covB_{im} \mid \latsB_i)) = \beta_{m0} +  \latsB_i' \bm \beta_m, 
& \hspace{5pt} \text{for } \beta_{m0} \in \mathbb{R}, \bm \beta_m \in \mathbb{R}^H,
& \hspace{5pt} m = 1, 2, \dots, p_\birds, & \hspace{5pt} \text{and} \\
& g_l^{-1}(E(\covP_{jl} \mid \latsP_j)) = \gamma_{l0} +  \latsP_j' \bm \gamma_l , 
& \hspace{5pt} \text{for } \gamma_{l0} \in \mathbb{R}, \bm \gamma_l \in \mathbb{R}^H,
& \hspace{5pt} l = 1, 2, \dots, p_\plants. & \hspace{5pt}
\end{aligned}
\label{eq:model_covariates}
\end{equation}
We adopt logistic link functions for binary traits. For continuous traits, we use the identity link function, and we incorporate a parameter for the residual variance. Therefore, the latent factors are specified to be the driving force of birds' and plants' physical traits, and they can be conceived as low-dimensional summaries of the species' traits. As long as the important information in the measured traits for detectability and species interactions is captured by the lower-dimensional latent factors, then the conditional distributions for $p_i, q_j$ and $L_{ij}$ in \cref{eq:latent_trait_likelihood} can be specified to depend only on the latent factors. We do so below.

We specify the {\it interaction submodel} as
\begin{equation} 
\mathrm{logit} P(L_{ij} = 1 \mid \covsB_i, \latsB_i, \covsP_j, \latsP_j) = \lambda_0 + \sum_{h=1}^H \lambda_h \latB_{ih} \latP_{jh}, \quad \text{for } \lambda_h \in \mathbb{R}, h= 0, 1, \dots, H.
\label{eq:model_probinter}
\end{equation}
In \cref{eq:model_probinter}, the latent factors are used as in classic bipartite network models \citep[e.g.][]{Hoff2011hierarchical}.
%, and they represent the species' locations in the latent space, where species in nearby locations are more likely to interact.
Alternatively, one could allow for a different number of latent factors for each set of species and include them linearly in the interaction submodel. However, using the same number of factors $H$ allows us to conceive the interaction submodel \cref{eq:model_probinter} as a flexible representation of species' interactions driven by interactions among the species' ``effective'' traits. Since the role of traits in an ecological network is believed to be interactive \citep{Fenster2015quantifying}, we prefer this over the alternative.

The detection of species is believed to depend on species traits such as size and behavior \citep{Garrard2013general, Troscianko2017quantifying}. A bird's body mass, whether they are solitary or gregarious, and a plant's height might affect whether their interactions are easily detected or not.
For that reason, we specify the {\it detection submodel} to depend on the species' latent factors (which act as a low-dimensional summary of the covariates) as:
\begin{equation}
\begin{aligned}
E[\logit[p_i] \mid \latsB_i, \covsB_i] = \delta_0 + \latsB_i^T \bm \delta, \quad \text{ and } \quad
E[\logit[q_j] \mid \latsP_j, \covsP_j] = \zeta_0 + \latsP_j^T \bm \zeta,
\end{aligned}
\label{eq:model_probobs}
\end{equation}
for $\delta_0, \zeta_0 \in \mathbb{R}$, and $\bm \delta, \bm \zeta \in \mathbb{R}^H$. 
We assume that $\logit[p_i]$ and $\logit[q_j]$ have conditional normal distributions with mean as in \cref{eq:model_probobs} and residual variance $\ssq_{p,\birds}$ and $\ssq_{q,\plants}$, respectively.

Even though all latent factors are allowed to be drivers of traits in \cref{eq:model_covariates}, true interactions in \cref{eq:model_probinter}, and detectability in \cref{eq:model_probobs}, different factors can be more or less important in each model component, and they might effectively contribute to only a subset of them if their corresponding coefficient is small (Supplement \ref{supp_subsec:same_latent}.) 
% For example, some latent factors might inform only the interaction probability, only the detection probability, both or none.
We discuss this further in \cref{subsec:bayesian}.

% \begin{figure}[!t]
% \centering
% \resizebox{0.35\textwidth}{0.2\textheight}{
% 	\centering
% 	\begin{tikzpicture}
% 	\tikzstyle{every node} = [draw, shape=circle, inner sep=0pt, text width=6mm, align=center]
	
% 	% nodes %
% 	\node[fill = gray!30, text centered] (U) {$\latsB_i$};
% 	\node[fill = gray!30, right = 2.6 of U, text centered] (V) {$\latsP_j$};
% 	\node[below left= 0.8 of U, text centered] (X) {$\covsB_i$};
% 	\node[below right=0.8 of V, text centered] (W) {$\covsP_j$};
% 	\node[fill = gray!30, ] (L) at ($(X)!0.5!(W)$) {$L_{ij}$};
% 	\node[fill = gray!30, below = 1 of U, text centered] (pi) {$p_i$};
% 	\node[fill = gray!30, below = 1 of V, text centered] (pj) {$q_j$};
%     \node[below = 0.8 of L, text centered] (A) {$A_{ij}$};
%     \node[below = 0.5 of A, text centered] (n) {$n_{ij}$};

% 	\draw[->] (U) -- (X);
% 	\draw[->] (U) -- (pi);
% 	\draw[->] (U) -- (L);
	
% 	\draw[->] (V) -- (W);
% 	\draw[->] (V) -- (pj);
% 	\draw[->] (V) -- (L);

%     \draw[->] (L) -- (A);
% 	\draw[->] (pi) -- (A);
% 	\draw[->] (pj) -- (A);
% 	\draw[->] (n) -- (A);

% 	\end{tikzpicture}}
% \caption{Graphical representation of the model. Shaded nodes represent latent variables. Bird and plant covariates are denoted by $\covsB_i, \covsP_j$, and latent factors by $\latsB_i, \latsP_j$, respectively. A recorded interaction is denoted by $A_{ij}=1$ and a possible interaction by $L_{ij} = 1$. The parameters $p_i, q_j$ represent the probability that a species is detected, and $n_{ij}$ denotes the number of studies that have recorded interactions of both species.}
% \label{fig:model}
% \end{figure}

\subsection{Bayesian inference}
\label{subsec:bayesian}

Our approach is placed within the Bayesian paradigm which allows for uncertainty quantification on the probability of truly possible interactions.
The prior on the latent factors specifies that
\begin{enumerate*}[label=\alph*)]
\item the marginal variance of the latent factors is equal to 1,
\item a {\it given} species' $H$ latent factors are independent, and
\item the latent factors {\it across} species are dependent with correlation that depends on their phylogeny.
\end{enumerate*}
Parts a) and b) are common in latent factor models: since the latent factors are not identifiable parameters, restricting their scale does not affect model fit. Assuming that latent factors are a priori independent across $h$ allows them to capture different aspects of the species' latent features, though it does not restrict them to being independent a posteriori. Latent factors are instead specified to be dependent across species: If $\latsB_{.h} = (\latB_{1h}, \dots, \latB_{n_\birds h})^T$ and  $\latsP_{.h} = (\latP_{1h}, \dots, \latP_{n_\plants h})^T$ represent the collection of the $h^{th}$ factor across species, we specify
%\begin{equation}
$\latsB_{.h} \sim \mathcal{N}(\bm 0, \bm \Sigma_\latB)$ and
%\quad \text{ and } \quad
$\latsP_{.h} \sim \mathcal{N}(\bm 0, \bm \Sigma_\latP),$
%\label{eq:prior_latfac}
%\end{equation}
independently across $h$, but for $\bm \Sigma_\latB = \rho_\latB \bm C_\latB + (1-\rho_\latB)\bm I$, and similarly for $\bm \Sigma_\latP$, where $\bm I$ is diagonal and $\bm C_\latB, \bm C_\latP$ are the phylogenetic correlation matrices discussed in \cref{sec:data}. We specify $\rho_\latB, \rho_\latP \sim \text{Beta}(a_\rho, b_\rho)$ with values near 0 or 1 representing close-to-independence and almost perfect phylogenetic dependence of the species' latent factors, respectively.

Prior distributions need to be adopted for the remaining parameters which include the intercept, variance terms, and the coefficients of the latent factors in the models \cref{eq:model_covariates}, \cref{eq:model_probinter} and \cref{eq:model_probobs}.
Due to the complete model's high dimensionality for a moderate value of $H$, 
%efficient estimation of model parameters requires either a {\it small} pre-specified value for $H$, or a moderate value of $H$ with sufficient shrinkage of model parameters for increasing $h$. We follow the latter option and 
we adopt a prior distribution on model parameters which assigns increasing weight to values close to zero as the index $h$ increases. Specifically, we specify
\begin{equation}
\begin{aligned}
\beta_{mh} |\tau_{mh}^\beta, \theta_h & \sim N(0, \tau_{mh}^\beta \theta_h), \quad &
\gamma_{lh} |\tau_{lh}^\gamma, \theta_h & \sim N(0, \tau_{lh}^\gamma \theta_h) \\
\lambda_h |\tau_h^\lambda, \theta_h & \sim N(0, \tau_h^\lambda \theta_h), 
\quad & \delta_h |\tau_h^\delta, \theta_h & \sim N(0, \tau_h^\delta \theta_h),
\quad & \zeta_h |\tau_h^\zeta, \theta_h & \sim N(0, \tau_h^\zeta \theta_h)
\end{aligned}
\label{eq:prior_parameters}
\end{equation}
where $\tau_{mh}^\beta, \tau_{lh}^\gamma, \tau_h^\lambda, \tau_h^\delta, \tau_h^\zeta \sim IG(\nu/2, \nu/2)$, and
\begin{equation}
\begin{aligned}
\theta_h \mid \pi_h & \sim (1 - \pi_h) P_0 + \pi_h \delta_{\theta_\infty},
\quad & \pi_h &= \sum_{l = 1}^h \omega_l,
\quad & \omega_l & = v_l \prod_{t = 1}^{l - 1}(1 - v_t) \\
v_t & \sim \text{Beta}(1, \alpha), \ t<H & & \text{and}
& v_H & = 1.
\end{aligned}
\label{eq:prior_shrinkage}
\end{equation}
In \cref{eq:prior_parameters}, the prior variance of model coefficients is specified using parameter-specific variance terms $\tau$ and overall variance terms $\theta$.
Equation (\ref{eq:prior_shrinkage}) specifies the truncated increasing shrinkage prior of \cite{Legramanti2020}, which uses a stick-breaking specification to define the mixing probabilities of a spike-and-slab prior distribution on $\theta_h$, where $P_0$ is a slab distribution, and $\delta_{\theta_\infty}$ represents a point-mass at $\theta_\infty$.
We set $P_0$ to be an inverse gamma distribution, and set $\theta_\infty$ close to zero.
This specification results in prior distributions for $\theta_h$ which assign larger weight to the point-mass rather than the slab distribution and are therefore concentrated closer to zero for larger values of $h$.
The parameter-specific variance terms $\tau$ are centered at 1 and provide flexibility to the $h^{th}$ coefficient from each model to deviate from a $N(0, \theta_h)$ prior if this prior would lead to over-shrinkage of the corresponding coefficient. 
Therefore, the prior on $\theta_h$ is used to penalize more heavily the contribution of latent factors corresponding to a higher index $h$, essentially implying that not all the species' information represented in the species' latent factors will be important for detectability and species' interactions, while the parameters $\tau$ adjust the prior variance to allow for additional flexibility in the coefficient of the latent factors across submodels. Inverse-gamma prior distributions are also assumed for the remaining residual variance parameters. Hyperparameter values are reported in \cref{app_tab:hyper}.

%, including the residual variances of the trait models and the probability of detecting a given species.

\subsection{Posterior computation}
\label{subsec:MCMC}

We sample from the posterior distribution of model parameters using Markov Chain Monte Carlo (MCMC). Here we describe the algorithm at a high-level, but all details are included in Supplement \ref{supp_sec:MCMC}. 
At each MCMC step, the entries of the true interaction matrix for pairs with a recorded interaction are set to 1. The remaining entries are set to 1 or 0 with weights resembling the current values of \cref{eq:model_probinter} while reflecting that an unrecorded interaction among species that co-existed in multiple studies is more likely to be impossible. The parameters of the interaction model in \cref{eq:model_probinter} are updated using the P\'olya-Gamma data augmentation scheme under which P\'olya-Gamma random variables are drawn for all $n_\birds \times n_\plants$ pairs, conditional on which the posterior distributions of model parameters are normally distributed \citep{Polson2013bayesian}.
Parameters of the models for binary traits are updated similarly.
% To update the parameters of the models for binary traits, we again employ that P\'olya-Gamma data augmentation, draw values from P\'olya-Gamma distributions for each unit and each binary covariate, and sample the parameters from their normal conditional posterior distributions. 
% The latent factors from each set of species inform the probability of interaction, the species' covariates, and their probability of being detected, and are therefore included in a number of continuous and binary models. 
Despite the involvement of the latent factors in all submodels, the latent factors have normal posterior distributions conditional on all other quantities. Updates for the parameters in the increasing shrinkage prior are adapted to our setting from \cite{Legramanti2020}. Species' detectability scores are updated employing Metropolis-Hastings steps with a Beta proposal distribution centered at the current value. Despite the large number ($n_\birds + n_\plants$) of parameters updated this way, these updates required minimal tuning. The parameters $\rho_\latB, \rho_\latP$ are updated similarly. Imputation of missing covariate values is based on \cref{eq:model_covariates}.

We investigated the impact of out-of-sample species, and developed an algorithm which combines samples from the posterior distribution using the original data and an importance sampling step to predict interactions for these species. To avoid distraction from our main focus, we refer interested readers to Supplement \ref{subsec:out_of_sample}.

\section{Variable importance in latent interaction models}
\label{subsec:variable_importance}

We propose a permutation-based approach to measure a covariate's importance in latent factor network models. In our study, this procedure will inform us of the relative importance of species traits for forming interactions. We briefly discuss the approach here, though further details are included in Supplement \ref{supp_sec:variable_importance}.
We are interested in studying the importance of the $k^{th}$ bird trait. We use $\covsB_{.k}$ to denote the vector of the $k^{th}$ covariate across all bird species.
% Our approach uses the posterior samples of the probability of interaction in \cref{eq:model_probinter}.
For each $(i,j)$ pair of species, let $l_{ij}^{(r)}$ be the logit of the $r^{th}$ posterior sample for the probability of interaction in \cref{eq:model_probinter}, and $\bm l_{.j}^{(r)}$ be the vector of these probabilities across $i$, $\big(l_{1j}^{(r)}, l_{2j}^{(r)}, \dots, l_{n_\birds j}^{(r)} \big)^T$.
For each posterior sample $r$ and plant species $j$, we calculate the squared correlation between the predicted interaction probabilities $\bm l_{.j}^{(r)}$ and the covariate $\covsB_{.k}$. We average these values over all plant species $j$ and posterior samples. 
For a large number of permutations $B$, we reorder the entries in $\covsB_{.k}$ and repeat this process. %, resulting in the distribution of our statistic under the null.
We use the number of standard deviations away from the mean of the permuted test statistics that the observed test statistic falls as a measure of variable importance.
% In all the steps, we only consider bird species with observed values of the covariate. The reason is that imputed covariate values are based on the latent factors which also drive the interaction model, and using them in a variable importance measure could lead to misleading conclusions on the covariate importance metric.
A similar approach is followed for the plant species $\covsP$.

In our latent factor model, the latent factors and their coefficients are not identifiable parameters, and as a result we cannot interpret the magnitude of these coefficients as a variable importance metric. Even though the interaction probabilities are conditionally defined, and resampling methods belong generally outside the Bayesian paradigm, we find this resampling procedure to perform well in practice.

\section{Simulations}
\label{sec:simulations}

\subsection{The setup: Data generative mechanisms imitating the observed data}
\label{subsec:data_generation}

We perform simulations to study the impact of ignoring the taxonomic and geographical biases, to evaluate our approach under a variety of data generative mechanisms (DGMs), and compare its performance to that of alternative approaches.
We consider 24 scenarios that are combinations of the following: 
\begin{enumerate*}[label = \alph*)]
\item the same or different covariates drive interactions and detectability,
\item the important covariates are observed, some are observed and some unobserved, or all are unobserved,
\item the correlation among covariates is 0 or 0.3, and
\item there is low or high information, corresponding to species co-occurrence and recorded interactions that are more or less sparse.
\end{enumerate*}
Choices of a) allow us to evaluate whether the performance of our model is hindered by the fact that it uses the same latent factors in all submodels. Detailed information on the DGMs and additional simulation results including simulations on the variable importance metric are included in Supplement \ref{supp_sec:simulations}, and are summarized below.

Our simulations are based on our data on recorded bird-plant interactions in terms of the observed number of species and studies, and the structure of measured covariates. We generate covariates $\widetilde{\covsB}, \widetilde{\covsP}$ from a matrix-normal distribution, with correlation across covariates equal to 0 or 0.3, and correlation across species resembling the species' phylogenetic correlation matrices. Some of the covariates were then transformed to binary variables using their initial values as linear predictors in a Bernoulli distribution with a logistic link function. Only a subset of the generated covariates are available in the simulated data, and the rest are considered unmeasured. For the measured covariates, we maintain the same structure and proportion of missingness as in the observed data: 2 continuous and 3 binary covariates with proportion of missing values varying from 0--32\% for bird species, and 4 continuous and 8 binary covariates with proportion of missing values varying from 0--80\% for plant species.
The interaction submodel and the detectability submodels are specified as multiplicative and linear in $\widetilde{\covsB}, \widetilde{\covsP}$, respectively. 
The important covariates in the models can be the same or different, measured or unmeasured, and measured covariates might be interacting with unmeasured covariates. For example, in DGM2 the measured $\widetilde{\covB}_{i1}$ interacts with the unmeasured $\widetilde{\covP}_{j13}$.
The set of unmeasured covariates includes the same number of binary and continuous covariates as the set of measured ones. Across all scenarios, the true interaction model achieves AUROC equal to 0.78.
The 6 combinations of a)--b) correspond to DGM1--6 shown in \cref{tab:sims_important_variables}, where we also show which covariates are included in the interaction and detactability submodels.

\begin{table}[!t]
\centering
\caption{Simulations Setup. Variables included in the interaction and detectability submodels. $\checkmark$ indicates that the covariate was used in the interaction model. Shaded cells indicate that the covariate was used in the model for species detectability.}
\resizebox{0.8\textwidth}{!}{%
\begin{tabular}{ccc | cc | ccc || cccc | ccc | }
& & & \multicolumn{5}{c||}{\textbf{Bird covariates}} & \multicolumn{7}{c|}{\textbf{Plant covariates}} \\ 
\multicolumn{3}{c|}{Description} & \multicolumn{2}{c|}{{\footnotesize Cont.}} & \multicolumn{3}{c||}{{\footnotesize Binary}} & \multicolumn{4}{c|}{{\footnotesize Cont.}} & \multicolumn{3}{c|}{{\footnotesize Binary}}  \\
& & & 1 & 2 & 3 & 4 & 5 & 1 & 2 & 3 & 4 & 5 & 6 & 7--12 \\ \hline
DGM1 & same \& & {\it meas.} & \cellcolor[HTML]{C0C0C0} \checkmark &  & \cellcolor[HTML]{C0C0C0} \checkmark & \cellcolor[HTML]{C0C0C0} \checkmark & \checkmark &
\cellcolor[HTML]{C0C0C0} \checkmark & & \checkmark & \cellcolor[HTML]{C0C0C0} \checkmark & \cellcolor[HTML]{C0C0C0} \checkmark & &  \\
& measured & {\it unmeas.} & & & & & & & & & & & &  \\
\hline
DGM2 & same \& & {\it meas.} & \cellcolor[HTML]{C0C0C0} \checkmark &  &  & \cellcolor[HTML]{C0C0C0} \checkmark & \checkmark & & & \checkmark & \cellcolor[HTML]{C0C0C0} \checkmark & \cellcolor[HTML]{C0C0C0} \checkmark & &  \\
& mixed & {\it unmeas.} & & & \cellcolor[HTML]{C0C0C0} \checkmark & & & \cellcolor[HTML]{C0C0C0} \checkmark & & & & & &  \\
\hline
DGM3 & same \& & {\it meas.} & & & & & & &  & & & & &  \\
& unmeasured & {\it unmeas.} & \cellcolor[HTML]{C0C0C0} \checkmark &  & \cellcolor[HTML]{C0C0C0} \checkmark & \cellcolor[HTML]{C0C0C0} \checkmark & \checkmark &
\cellcolor[HTML]{C0C0C0} \checkmark & & \checkmark & \cellcolor[HTML]{C0C0C0} \checkmark & \cellcolor[HTML]{C0C0C0} \checkmark & & \\
\hline
%%%%
DGM4 & different \& & {\it meas.} & \checkmark & \cellcolor[HTML]{C0C0C0} \phantom{\checkmark} & \cellcolor[HTML]{C0C0C0} \checkmark & \cellcolor[HTML]{C0C0C0} \checkmark & \checkmark &
\checkmark & \cellcolor[HTML]{C0C0C0} & \checkmark & \cellcolor[HTML]{C0C0C0} \checkmark & \checkmark & \cellcolor[HTML]{C0C0C0} & \\
& measured & {\it unmeas.} & & & & & & & & & & & &  \\
\hline
DGM5 & different \& & {\it meas.} & \checkmark & & & \cellcolor[HTML]{C0C0C0} \checkmark & \checkmark & & & \checkmark & \cellcolor[HTML]{C0C0C0} \checkmark & \checkmark & \cellcolor[HTML]{C0C0C0} & \\
& mixed & {\it unmeas.} & & \cellcolor[HTML]{C0C0C0} & \cellcolor[HTML]{C0C0C0} \checkmark & & & \checkmark & \cellcolor[HTML]{C0C0C0} & & & & &  \\
\hline
DGM6 & different \& & {\it meas.} & & & & & & &  & & & & & \\
& unmeasured & {\it unmeas.} & \checkmark & \cellcolor[HTML]{C0C0C0} \phantom{\checkmark} & \cellcolor[HTML]{C0C0C0} \checkmark & \cellcolor[HTML]{C0C0C0} \checkmark & \checkmark &
\checkmark & \cellcolor[HTML]{C0C0C0} & \checkmark & \cellcolor[HTML]{C0C0C0} \checkmark & \checkmark & \cellcolor[HTML]{C0C0C0} & \\
\hline
\end{tabular}
}
\label{tab:sims_important_variables}
\end{table}

\subsection{The setup: Alternative approaches}

We focus on comparing the proposed approach to an alternative approach which uses covariates directly.
We also considered alterations of our model where versions of it
\begin{enumerate*}[label=\alph*)]
\item fix the number of latent factors $H$,
\item exclude the parameters $\tau$, and
\item allow the covariates to inform the latent factors only through \cref{eq:model_covariates}, cutting the feedback from the interaction and detectability submodels \citep{jacob2017better}.
\end{enumerate*}
We also considered an approach that uses both covariates and latent factors in the interaction submodel, and our approach and the covariates approach while assuming that the observed interaction network is not measured with error and bias correction is not performed. 
We present all models in Supplement \ref{supp_sec:other_methods}. Due to space constraints, we include the results from these approaches in detail in the supplement, and we summarize them below.
Note that the competing method presented here is our own construction and it does not exist in the literature, and that the models that are based directly on the covariates do not incorporate phylogenetic information.
% Computationally, the MCMC for the model depending on latent factors is complicated by the update of the latent factors, whereas the complicating computational aspect for the MCMC for the model depending directly on covariates is the imputation of missing covariate values.

\subsection{Simulation results}

\begin{figure}[!t]
\centering
\begin{minipage}{0.035\textwidth}
\includegraphics[width=\textwidth,trim=0 20 522 0, clip]{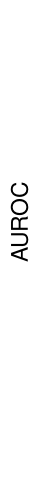} \\[2pt]
\end{minipage}
\begin{minipage}{0.8\textwidth}
\includegraphics[width=\textwidth,trim=19 61 0 0, clip]{sims_high.pdf} \\[2pt]
\includegraphics[width=\textwidth,trim=19 0 0 22, clip]{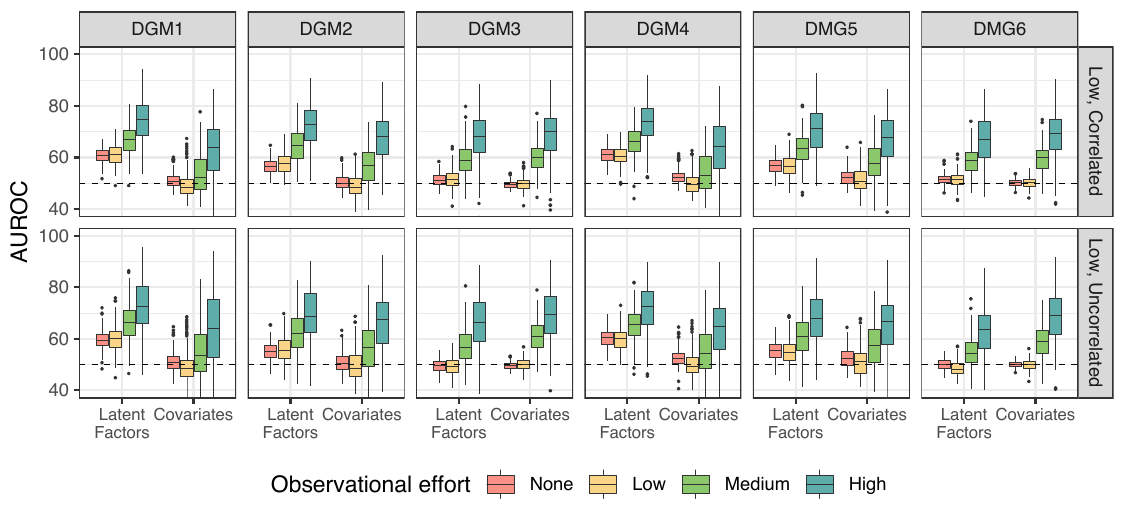}
\end{minipage}
\vspace{-15pt}
\caption{Predictive Performance in Simulations. The methods considered use latent factors or observed covariates (horizontal axis). The method using latent factors is the proposed approach. The columns represent the 6 DGMs in \cref{tab:sims_important_variables}. The rows correspond to combinations of the high and low signal scenarios and the two correlation values. Results are shown by observational effort for pairs of species by color.}
\label{fig:sim_res_A0yes}
\end{figure}

Methods were evaluated in terms of their predictive power in identifying true interactions. \cref{fig:sim_res_A0yes} shows the simulation results in terms of the AUROC (area under the receiver operating characteristics curve) when predicting the values of $\bm L$ among in-sample pairs with unrecorded interactions, separately by DGM, and amount of observational effort defined as the number of studies that could have recorded the interaction if it was observed, $\sum_s F_{ijs} O_{is}^\birds O_{js}^\plants$. Even though the AUROC is not a Bayesian criterion, it has been used before in a related setting \citep{sosa2022}, and it is not clear how one could use alternatives like the WAIC \citep{Watanabe2010asymptotic} since its computation for network data is complicated \citep{Gelman2014} and our network of interest (the matrix $\bm L$) is latent.

First, we notice that the performance of both methods improves with higher observational effort, implying that both models accommodate that an unrecorded interaction that was possible to be recorded across many studies is most likely not possible. Therefore, focusing on pairs of species on the lower end of observational effort compares the model structure more directly.
Across all 24 scenarios considered and across the spectrum of observational effort, our approach that uses latent factors performs better than or comparably to the method that uses covariates directly, and all alternatives considered in the supplement. The improvement by using the latent factors is most visible for the low observational effort pairs.
The performance of our approach is essentially unaltered by whether the same or different covariates drive the different submodels, irrespective of whether these covariates are measured or correlated (DGM 1 vs 4, 2 vs 5, and 3 vs 6).
Also, its performance is very similar when the covariates are correlated or not (comparing rows 1-2, and 3-4). The only exception is in the low information setting when all important covariates are unmeasured (rows 3 and 4, DGM 3 and 6), in which case correlation among covariates might improve the performance of the latent factor model.
When the important covariates are measured, the performance of our model improves (comparing DGMs 1 through 3, and 4 through 6), though in the high information setting the model can learn interaction profiles even when the important covariates are all unmeasured and uncorrelated with the measured ones (DGM6, High, Uncorrelated). These results inform us that the latent factor model performs better than using covariates directly across a variety of scenarios, and agrees with prior work on this topic which illustrated that flexible approaches perform better than including the traits directly into the model \citep{Pichler2020machine}.

\subsection{Results from additional simulation studies}

Here, we summarize some additional simulation results, all of which are shown in the Supplement. We have found that our approach performs better (or equally as well) compared to all other approaches considered. In all scenarios, approaches that ignore the taxonomic and geographical biases lead to very poor performance for predicting missing interactions which deteriorates for species with a higher observational effort, and a smaller number of predicted possible interactions compared to their counterparts with bias correction (also noted by \cite{Weinstein2017comparing} and \cite{Graham2018towards}). Using a higher number of latent factors $H$ with sufficient shrinkage, and incorporating the variance parameters $\tau$ improve the performance of the proposed approach. Cutting the feedback among the submodels for informing the latent factors performed better only in sparse settings and for pairs of species with high observational effort, indicating that the measured interactions can be helpful in informing the latent factors for predicting missing interactions.

We also found that the variable importance metric introduced in \cref{subsec:variable_importance} accurately identifies the covariates that are important for forming interactions, without specifying the functional form in which covariates drive interactivity. We find that variable importance should be interpreted separately for continuous and binary covariates. Our approach to variable importance is based on resampling techniques, and as a result it is arguably not-fully Bayesian. As an alternative we investigated variable importance for the model that includes covariates and latent factors. Apart from requiring a parametric specification of how covariates are included in the model, we find that using the coefficients of the covariates from this model for variable importance is flawed. This issue is related to spatial confounding in the spatial literature and arises due to collinearity of the phylogenetically-correlated covariates and latent factors \citep[see][for a discussion on spatial confounding in ecology, and references therein]{VanEe2021community}.

Across our simulations, we found that 1,000 MCMC iterations took on average 89 minutes. In the supplement, we also investigate the computational time of the proposed approach when varying the number of species and number of individual studies.

\section{Bird--plant interactions in the Atlantic Forest}
\label{sec:application}

We considered the two approaches that correct for taxonomic and geographic bias discussed in \cref{sec:simulations}, and we use the increasing shrinkage prior on the latent factor coefficients.
We specify $O_{ijs} = O_{is}^\birds O_{js}^\plants$, where $\bm O^\birds$ is an $n_\birds \times S$ binary occurrence matrix for birds, and similarly for $\bm O^\plants$. These are assumed known with entries equal to 1 if the species has a recorded interaction in the study and 0 otherwise. Due to a large number of recorded interactions with missing coordinate information, we are unable to include environmental or geographical covariates, though we discuss extensions in that direction in Supplement \ref{supp_subsec:latent_likelihood}.
We ran four chains of 80,000 iterations each, with a 40,000 burn in, and kept every 40$^{th}$ iteration. For our approach, 1,000 iterations took on average 86 minutes. MCMC convergence was investigated by studying traceplots and running means for identifiable parameters. Convergence diagnostics are shown in Supplement \ref{supp_sec:diagnostics}. Based on similar diagnostics, we found that the MCMC of the alternative approach failed to converge based on the same number of iterations. For that reason, we excluded from this analysis the two traits of the plant species with the largest amounts of missingness (seed length and whether the species is threatened for extinction) which led to no detectable lack of convergence. % Here, we show results for the probability of interaction for each pair of species. We discuss model performance in \cref{subsec:model_comparison} and trait matching in \cref{subsec:trait_matching}.

\begin{figure}[!t]
\centering
\subfloat[\textsf{Latent, Bias corrected}]{
\includegraphics[height=0.25\textheight, trim=1cm 3cm 1cm 1cm, clip]{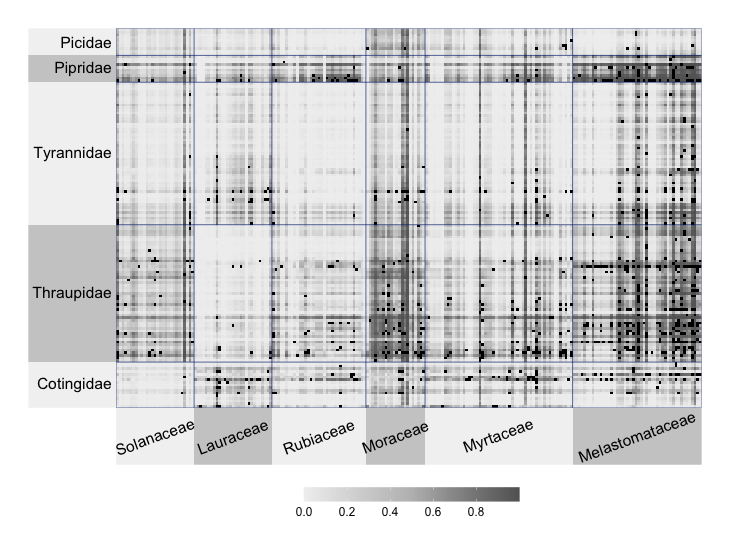} \label{fig:results_short_ours}}
\hspace{10pt}
\subfloat[\textsf{Covariates, Bias corrected}]{\includegraphics[height=0.25\textheight, trim=4cm 3cm 1cm 1cm, clip]{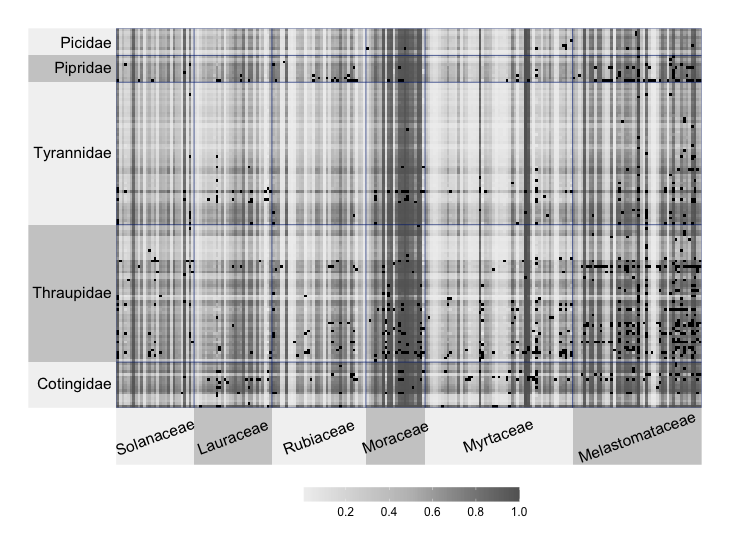} \label{fig:results_short_covs}}
\hspace{10pt}
\includegraphics[width=0.32\textwidth, angle=90, trim=11.3cm 1cm 7.5cm 19cm, clip]{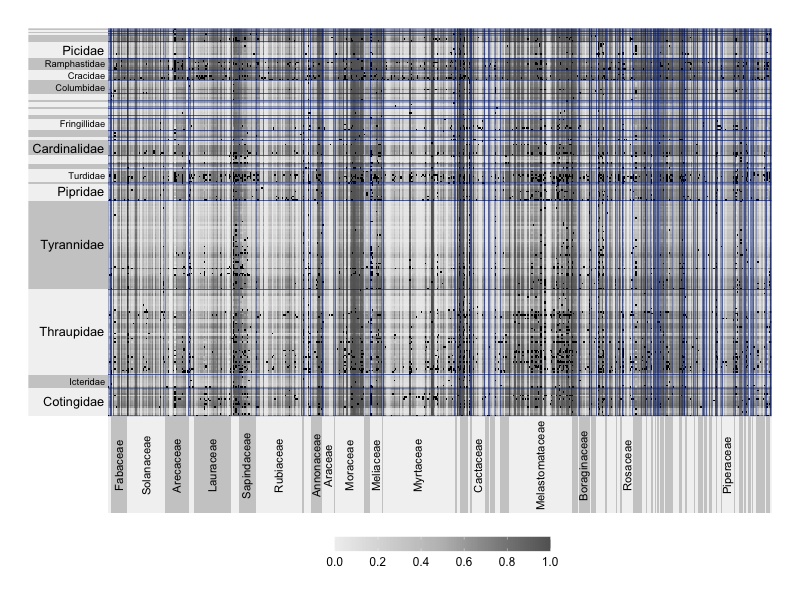}
\vspace{-2pt}
\caption{Posterior Probability of Interactions. Posterior probability that bird species (y-axis) and plant species (x-axis) interact according to (a) the proposed method and (b) the alternative method. Species are organized in taxonomic families separated by blue lines. Only taxonomic families with at least 10 bird and 20 plant species are shown to ease visualization. Black color is used to represent recorded interactions.}
\label{fig:results_short}
\end{figure}

In \cref{fig:results_short}, we show estimates for the probability of interaction for species in the largest taxonomic families. According to our model (\cref{fig:results_short_ours}), species in the same family form similar interactions as evidenced by the taxonomically-structured posterior interaction probabilities where the blue lines separate them in clusters with similar values.
In contrast, results from the alternative approach that employs covariates directly (\cref{fig:results_short_covs}) indicate that some species interact with most other species and some species with none, as evidenced by {\it rows and columns} that are mostly close to one or zero. Since we do not expect this ``all or none'' structure in species interactions, results from the covariate approach seem untrustworthy, and indicate that it might rely on covariates too heavily.
Species within the same family can belong to different genera, though genera are not shown in the figure to ease visualization. However, we observed that clusters of posterior interaction probabilities from the latent factor model within the depicted taxonomic families generally correspond to species organization by genera, supporting that interactions are taxonomically structured. The taxonomic structure is further supported by posterior means (95\% credible intervals) for $\rho_\latB$ and $\rho_\latP$ which were 0.97 $(0.947, 0.988)$ and 0.95 $(0.935, 0.97)$, respectively.
In Supplement \ref{supp_sec:app_results} we show that the results remain unchanged when using an alternative specification of the intra-species correlation matrix based on species' taxonomic relationships.

% \begin{table}[!b]
% \centering
% \caption{Comparison of Model Predictions. Pairs with posterior probability between 0--0.1 and between 0.9--1 are labelled as likely and unlikely to interact, respectively. The table shows the average posterior probability of interaction and median posterior deviation of interaction probabilities (in parentheses) based on both models, for pairs that are identified as likely and unlikely to interact according to each of the models.}
% \label{tab:app_compare}
% \begin{tabular}{lrrr}
%   \hline
% Identified based on &  & Latent Factors & Covariates \\ 
%   \hline
% Latent Factors & unlikely to interact & 0.04 (0.19) & 0.37 (0.40) \\ 
%               & likely to interact & 0.95 (0.23) & 0.91 (0.21) \\ 
% \hline
% Covariates & unlikely to interact & 0.06 (0.16) & 0.05 (0.22) \\ 
%          & likely to interact & 0.44 (0.43) & 0.96 (0.18) \\ 
%   \hline
% \end{tabular}
% \end{table}

\subsection{Comparison of model results and performance}
\label{subsec:model_comparison}

The two approaches often return opposite conclusions about species' interactions. The latent factor approach almost always returns probabilities of interaction that are lower than those from the covariate approach. The covariate approach predicts that 18\% of pairs interact (posterior probability above 80\%), and only 9\% of pairs do not (posterior probability below 10\%), both unrealistic. In contrast, the latent factor model predicts that 5\% of pairs interact, and 41\% do not. The vast majority of pairs that are predicted to not interact under the covariate model are also predicted to not interact based on the latent factor model, though the latent factor approach has substantially lower posterior standard deviation in these predictions. A more in-depth comparison  is given in Supplement \ref{supp_sec:app_results}.

To compare model performance directly, we applied a variant of cross-validation.
%, by holding out a subset of the recorded interactions and studying model predictions for these pairs.
We randomly choose 100 recorded interactions, we set their corresponding values in the observed interaction matrix equal to 0, and we predict their probability of interaction. We repeat this procedure 30 times, each time holding out a different subset of recorded interactions.
Our setting forbids us from comparing model performance based on {\it un}recorded interactions, since those interactions are not certainly impossible.
Our comparison is based on how well each approach can differentiate the held-out pairs of species that truly interact from the group of all pairs, which necessarily includes pairs that do not interact.
Since the two approaches return drastically different prevalence of interactions, we evaluate the relative magnitude of posterior interaction probabilities in the held-out and in the overall data.
For the covariate approach, the mean and median posterior probability of interaction for the held out pairs was on average 1.21 and 1.36 times higher than the corresponding value across all pairs of species. In contrast, those numbers where substantially higher and equal to 1.85 and 3.19 for the mean and median, respectively, for our approach.
%, these numbers were substantially higher, and the held-out pairs had mean and median posterior probability of interaction that was on average 1.8 and 3.2 times higher than the value across all pairs of species.
Therefore, our approach is much more effective in differentiating the pairs that are truly interactive from the set of all pairs compared to the approach that uses covariates directly.

% Specifically, for each of the 30 iterations of our cross-validation procedure, we calculate the ratio of the mean and median posterior probability of interaction in the held-out pairs to the corresponding quantity among all pairs. The ratio for the mean ranges over 1.66--2.06 and 1.1--1.37 for the latent factor model and the covariates model, respectively. The difference is even more prevalent when considering the median with ranges of 2.67--3.75 for the latent factor model and 1.1--1.6 for the covariates model (Figure included in Supplement \ref{supp_sec:app_results}). 

\subsection{The importance of traits and phylogeny for species interactivity}
\label{subsec:trait_matching}

\begin{figure}[!b]
\centering
\begin{minipage}{0.32\textwidth}
\centering
\subfloat[Bird Traits Importance\hspace*{\fill}]{
\includegraphics[width=0.9\textwidth, trim=15 4 12 11, clip]{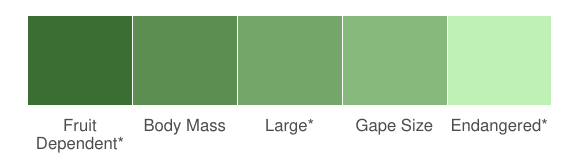}} \\
\subfloat[Plant Traits Importance \hspace*{\fill}]{
\stackunder{\includegraphics[width=0.9\textwidth, trim=15 5 280 10, clip]{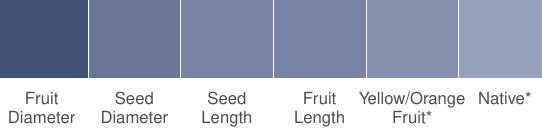}}{\includegraphics[width=0.9\textwidth, trim=280 5 10 15, clip]{app_sd_awayW.pdf}}}
\end{minipage}
\hspace{25pt}
\begin{minipage}{0.48\textwidth}
\centering
\subfloat[Interaction matrix ordered by traits \hspace*{\fill}]{\includegraphics[width=0.81\textwidth]{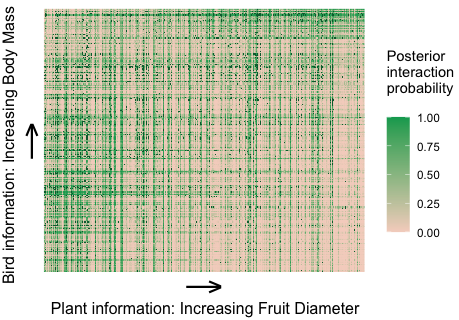}}
\end{minipage}
\caption{Figures (a) and (b) show the variable importance metric of \cref{subsec:variable_importance} for bird and plant species, which should be interpreted separately for continuous and binary traits. Traits are ordered from most important (dark color) to least important (light color), and $^\ast$ is used for binary traits. Figure (c) shows the matrix of posterior probabilities of interaction where the species are re-ordered in increasing order of body mass for birds and fruit diameter for plants. Pink and green are used for low and high probability of interaction, respectively, and dark green is used for recorded interactions. \\[-20pt]}
\label{fig:app_trait_importance}
\end{figure}

Apart from understanding which pairs of species are interactive, ecologists are also interested in understanding the traits which make species interactions possible \citep{Garrard2013general, Bastazini2017untagling, Troscianko2017quantifying}.
% Towards that goal, \cite{Bastazini2017untagling} studied how traits and phylogenetic information drive species interactions, and accounted for unrecorded interactions due to lack of overlap in species distributions. \cite{Pichler2020machine} showed that flexible models perform better than generalized linear models in both predicting interactions and identifying the important trait for these interactions.
%In \cref{fig:app_trait_importance} we show the results of trait matching in our study.
Figures \ref{fig:app_trait_importance}(a-b) show the variable importance metric described in \cref{subsec:variable_importance} for bird and plant traits.
%Darker colors are used to indicate higher importance. %We find that whether the species are endangered plays a minimal role in whether an interaction is possible. 
We identify a bird's body mass and a plant's fruit diameter as the most important continuous traits in forming interactions. In \cref{fig:app_trait_importance}(c) we plot the posterior probabilities of interaction reordering the species in increasing values of the two covariates. High posterior probabilities are concentrated on the upper left triangle, indicating that a given bird would interact with most plant species that are smaller than some threshold size,
% whereas low posterior probabilities are concentrated on the bottom right triangle.
%Therefore, our approach estimates that small birds mostly interact with plants that produce small fruits, whereas large birds can consume fruits of all sizes,
in line with the current ecological literature \citep{Fenster2015quantifying}. At the same time, there seems to be some preference for larger birds to not consume fruits that are too small, indicating a matching-size type of behavior for forming interactions. These results illustrate that our approach can identify complicated interactive relationships without having to specify these trends parametrically.

We studied the overall importance of these traits and phylogenetic information using the cross-validation technique discussed in \cref{subsec:model_comparison}. For trait importance, we excluded each of the traits from the available information, separately, and for the phylogenetic information we set $\rho_\latB, \rho_\latP$ to 0 which forces the latent factors to not be phylogenetically structured a priori. Excluding bird mass or fruit diameter returned on average a median posterior probability of interaction for the held out pairs 3.17 and 3.06 times higher, respectively, than the corresponding value across all pairs, compared to 3.19 when all traits are included. Therefore, fruit diameter can be an important covariate to measure for predicting species interactions. When ignoring phylogenetic information the corresponding value was 1.66 illustrating that phylogenetic information is crucial for predicting missing interactions.

\section{Discussion}
\label{sec:discussion}

%\commentG{say somewhere that by setting occurrence to 0 we protect ourselves from inflated certainty}
We introduced an approach based on latent factors that uses species traits and recorded interactions to complete the bipartite graph of species interdependence accounting for the taxonomic and geographic biases of individual studies, and we proposed an approach to study variable importance in latent network models.
We found that using covariates to inform the latent factors performs better in predicting pairs of species that do not interact and 
separating those that interact from the rest, compared to using the covariates directly.  Even though using the covariates in the proposed manner complicates the investigation of variable importance, we proposed a variable importance metric which performed well in simulations and identified important physical traits for species interdependence that are in line with ecological knowledge. 
% Open questions remain about how one can investigate the information content in species' phylogeny for predicting interactions based on approaches like ours that use species' traits which manifest phylogenetic structure themselves.

A possible extension to our model could accommodate simultaneous modeling of species co-occurrence, that would allow us to incorporate geographic information and other environmental variables that define the environmental niche of the species such as temperature, precipitation, and evapotraspiration \citep{Gravel2016bringing}.
Even though we provide an overview of such an approach in the supplement, studying the co-existence of species across space is a hard problem in itself and it is the topic of joint species distribution modeling in ecology \citep{Ovaskainen2020joint}. Importantly, modeling species co-occurrence and interactivity simultaneously and allowing for different interaction profiles based on environmental and geographical covariates would open the road to investigating the importance of species abundance, co-occurrence and competition in forming interactions. We find this to be an exciting line of future work.

% {Benadi2021quantitative does something related though they use abundances, observed trait differences based on a gaussian kernel, and the same for eigenvectors of phylogenetic correlation matrix -- if this paper is doing any bias correction it is only for abundance}

% One of the key aspects of our approach is that we assume that recorded interactions are necessarily possible. Falsely recorded interactions could be accommodated by altering the model component in \cref{eq:model_probA_givenL} to reflect
% \( \displaystyle P(A_{ijs} = 1 \mid L_{ij} = 0) = 1 - \prod_s (1 - p_s^{\text{error}}),\)
% where $p_s^{\text{error}}$ is a study-specific probability of mis-recording an interaction. Even though we consider this an interesting extension, we suspect that accommodating false positives could drastically affect model efficiency, especially in our very sparse scenario with only $\sim$3.2\% of all possible pairs having a recorded interaction. An alternative approach to investigating the presence of false positives could assess recorded interactions post-hoc by examining cases where the posterior probability of interaction is low even though the interaction is recorded.

\if1\blind{
\section*{Acknowledgements}
This project has received funding from the European Research Council (ERC) under the European Union Horizon 2020 research and innovation programme (grant agreement No 856506; ERC-synergy project LIFEPLAN). 
Otso Ovaskainen was funded by Academy of Finland (grant no. 309581), Jane and Aatos
Erkko Foundation, Research Council of Norway through its Centres of
Excellence Funding Scheme (223257).
Carolina Bello acknowledges funding support from the European Research Council (ERC) under the European union's Horizon 2020 research and innovation programme (grant agreement No 787638) and the Swiss National Science Foundation (grant No. 173342), both granted to Catherine Graham.
}\fi

\bibliographystyle{agsm}
\bibliography{Birds,Networks}
\clearpage

\doparttoc % Tell to minitoc to generate a toc for the parts
\faketableofcontents % Run a fake tableofcontents command for the partocs

\part{} % Start the document part
%\parttoc % Insert the document TOC

\begin{center} \Large
\if0\blind{Supplementary materials for \\ \usetitle}\fi
\if1\blind{Supplementary materials for \\ \usetitle \\ by \\ \useauthorshort}\fi
\end{center}

\allowdisplaybreaks
\setstretch{1.4}

\appendix
\setcounter{page}{1}
\setcounter{section}{0}    
\renewcommand{\thesection}{\Alph{section}}
\setcounter{equation}{0}    
\renewcommand{\theequation}{S.\arabic{equation}}
\setcounter{table}{0}    
\renewcommand{\thetable}{S.\arabic{table}}
\setcounter{figure}{0}    
\renewcommand{\thefigure}{S.\arabic{figure}}
\titleformat{\section}
{\normalfont\bfseries\large\centering}{Supplement \thesection.}{1em}{}
\titleformat{\subsection}
{\normalfont\normalsize\bfseries}{\Alph{section}.\arabic{subsection}}{1em}{}

\vspace{-40pt}

\addcontentsline{toc}{section}{Supplement} % Add the appendix text to the document TOC
\part{ } % Start the appendix part
\parttoc

\clearpage

\section{Notation}
\label{supp_sec:notation}

{\small
\renewcommand{\arraystretch}{1.3}
\begin{longtable}{R{0.14\textwidth}|L{0.79\textwidth}}
\caption{Glossary of notation.} \\
$n_\birds, n_\plants, S$ & Number of bird species, plant species, and ecological studies, respectively. \\
$i, j, s$ & Index for bird species, plant species, and study, respectively.  \\
$p_\birds, p_\plants$ & Number of measured traits for bird and plant species, respectively. \\
$\covsB_i$, $\covsP_j$ & Vectors of length $p_\birds$ and $p_\plants$ including the measured covariates of bird $i$ and plant $j$, respectively. \\
$\covsB_{.k}$ & Vector of length $n_\birds$ including the entries of the $k^{th}$ measured bird trait. \\
$p_i, q_j$ & Detectability score for bird $i$ and plant $j$, respectively. Values in (0, 1). \\
$\bm A, \bm F, \bm O$ & Arrays of dimension $n_\birds \times n_\plants \times S$ with binary entries.
Entry $A_{ijs}$ is equal to 1 if study $s$ recorded an interaction between bird $i$ and plant $j$, and 0 otherwise.
Entry $F_{ijs}$ reflects whether species $i,j$ were part of the taxonomic focus of study $s$, it is equal to 1 if study $s$ would have recorded the $ij$ interaction if observed, and 0 otherwise.
Entry $O_{ijs}$ is equal to 1 if species $i,j$ co-exist in the geographical area covered by study $s$, and 0 otherwise. \\
$\bm L$ & Matrix of dimension $n_\birds \times n_\plants$ with binary entries. Entry $L_{ij}$ is equal to 1 if bird $i$ is possible to interact with plant $j$ if given the opportunity. \\
$l_{ij}^{(r)}, \bm l_{.j}^{(r)}$ & Logit of the probability of interaction from \cref{eq:model_probinter} for pair $ij$ at the $r^{th}$ MCMC iteration, and vector of length $n_\birds$ including the $l_{ij}^{(r)}$ values for all $i$. \\
% $\bm F$ & Array of dimension $n_\birds \times n_\plants \times S$ with binary entries. Entry $F_{ijs}$ is equal to 1 if study $s$ would have recorded the $ij$ interaction, if observed, and 0 otherwise. Reflects whether species $i,j$ were part of the study's taxonomic focus. \\
% $\bm O$ & Array of dimension $n_\birds \times n_\plants \times S$ with binary entries. Entry $O_{ijs}$ is equal to 1 if species $i,j$ co-exist in the geographical area covered by study $s$, and 0 otherwise. \\
$H$ & Number of latent factors for both bird and plant species. \\
$\latsB_i, \latsP_j$ & Vector of length $H$ including bird $i$'s and plant $j$'s latent factors, respectively. \\
%
% Now, parameters:
$\beta_{m0}, \bm \beta_m$ & Intercept and vector of length $H$ including the coefficients of the $H$ bird latent factors in the regression model of bird species' $m^{th}$ measured trait. \\
$\gamma_{l0}, \bm \gamma_l$ & Same as above for the plant species and their $l^{th}$ measured trait. \\
$\lambda_0, \lambda_h$ & Parameters in the interaction submodel, intercept and coefficient of the product of the $h^{th}$ latent factors. \\
$\delta_0, \bm \delta$ & Intercept and coefficients of the $H$ bird latent factors in the bird detectability submodel. \\
$\zeta_0, \bm \zeta$ & Same as above for the plant detectability submodel. \\
$\ssq_{p, \birds}, \ssq_{q,\plants}$ & Residual variance for the detectability submodels for bird and plant species, respectively. \\
$\Sigma_\latB, \Sigma_\latP$ & Covariance matrix for the bird and plant latent factors of dimensions $n_\birds \times n_\birds$ and $n_\plants \times n_\plants$, respectively. \\
$\bm C_\latB, \bm C_\latP$ & Phylogenetic correlation matrices for bird and plant species, respectively. \\
$\rho_\latB, \rho_\latP$ & Parameters in (0, 1) for bird and plant species, respectively, deciding how much weight to give to phylogentic correlation matrix and the identity matrix in the definition of $\Sigma_\latB, \Sigma_\latP$. \\
$\theta, \tau$ & Global and parameter-specific variance terms in the increasing shrinkage prior used for the coefficients of the latent factors in the various submodels. \\
$\pi$, $\omega$, $v$, $\alpha$, $\alpha_\theta$, $\beta_\theta$, $\theta_\infty$ & Additional parameters of the increasing shrinkage prior.
\label{supp_tab:glossary}
\end{longtable}
}

Below we introduce some additional notation that is used throughout the Supplement.

\begin{enumerate}[leftmargin=*]
\item \textbf{Outer product:} We use $\bm A \otimes \bm B$ to denote the outer product of vector $\bm A$ of length $l_A$ and vector $\bm B$ of length $l_B$, where $\bm A \otimes \bm B$ is a matrix of dimension $l_A \times l_B$ with $(i_1, i_2)$ entry equal to $A_{i_1}B_{i_2}$.
\item \textbf{Vectorization:} For a matrix $\bm M$ of dimension $r \times c$, denote the vectorization of $\bm M$ as $\text{vec}(\bm M)$ where $\text{vec}(\bm M)$ is a vector of length $rc$ with entries \[ M_{11}, M_{12}, \dots, M_{1c}, M_{21}, \dots, M_{2c}, \dots, M_{rc},\]
hence unpacking first across the columns and then across the rows. 
\item \textbf{Conditional distributions:} We use $p(x_1 \mid x_2, x_3)$ to denote the distribution of $x_1$ given $x_2$ and $x_3$, $p(x \mid \cdot)$ to denote the distribution of $x$ given everything else, and $p(x \mid \cdot, -y) $ to denote the distribution of $x$ given everything except $y$.

\item We often deal with matrices of dimension $n_\birds \times n_\plants$ (such as $\bm L$) or with 3-dimensional arrays with dimension $n_\birds \times n_\plants \times S$ (such as $\bm A$). We always denote the entries corresponding to bird species by $i$, to plant species with $j$, and studies with $s$. So an entry in $\bm A$ is $A_{ijs}$ and an entry in $\bm L$ is $L_{ij}$.

To avoid heavy notation, in what follows, we use $\{L\} = \{L_{ij}\}_{i,j}$ to denote the collection of all elements in $\bm L$ across all indices $i$ and $j$, $\{L_i\} = \{L_{ij}\}_j$ to denote the collection of elements in $\bm L$ corresponding to bird index $i$ only and all indices $j$, $\{A_s\} = \{A_{ijs}\}_{i,j}$ to denote all elements in $\bm A$ corresponding to study $s$, $\{ \covsB \}$ the collection of covariates for all bird species, etc. 

\end{enumerate}

\section{Observed data likelihood}
\label{supp_sec:likelihood}

Our observed data include the measured networks across studies $\bm A$, and the measured covariate information $\covsB, \covsP$. We also know the focus of each study $\bm F$. We would like to acquire a model for the observed data corresponding to the measured networks and measured covariate values conditional on the studies' focus, $P (\{ A \}, \{ \covsB \}, \{ \covsP \} \mid \{ F \})$.

\subsection{Conditional distribution of measured networks}
\label{supp_subsec:network_likelihood}

We first focus on the likelihood of the measured networks conditional on covariates, $P (\{ A \} \mid \{ \covsB \}, \{ \covsP \}, \{ F \})$. We believe that correlations across $A_{ijs}$ for different values of $i,j,s$ can arise due to the following reasons:
\begin{enumerate}[label=(\alph*),leftmargin=*]
\item Studies that are geographically close are more likely to record similar interactions since the same species might co-occurring in both study areas. Conditioning on species co-occurrence should account for this type of dependence.

\item Species-oriented studies that focus on the same species will exhibit correlated records of interactions. Conditioning on the study's focus should account for this type of dependence.

\item If an interaction is truly \textit{im}possible, it will induce correlation across records, since none of the studies will record the specific interaction. Conditioning on the true interaction indicator should account for this type of dependence in the measured networks.

\item Species that have similar covariate profiles (either observed covariates or latent features) might exhibit similar profiles for which interactions are possible, and might have related co-occurrence patterns. Conditioning on the true interaction indicators and the occurrence of the species should account for this type of dependence as well.

\item Species that are hard to detect because of their behavior or physical characteristics will be hard to detect across studies, and their records of interactions will be correlated across studies. Conditioning on the probability of detecting a species should account for this type of dependence.

\end{enumerate}
To summarize the points above, we assume that, conditional on studies' focus, species occurrence across study sites, the true matrix of possible interactions, and species detectability, the records of interactions are independent across studies and across pairs of species, and they no longer depend on their measured or latent covariates. Therefore, for $a_{ijs} \in \{0, 1\}$ representing a possible realization of the $(i,j)$ interaction in the $s^{th}$ measured networks and $\{ a \}$ representing its collection across $i,j$ and $s$, we can write the recorded networks' conditional likelihood as
\begin{equation}
\begin{aligned}
P & \left( \{ A \} = \{ a \} \mid \{ \covsB \}, \{ \covsP \}, \{ \latsB \}, \{ \latsP \}, \{ L \}, \{ F \}, \{ O \}, \{ p \}, \{ q \} \right) = \\
&= P \left( \{ A \} = \{ a \} \mid \{ L \}, \{ F \}, \{ O \}, \{ p \}, \{ q \} \right) \\
&= \prod_{i,j,s} P \left( A_{ijs} = a_{ijs} \mid \{ L \}, \{ F \}, \{ O \}, \{ p \}, \{ q \} \right) \\
&= \prod_{i,j,s} P \left( A_{ijs} = a_{ijs} \mid L_{ij}, F_{ijs}, O_{ijs}, p_i, q_j \right),
\end{aligned}
\label{app_eq:cond_likelihood}
\end{equation}
where the first and second equality stem from the conditional independence assumptions described above, and the third equality stems from a type of ``individuality'' assumption that the only study focus, species occurrence, and the possibility of an interaction that matter are those that involve the given pair and the specific study.

In \cref{app_eq:cond_likelihood}, we have one term for each $(i,j,s)$ combination. The $(i,j,s)$ triplets can be split into two groups: those for which $F_{ijs}O_{ijs}L_{ij} = 1$ and an interaction between the species in the given pair in the specific study is possible to be recorded, and those for which $F_{ijs}O_{ijs}L_{ij} = 0$ and an interaction is impossible to be recorded. Therefore, \cref{app_eq:cond_likelihood} can be re-written as
{\small
\begin{align*}
& \prod_{\substack{{i,j,s} \\ F_{ijs}O_{ijs}L_{ij} = 1}}
P \left( A_{ijs} = a_{ijs} \mid L_{ij}, F_{ijs}, O_{ijs}, p_i, q_j \right) 
\prod_{\substack{{i,j,s} \\ F_{ijs}O_{ijs}L_{ij} = 0}}
P \left( A_{ijs} = a_{ijs} \mid L_{ij}, F_{ijs}, O_{ijs}, p_i, q_j \right) \\
&=
\prod_{\substack{{i,j,s} \\ F_{ijs}O_{ijs}L_{ij} = 1}}
(p_i q_j)^{a_{ijs}} (1 - p_i q_j)^{1 - a_{ijs}}
\prod_{\substack{{i,j,s} \\ F_{ijs}O_{ijs}L_{ij} = 0}} I \left( a_{ijs} = 0 \right).
\end{align*}}
For the first term, it is assumed that a study will record a possible interaction with probability that is equal to the product of the individual species detectability probabilities ($p_i q_j$). For the second term, since the study could not have recorded the given interaction, the only allowed value for whether the interaction is recorded is $a_{ijs} = 0$.

\subsection{Distribution of latent parameters}
\label{supp_subsec:latent_likelihood}

To go from the distribution in \cref{app_eq:cond_likelihood} to $P(\{ A \} = \{ a \} \mid \{ \covsB \}, \{ \covsP \}, \{ F \} )$ one would need to integrate over the distribution of
\begin{align*}
p & \left( \{ \latsB \}, \{ \latsP \}, \{ O \}, \{ L \}, \{ p \}, \{ q \} \mid \{ \covsB \}, \{ \covsP \}, \{ F \} \right) = \\
&=
p \left( \{ q \} \mid  \{ \latsB \}, \{ \latsP \}, \{ O \}, \{ L \}, \{ p \}, \{ \covsB \}, \{ \covsP \}, \{ F \} \right) \\
& \hspace{20pt} p \left( \{ p \} \mid  \{ \latsB \}, \{ \latsP \}, \{ O \}, \{ L \}, \{ \covsB \}, \{ \covsP \}, \{ F \} \right) \\
& \hspace{20pt} p \left( \{ L \} \mid \{ \latsB \}, \{ \latsP \}, \{ O \}, \{ \covsB \}, \{ \covsP \}, \{ F \} \right) \\
& \hspace{20pt} p \left( \{ O \} \mid \{ \latsB \}, \{ \latsP \}, \{ \covsB \}, \{ \covsP \}, \{ F \} \right) \\
& \hspace{20pt} p \left( \{ \latsB \}, \{ \latsP \} \mid \{ \covsB \}, \{ \covsP \}, \{ F \} \right) \\
&=
\Big\{ \prod_j p \left( q_j \mid \latsP_j, \covsP_j \right) \Big\}
\Big\{ \prod_i p \left( p_i \mid \latsB_i, \covsB_i \right) \Big\}
\Big\{ \prod_{i,j} p \left( L_{ij} \mid \latsB_i, \latsP_j, \covsB_i, \covsP_j \right) \Big\}
\numberthis{} \label{app_eq:measured_latent}
\\
& \hspace{20pt} \Big\{ \prod_{i,s} p(O^B_{is}) \Big\} \Big\{ \prod_{j,s} p(O^P_{js}) \Big\} 
p \left( \{ \latsB \}, \{ \latsP \} \mid \{ \covsB \}, \{ \covsP \}, \{ F \} \right).
\end{align*}
The last equation holds because we assume that (1) species detectability depends on their individual characteristics (measured or latent), is independent across species, and does not depend on species co-occurrence or study focus, (2) whether an interaction is possible depends solely on the covariates of the species involved, is independent across pairs, and does not depend on species co-occurrence which essentially limits the possibility of competition among species, and (3) we have access to the probability that species occur in a study area, so $O_{is}^B, O_{js}^P$ are probabilistic draws from this distribution and they do not depend on remaining information.

\paragraph{Note on species co-occurrence and environmental covariates} A future direction could combine modeling of possible interactions with modeling species co-occurrence, a hard problem in its own right. In that situation, one could specify in the equality above that
\begin{align*}
p & \left( \{ O \} \mid \{ \latsB \}, \{ \latsP \}, \{ L \}, \{ \covsB \}, \{ \covsP \}, \{ F \} \right) \\
&= \Big\{ \prod_{i,s} p \left( O_{is}^B \mid \latsB_i, \covsB_i, \{ L_i \}, \{ O_s^P \} \right) \Big\}
\Big\{ \prod_{j,s} p \left( O_{js}^P \mid \latsP_i, \covsP_i, \{ L_j \} \right) \Big\}.
\end{align*}
This would entail an assumption that a species' occurrence depends on its individual characteristics, and the set of species it can interact with, and the species of the other set that occur in each study area. To satisfy this assumption or to improve precision of these models, one might also include environmental covariates in the occurrence models, such as altitude, temperature and precipitation. Such covariates are believed to influence species co-occurrence more than whether species truly interact \citep{Gravel2016bringing}, so they would be most useful to be included when species co-occurrence is modeled than when considered known.

\subsection{Distribution of measured and latent covariates}
\label{supp_subsec:covariates_likelihood}

The last term in the distribution for all latent variables correspond to the conditional distribution of latent covariates given measured covariates and the studies' focus. This term is combined with the likelihood for the covariates and we write
\begin{align*}
p & \left( \{ \latsB \}, \{ \latsP \} \mid \{ \covsB \}, \{ \covsP \}, \{ F \} \right) p \left( \{ \covsB \}, \{ \covsP \} \mid \{ F \} \right) \\
&= p \left( \{ \covsB \}, \{ \covsP \} \mid \{ \latsB \}, \{ \latsP \}, \{ F \} \right) p \left( \{ \latsB \}, \{ \latsP \} \mid \{ F \} \right) \\
&= p \left( \{ \covsB \}, \{ \covsP \} \mid \{ \latsB \}, \{ \latsP \} \right) p \left( \{ \latsB \}, \{ \latsP \} \right) \\
&= \Big\{ \prod_i p(\covsB_i \mid \latsB_i) \Big\} \Big\{ \prod_j p(\covsP_j \mid \latsP_j) \Big\} p(\{ \latsB \}) p(\{ \latsP \}).
\end{align*}
This representation holds because we assume that studies' focus is not related to species' measured or latent covariates, and that a species' latent or measured covariates are not informative of another species' covariates.

\subsection{Using the same latent covariates in all the models}
\label{supp_subsec:same_latent}

In \cref{app_eq:measured_latent} we see that the measured and latent covariates co-exist in the detection and interaction submodels.
In their most generality, the latent covariates ($\latsB$ for bird species and $\latsP$ for plant species) can be arbitrarily close to the measured covariates ($\covsB$ for bird species and $\covsP$ for plant species). If the latent covariates capture the information in the measured covariates sufficiently well, then the measured covariates can be excluded from the detectability and interaction submodels in \cref{app_eq:measured_latent}, and we could allow
\begin{align*}
p \left( q_j \mid \latsP_j, \covsP_j \right) &= p \left( q_j \mid \latsP_j \right) \\
p \left( p_i \mid \latsB_i, \covsB_i \right) &= p \left( p_i \mid \latsB_i \right), \quad
\text{and} \\
p \left( L_{ij} \mid \latsB_i, \latsP_j, \covsB_i, \covsP_j \right) & =
p \left( L_{ij} \mid \latsB_i, \latsP_j \right).
\end{align*}
In addition, the latent covariates could be allowed to be higher-dimensional than the measured covariates capturing different features of the species that are not immediately available in the measured covariates. These features could be informative in one of the submodels without necessarily being informative in the others. For example, one of the variables in $\latsB$ can be informative for detectability and not informative for forming interactions. This covariate could still be included in the interaction submodel without creating any issues. Therefore, if latent features are sufficiently high-dimensional and resemble the measured covariates, we can allow all latent features to be included in each model component and measured covariates would no longer be directly necessary.

\subsection{Distribution of measured and latent variables}

Measured variables correspond to the measured networks and measured covariate information for the species. Latent variables correspond to every other variable discussed above: latent covariate information, detectability probabilities, the true interaction matrix, and the occurrence indicators. Combining the discussion above, we have the joint distribution of all these variables as
\begin{align*}
p & \left( \{ A \} = \{ a \}, \{ \covsB \}, \{ \covsP \}, \{ \latsB \}, \{ \latsP \}, \{ L \}, \{ O \}, \{ p \}, \{ q \} \mid \{ F \} \right) = \\
&= \Big\{\prod_{\substack{{i,j,s} \\ F_{ijs}O_{ijs}L_{ij} = 1}}
(p_i q_j)^{a_{ijs}} (1 - p_i q_j)^{1 - a_{ijs}} \Big\}
\Big\{ \prod_{\substack{{i,j,s} \\ F_{ijs}O_{ijs}L_{ij} = 0}} I \left( a_{ijs} = 0 \right) \Big\} \times \\
& \hspace{20pt}
\Big\{ \prod_i p \left( p_i \mid \latsB_i \right) \Big\}
\Big\{ \prod_j p \left( q_j \mid \latsP_j \right) \Big\}
\Big\{ \prod_{i,j} p \left( L_{ij} \mid \latsB_i, \latsP_j \right) \Big\}
\Big\{ \prod_{i,s} p(O^B_{is}) \Big\} \Big\{ \prod_{j,s} p(O^P_{js}) \Big\} \times \\
& \hspace{20pt}
\Big\{ \prod_i p(\covsB_i \mid \latsB_i) \Big\} \Big\{ \prod_j p(\covsP_j \mid \latsP_j) \Big\} p(\{ \latsB \}) p(\{ \latsP \})
\end{align*}

\section{MCMC scheme}
\label{supp_sec:MCMC}

\subsection{List of model parameters to be updated in an MCMC}

Model parameters to be updated include
\begin{itemize}[itemsep = 0pt, label=--]
\item  the $(n_\birds \times n_\plants)$ true interaction matrix $\bm L$,
\item the parameters of the interaction model $\bm \lambda$ where $\bm \lambda =  (\lambda_1, \lambda_2, \dots, \lambda_H)^T$
\item the latent factors $\latsB, \latsP$ of dimension $(n_\birds \times H)$ and $(n_\plants \times H)$ respectively,
\item the parameters of the trait models: $\bm B$ and $\bm \Gamma$, where $\bm B = (\bm \beta_1 \ \bm \beta_2 \ \dots \ \bm \beta_{p_\birds})$ is of dimension $H \times p_\birds$, and $\bm \Gamma = (\bm \gamma_1 \ \bm \gamma_2 \ \dots \ \bm \gamma_{p_\plants})$ is of dimension $H \times p_\plants$, and the residual variances $\ssq_m$ and $\ssq_l$ of continuous traits,
\item the parameters of the models for the probability of observing a true interaction of a given species $\bm \delta$ and $\bm \zeta$ and the residual variances $\ssq_{p, \birds}, \ssq_{q,\plants}$,
\item the probabilities themselves $\bm p_\birds = (p_1, p_2, \dots, p_{n_\birds})$, and $\bm q_\plants = (q_1, q_2, \dots, q_{n_\plants})$,
\item the matrices representing species occurrence across studies
$\bm O^B \in \{0, 1\}^{n_\birds \times S}$ and 
$\bm O^B \in \{0, 1\}^{n_\plants \times S}$
based on which the array of co-occurrences $\bm O \in \{0, 1\}^{n_\birds \times n_\plants \times S}$ is defined as $O_{ijs} = O^B_{is} O^P_{js}$,
\item the parameter in the latent factor covariance matrices $\rho_\latB, \rho_\latP$,
\item the variance scaling parameters $\tau$ across all models,
\item the parameters $\bm \theta, \bm \pi, \bm \omega$ and $\bm v$ controlling the increasing shrinkage prior, and
\item covariate missing values, if applicable.
\end{itemize}

\subsection{The posterior distribution}

The posterior distribution of all model parameters (assuming no missing values of covariates) is
\begin{align*}
p(\text{parameters} \mid & \text{Data}) \propto  \\
& \propto
% Models for parameters
\prod_{m = 1}^{p_\birds} p(\covsB_{.m} \mid \bm \beta_{m.}, \latsB, \ssq_m)
\times \prod_{l = 1}^{p_\plants} p(\covsP_{.l} \mid \bm \gamma_{l.}, \latsP, \ssq_l) \\
% Model for the observed interaction matrix
& \hspace{40pt} \times
\Big\{ \prod_{s = 1}^S \prod_{i = 1}^{n_\birds} \prod_{j = 1}^{n_\plants}
 p(A_{ijs} \mid p_i, q_j, L_{ij}, F_{ijs}, O_{ijs})  \Big\} \\
% Model for the true interaction matrix
& \hspace{40pt} \times 
\Big\{ \prod_{i = 1}^{n_\birds} \prod_{j = 1}^{n_\plants}
p(L_{ij} \mid \bm \lambda, \latsB_{i.}, \latsP_{j.}) \Big\} \\
% Model for observing each species
& \hspace{40pt} \times
\Big\{ \prod_{i = 1}^{n_\birds} p(\logit[p_i] \mid \bm \delta, \latsB_{i.}, \ssq_{p,\birds}) \Big\}
\times
\Big\{ \prod_{j = 1}^{n_\plants} p(\logit[q_j] \mid \bm \zeta, \latsP_{j.}, \ssq_{q,\plants}) \Big\} \\
% Latent factors
& \hspace{40pt} \times \prod_{h = 1}^H \Big\{ p(\latsB_{.h} \mid \bm \Sigma_\latB) \ p(\latsP_{.h} \mid \bm \Sigma_\latP) \Big\} \\
% Occurrence indicators
& \hspace{40pt} \times
\Big\{ \prod_{s = 1}^S \prod_{i = 1}^{n_\birds} \prod_{j = 1}^{n_\plants}
 p(O_{ijs})  \Big\} \\
% Other parameters
& \hspace{40pt} \times p(\rho_\latB) p(\rho_\latP) \\
& \hspace{40pt} \times \prod_{m = 1}^{p_\birds} p(\beta_{m0}) \prod_{h = 1}^H p(\beta_{mh} \mid \tau_{mh}^\beta, \theta_h) \\
& \hspace{40pt} \times \prod_{l = 1}^{p_\plants} p(\gamma_{l0}) \prod_{h = 1}^H p(\gamma_{lh} \mid \tau_{lh}^\gamma, \theta_h) \\
& \hspace{40pt} \times
p(\lambda_0)p(\delta_0)p(\zeta_0) \prod_{h = 1}^H p(\lambda_h \mid \tau_h^\lambda, \theta_h) p(\delta_h \mid \tau_h^\delta, \theta_h) p(\zeta_h \mid \tau_h^\zeta, \theta_h) \\
% tau priors
& \hspace{40pt} \times
\prod_{h=1}^H \left[ p(\tau_h^\delta) p(\tau_h^\zeta) p(\tau_h^\lambda) \prod_{m = 1}^{p_\birds} p(\tau_{mh}^\beta) \prod_{l = 1}^{p_\plants} p(\tau_{lh}^\gamma) \right] \\
% increasing shrinkage
& \hspace{40pt} \times \prod_{h = 1}^H p(\theta_h \mid \pi_h) p(\pi_h \mid \omega_1, \omega_2, \dots, \omega_h) p(\omega_h \mid v_1, v_2, \dots, v_h) p(v_h),
\end{align*}
where $p(\pi_h \mid \omega_1, \omega_2, \dots, \omega_h)$ and $p(\omega_h \mid v_1, v_2, \dots, v_h)$ are point mass distributions satisfying the equations in \cref{eq:prior_shrinkage}.

\subsection{MCMC updates}

% ----- MCMC: Update L ----- %
\paragraph{\underline{Updating the true interaction matrix $\bm L$}}
In deriving the posterior distribution of $L_{ij}$ we found that for pairs $(i,j)$ for which there exists a study $s$ that recorded their interaction, $A_{ijs} = 1$, we have that $P(L_{ij} = 1 \mid \cdot) = 1$. Therefore, if $A_{ijs} = 1$ for at least one $s$, then $L_{ij}$ is set to 1. Now, in the case where the interaction is unrecorded across all studies, $A_{ijs} = 0$ for all $s$, $L_{ij}$ is sampled using a Bernoulli distribution with
\[
p(L_{ij} = l \mid \cdot) \propto \begin{cases}
1 - p_{ij}^L, & \text{if } l = 0 \\
p_{ij}^L (1 - p_iq_j)^{\sum_s F_{ijs} O_{ijs}} & \text{if } l = 1,
\end{cases} 
\]
where $p_{ij}^L = \text{expit} \left\{\lambda_0 + \sum_{h = 1}^H \lambda_h \latB_{ih} \latP_{jh} \right\}$, and $p_i, q_j$ are the probabilities of observing bird $i$ and plant $j$ in \cref{eq:model_probobs}. Notice that when $A_{ijs} = 0$ for all $s$, the probability that the interaction is possible is smaller when $\sum_s F_{ijs} O_{ijs}$ is larger. This makes intuitive sense as are under the scenario where the interaction was not recorded across any study, and this quantity counts the number of studies for which the species $i,j$ co-occur, and the $(i,j)$ interaction would have been recorded if observed.

% ----- MCMC: Update lambda ----- %
\paragraph{\underline{Updating the parameters $\bm \lambda$ of the interaction model}}
We update these parameters using the P\'olya-Gamma data-augmentation of \cite{Polson2013bayesian} in the following manner:
\begin{enumerate}[leftmargin=*]
\item For each $(i,j)$ pair, draw latent variables $\omega_{ij}^L \sim \text{PG}(1, \lambda_0 + \sum_h \lambda_h \latB_{ih} \latP_{jh})$. Conditional on $\omega_{ij}^L$ the contribution of $L_{ij}$ to the likelihood is
\[
p(L_{ij} \mid \omega_{ij}^L, \bm \lambda, \latsB_{i.}, \latsP_{j.}) \propto \exp\left\{ - \frac{\omega_{ij}^L}2 \left[ \frac{L_{ij} - 1 / 2}{\omega_{ij}^L} - \left(\lambda_0 + \sum_{h = 1}^H \lambda_h \latB_{ih} \latP_{jh} \right) \right]^2 \right\},
\]
which is the kernel of a normal distribution, and can be combined with the normal prior distribution on $\bm \lambda$.

\item Sample $\bm \lambda \sim N_{H + 1}(\mu_{new}, \bm \Sigma_{new})$ for parameters
\[
\bm \Sigma_{new} = \left[\bm D_{UV}^T \bm \Omega^L \bm D_{UV} + (\bm \Sigma_0^\lambda)^{-1} \right]^{-1}, \]
and
\[
\mu_{new} = \bm \Sigma_{new} \left[\bm D_{UV}^T \left(\text{vec}(\bm L) - 1 / 2 \right) + (\bm \Sigma_0^\lambda)^{-1} \bm \mu_0^\lambda \right],
\]
where
\begin{itemize}[leftmargin=*]
\item $\bm D_{UV}$ is a matrix with $(n_\birds \times n_\plants)$ rows and $(H + 1)$ columns, with first column equal to 1, and $(h + 1)^{th}$ column equal to 
\[\text{vec}(\latsB_{.h} \otimes \latsP_{.h}) = (\latB_{1h}\latP_{1h}, \latB_{1h}\latP_{2h}, \dots, \latB_{1h}\latP_{n_\plants h}, \latB_{2h} \latP_{1h}, \dots, \latB_{2h}\latP_{n_\plants h}, \dots, \latB_{n_\birds h} \latP_{n_\plants h}), \]
\item $\bm  \Omega^L$ is a matrix of dimension $(n_\birds n_\plants \times n_\birds n_\plants)$ with the entries $\text{vec}(\omega_{ij}^L)$ on the diagonal and 0 everywhere else,
\item $\bm \Sigma_0^\lambda$ is a diagonal matrix with entries $\sigma_0^2, \tau_1^\lambda \theta_1, \tau_2^\lambda\theta_h, \dots, \tau_H^\lambda\theta_H$ on the diagonal ($\ssq_0$ is the prior variance of $\lambda_0$), and
\item $\bm \mu_0^\lambda$ is equal to $(\mu_0^{\lambda_0}, 0, 0, \dots, 0)^T$, where $\mu_0^{\lambda_0}$ is the prior mean of $\lambda_0$.
\end{itemize}
\end{enumerate}

\paragraph{\underline{Updating the variance scaling parameters $\tau$}}
Sample $\tau_{mh}^\beta$ from an inverse gamma distribution with parameters $(\nu + 1) / 2$ and $(\nu + \beta_{mh}^2 / \theta_h) / 2$. Similarly for $\tau_{lh}^\gamma, \tau_h^\delta, \tau_h^\zeta$ and $\tau_h^\lambda$.

\paragraph{\underline{Updating the parameters of continuous traits models}}
For a continuous trait $m$, the full conditional posterior distribution of $\bm \beta_{m.} = (\beta_{m0}, \beta_{m1}, \dots, \beta_{mH})^T$ is $N_{H + 1}(\bm \mu_{new}, \bm \Sigma_{new})$ for parameters
$$ \bm \Sigma_{new} = \left[ \bm D_\birds^T \bm D_\birds / \ssq_m + (\bm \Sigma_0^\beta)^{-1} \right]^{-1} $$
and
$$ \bm \mu_{new} = \bm \Sigma_{new} \left[\bm D_\birds^T \covsB_{.m} / \ssq_m + (\bm \Sigma_0^\beta)^{-1} \bm \mu_0^\beta \right],$$
where
\begin{itemize}[leftmargin=*]
\item $\bm D_\birds = \left(\bm 1 \mid \latsB_{.1} \mid \latsB_{.2} \mid \dots \mid \latsB_{.H} \right)$ matrix of dimension $(n_\birds \times (H+1))$,
\item $\covsB_{.m}$ vector of entries for the $m^{th}$ trait $(\covB_{1m}, \covB_{2m}, \dots \covB_{n_\birds m})^T$,
\item $\bm \Sigma_0^\beta$ diagonal matrix with entries $\ssq_0, \tau_{m1}^\beta \theta_1, \dots, \tau_{mH}^\beta\theta_H$ ($\ssq_0$ is the prior variance of $\beta_{m0}$), and
\item $\bm \mu_0^\beta = (\mu_0^{\beta_0}, 0, 0, \dots, 0)^T$ ($\mu_0^{\beta_0}$ is the prior mean of $\beta_{m0}$).
\end{itemize}

To update the residual variance of continuous trait $m$, we sample $\ssq_m$ from an inverse gamma distribution with parameters $a_\sigma + n_\birds / 2$ and $b_\sigma + \sum_{i = 1}^{n_\birds} (\covB_{im} - (1, \latsB_{i.}^T)^T \bm \beta_{m.})^2 / 2$.
Similarly we update parameters $\bm \gamma_{l.} = (\gamma_{l0}, \gamma_{l1}, \dots, \gamma_{lH})^T$ and $\ssq_l$ for continuous trait $\bm L$ of the other set of units.

\paragraph{\underline{Updating the parameters of binary traits models}}

To update the coefficients $\bm \beta_{m.}$ for a binary trait $m$ we again follow the P\'olya-Gamma data augmentation approach. %\cite{Polson2013bayesian}. 
Specifically,
\begin{enumerate}
\item We sample $\omega_{im}$ from $\text{PG}(1, (1, \latsB_{i.}^T)^T \bm \beta_{m.})$ for all $i = 1, 2, \dots, n_B$.
\item We draw $\bm \beta_{m.}$ from $N_{H + 1}(\bm \mu_{new}, \bm \Sigma_{new})$ for parameters
$$
\bm \Sigma_{new} = \left[\bm D_\birds^T \bm \Omega_m \bm D_\birds + (\bm \Sigma_0^\beta)^{-1} \right]^{-1}
$$
and
$$
\bm \mu_{new} = \bm \Sigma_{new} \left[ \bm D_\birds \left(\covsB_{.m} - 1 /2 \right) + (\bm \Sigma_0^\beta)^{-1} \bm \mu_0^\beta \right],
$$
where $\bm \Sigma_0^\beta, \bm \mu_0^\beta, \bm D_\birds$ and $\covsB_{.m}$ are as above, and $\Omega_m$ is a diagonal matrix with entries $\{\omega_{im}\}_{i = 1}^{n_\birds}$.
\end{enumerate}

Similarly we update the coefficients $\bm \gamma_{l.} = (\gamma_{l0}, \gamma_{l1}, \dots, \gamma_{lH})^T$ for the models of the binary traits for the other set of units.

\paragraph{\underline{Updating the parameters of the probability of observing an interaction}}
The parameters $\bm \delta$ and $\ssq_{p,\birds}$ are updated similarly to the updates for the parameters of the continuous trait models $\bm \beta$ and $\ssq_m$, using the same matrix $\bm D_\birds$, and setting \[\covsB_{.m} = (\logit[p_1], \logit[p_2], \dots, \logit[p_{n_\birds}])^T.\]
The update of $\bm \zeta$ and $\ssq_{q,\plants}$ proceeds similarly.

\paragraph{\underline{Updating the latent factors}}
We describe the update of the latent factors for the first set of units $\latsB_{.h}$ for $h = 1, 2, \dots, H$, and updates for $\latsP_{.h}$ are similar. Here, we will use the P\'olya-Gamma draws $\omega_{im}$ for binary traits $m$, and $\omega_{ij}^L$, described above. For each $h = 1,2, \dots, H$, $\latsB_{.h}$ is drawn from $\mathcal{N}_{n_\birds}(\bm \mu_{new}, \bm \Sigma_{new})$ for parameters
\[
\bm \Sigma_{new} = \left( \sum_{\substack{\text{m: }\covB_m \\ \text{continuous}}} \beta_{mh}^2 / \ssq_m \bm I_{n_\birds} + \delta_h^2 / \ssq_{p,\birds} \bm I_{n_\birds} + \sum_{\substack{\text{m: }\covB_m \\ \text{binary}}} \beta_{mh}^2 \bm \Omega_m + \sum_{j = 1}^{n_\plants} \lambda_h^2 \latP_{jh}^2 \bm \Omega_j^L + \bm \Sigma_\latB^{-1} \right)^{-1}
\]
and
\begin{align*}
\bm \mu_{new} = \bm \Sigma_{new} \Bigg\{ \sum_{\substack{\text{m: }\covB_m \\ \text{continuous}}} & \beta_{mh} / \ssq_m \ \textbf{part}(m,h) + \delta_h / \ssq_{p,\birds} \ \textbf{part}(p, h) + \\
&
\sum_{\substack{\text{m: } \covB_m \\ \text{binary}}} \beta_{mh} \bm \Omega_m \left[ \left(\frac{\covsB - 1/ 2}{\bm \omega} \right)_m - \left(1 \mid \latsB_{.-h} \right) \bm \beta_{m(-h)} \right] +\\
&
\sum_{j = 1}^{n_\plants} \lambda_h \latP_{jh} \bm \Omega_j^L \left[ \left(\frac{\bm L - 1/ 2}{\bm \omega} \right)_j - \left(1 \mid \latsB_{.-h}\latsP_{j(-h)} \right) \bm \lambda_{-h} ]\right]
\Bigg\}
\end{align*}
where
\begin{itemize}
\item $\bm \Omega_j^L$ is used to denote, with some abuse of notation, the diagonal matrix of dimension $n_\birds$ with entries representing the P\'olya-Gamma draws from the interaction model involving unit $j$: $(\omega_{1j}^L, \omega_{2j}^L, \dots, \omega_{n_\birds j}^L)$,
\item $\bm \Sigma_{\latB}$ is the covariance matrix of the latent factors in the prior distribution specified in \cref{subsec:bayesian}, %\cref{eq:prior_latfac},
\item $\textbf{part}(m,h)$ is used to denote the residuals from the model for $\covB_m$ when excluding the $h^{th}$ latent factor, and is the following vector of length $n_\birds$:
\begin{align*}
\textbf{part}(m,h) = \covsB_{.m} - \big(\beta_{m0} \bm1 + & \beta_{m1} \latsB_{.1} + \dots + \beta_{m(h-1)} \latsB_{.(h-1)} + \\
& \beta_{m(h+1)} \latsB_{.(h+1)} + \dots + \beta_{mH} \latsB_{.H} \big),
\end{align*}
\item Similarly, $\textbf{part}(p,h)$ is used to denote a vector of length $n_\birds$ including the residuals of the model for the probability of observing when excluding the $h^{th}$ latent factor:
\begin{align*}
[\text{part}(p,h)]_i = \logit[p_i] - \big( \delta_0 + \delta_1\latB_{i1} + \dots + \delta_{h-1}\latB_{i(h-1)} + \delta_{h+1}\latB_{i(h+1)} + \dots + \delta_H\latB_{iH} \big),
\end{align*}
\item \(\displaystyle \left(\frac{\covsB - 1/ 2}{\bm \omega} \right)_m \) is a vector of length $n_\birds$ with $i^{th}$ element equal to $(\covB_{im} - 1/2) / \omega_{im}$,
\item \( (1 \mid \latsB_{.-h}) \) is a matrix of dimension $n_\birds \times H$ representing a concatenation of a vector of 1 in the first column and the latent factors $\latsB$ excluding the $h^{th}$ one,
\item $\bm \beta_{m(-h)}$ is the vector $\bm \beta_m$ excluding the coefficient of the $h^{th}$ latent factor,
\item \( \displaystyle  \left(\frac{\bm L - 1/ 2}{\bm \omega} \right)_j \) is the diagonal matrix of dimension $n_\birds$ including the transformed versions of unit $j$'s interactions: $\big((L_{1j} - 1 /2) / \omega_{1j}^L, (L_{2j} - 1/2)/ \omega_{2j}^L, \dots, (L_{n_\birds j} - 1/2)/\omega_{n_\birds j}^L \big)$,
\item \(\displaystyle \left(1 \mid \latsB_{.-h}\latsP_{j(-h)} \right) \) is the $n_\birds \times H $ matrix with first column equal to 1, second column equal to $\latP_{j1} \latsB_{.1} = (\latB_{11} \latP_{j1}, \latB_{21} \latP_{j1}, \dots, \latB_{n_\birds 1} \latP_{j1})^T$, third column equal to $\latP_{j2} \latsB_{.2}$, up to the last column which is equal to $\latP_{jH} \latsB_{.H}$, {\it excluding} the $h^{th}$ vector $\latP_{jh}\latsB_{.h}$, and {\it always} using the same unit $j$'s latent factors, and
\item $\bm\lambda_{-h}$ includes the coefficients of the interaction model excluding $\lambda_h$.
\end{itemize}

\paragraph{\underline{Updating the parameters of the increasing shrinkage prior}}
In order to ease the updates of the increasing shrinkage prior parameters in \cref{eq:prior_shrinkage}, we introduce parameters $z_1, z_2, \dots, z_H$ with $z_h \sim \text{Multinomial}(\omega_1, \omega_2, \dots, \omega_H)$ and
$\theta_h | z_h \sim I(z_h \leq h) \delta_{\theta_{\infty}} + I(z_h > h) P_0$, similarly to \cite{Legramanti2020}. Then, updates of the parameters proceeds by updating the parameters $\{ v_h \}_h$ which deterministically set the values of $\{\omega_h\}_h$ and $\{\pi_h\}_h$, and updating the parameters $\{z_h\}_h$ and $\{\theta_h\}_h$.

First, updates for $v_h$ are performed conditional on $z_1, z_2, \dots, z_H$ by counting the number of $z$'s with values equal or greater than $h$: $v_h$ is sampled from a Beta distribution with parameters $\left( 1 + \sum_{h'=1}^H I(z_{h'} = h), \alpha + \sum_{h' = 1}^H I(z_{h'} > h) \right)$. Based on the sampled values for $v_1, v_2, \dots, v_H$, the values of $\omega_h$ are updated from their deterministic relationship in \cref{eq:prior_shrinkage}.

Then, the variance parameters $\theta_h$ are updated using the part of the prior that is the slab $P_0$ or the spike $\delta_{\theta_\infty}$ depending on the value of the corresponding $z_h$:
\begin{itemize}
\item If $z_h \leq h$ (which happens with prior probability $\sum_{l = 1}^h \omega_l = \pi_h$) the variance component $\theta_h$ belongs to the spike part of the prior, and it is set equal to $\theta_\infty$.
\item If $z_h > h$, then $\theta_h$ belongs to the $P_0$ part of the prior which is an inverse gamma distribution in our case, and $\theta_h$ is drawn from an inverse gamma with parameters $\alpha_\theta + (p_\birds + p_\plants + 3) / 2$ and $\beta_\theta + \left( \sum_m \beta_{mh}^2 / \tau_{mh}^\beta + \sum_l \gamma_{lh}^2 / \tau_{lh}^\gamma + \lambda_h^2 / \tau_h^\lambda + \delta_h^2 / \tau_h^\delta + \zeta_h^2/ \tau_h^\zeta \right) / 2$.
\end{itemize}

Lastly, the parameters $z_h$ are updated from a Multinomial distribution such that
\[
p(z_h = l \mid \cdot, - \bm \theta) \propto \begin{cases}
\omega_l \ \phi(\bm x; \theta_\infty \bm \Sigma) & \text{ for } l = 1, 2, \dots, h \\
\omega_l \ \tau(\bm x; 2\alpha_\theta, \beta_\theta/\alpha_\theta \bm \Sigma) & \text{ for } l = h+1, h+2, \dots, H,
\end{cases}
\]
where the vector $\bm x$ includes all coefficients of the $h^{th}$ latent factors: $\bm x = (\bm \beta_{.h}^T, \bm \gamma_{.h}^T, \lambda_h, \delta_h, \zeta_h)^T$, $\bm \Sigma$ is a diagonal matrix with entries $\big( (\tau_{.h}^\beta)^T, (\tau_{.h}^\gamma)^T, \tau_h^\lambda, \tau_h^\delta, \tau_h^\zeta \big)$, $\phi(\bm x;\theta_\infty \bm \Sigma)$ is the density of a normal distribution centered at 0 with covariance matrix $\theta_\infty \bm \Sigma$ evaluated at $\bm x$, and $\tau(\bm x;2 \alpha_\theta, \beta_\theta/\alpha_\theta \bm \Sigma)$ is the density of a multivariate $t$-distribution with $2\alpha_\theta$ degrees of freedom and covariance matrix $\beta_\theta / \alpha_\theta \bm \Sigma$ evaluated at $\bm x$. Note here that, even though the covariance matrix is diagonal, the density of the multivariate $t$-distribution is not the same as the sum of the densities from univariate $t$-distributions.

\paragraph{\underline{Updating the probability of observing an interaction}}
Since the conditional posterior distributions of $p_i$, and $q_j$ are not of known distributional form, we update them using Metropolis-Hastings. To update $p_i$:
\begin{itemize}
\item If $p_i^{(t)}$ is the value of $p_i$ at iteration $t$, propose new value $x$ from Beta$(np_i^{(t)}, n(1 - p_i^{(t)}))$.
\item Calculate the acceptance probability which is equal to
\begin{align*} AP & = \left\{ \prod_{j = 1}^{n_P} \left[ \frac{ xq_j}{ p_i^{(t)}q_j} \right]^{L_{ij} \sum_s (A_{ijs} F_{ijs} O_{ijs}) } \left[ \frac{1 - xq_j}{1 - p_i^{(t)}q_j} \right]^{L_{ij} \sum_{s} [(1 - A_{ijs}) F_{ijs} O_{ijs}]} \right\} \\
& \hspace{20pt} \times
\frac{\phi(\logit[x]; (1 \ \latB_i^T) \bm \delta, \ssq_{p, \birds})} {\phi(\logit[p_i^{(t)}]; (1 \ \latB_i^T) \bm \delta, \ssq_{p, \birds}) }
\times
\frac{b(p_i^{(t)}; nx, n(1 - x))}{b(x; np_i^{(t)}, n(1 - p_i^{(t)}))},
\end{align*}
where all other parameters are set to their most recent values, and $b(x; a, b)$ is the density of a Beta$(a,b)$ distribution evaluated at x.
\item Accept $x$ with probability $AP$, or stay at $p_i^{(t)}$ with probability $1 - AP$.
\end{itemize}
Similarly update the parameters $q_j$.

\paragraph{\underline{Updating the indicator of species occurrence in a study area}}

We assume that the entries in $\bm O^B$ are a priori independent, and we specify that each entry arises from a Bernoulli random variable with probability of success $\pi_{O^B_{is}}$, pre-specified. When deriving the conditional posterior distribution of $O^B_{is}$ we find that if $A_{ijs} = 1$ for some $j$, then $p(O^B_{is} = 1 \mid \cdot) = 1$. This makes sense since $A_{ijs} = 1$ for some $j$ means that study $s$ recorded an interaction of species $i$, which necessarily implies that the species exists in the area. Now, if $A_{ijs} = 0$ for all $j$, then $O^B_{is}$ is drawn from a Bernoulli distribution with
\[
p(O^B_{is} = o \mid \cdot) \propto \begin{cases}
\pi_{O^B_{is}} \prod_{j = 1}^{n_\plants} (1 - p_iq_j)^{L_{ij} F_{ijs}}
& \text{if } o = 1, \text{ and} \\
1 - \pi_{O^B_{is}},
& \text{if } o = 0.
\end{cases}
\]
This formula also makes intuitive sense. Essentially, if there are many species $j$ with which $i$ interacts and for which study $s$ could have observed an interaction, observing no interactions of $i$ must mean that the species do not occur.

\paragraph{\underline{Updating the latent factor covariance parameter}}

We update the parameters $\rho_\latB, \rho_\latP$ using a Metropolis-Hastings step (similarly to the probability of detecting species). Specifically:
\begin{itemize}
\item If $\rho_\latB^{(t)}$ is the value of $\rho_\latB$ at iteration $t$, propose new value $x$ from Beta$(n \rho_\latB^{(t)}, n(1 - \rho_\latB^{(t)}))$.
\item Calculate the current and proposed value for the correlation matrix and denote them by $\bm \Sigma_\latB^{(t)}$ and $\bm \Sigma_\latB^x$, respectively.
\item Calculate the acceptance probability which is equal to
$$AP = \frac{ \prod_{h = 1}^H \phi(\latsB_{.h}; \bm 0, \bm \Sigma_\latB^{(t)}) }
{\prod_{h = 1}^H \phi(\latsB_{.h}; \bm 0, \bm \Sigma_\latB^x)}
\frac{b(\rho_\latB^{(t)} ; \alpha_\rho, \beta_\rho)}{b(x ; \alpha_\rho, \beta_\rho)}$$ 
where all other parameters are set to their most recent values.
\item Accept $x$ with probability $AP$, or stay at $\rho_\latB^{(t)}$ with probability $1 - AP$.
\end{itemize}
Update $\rho_\latP$ is a similar manner.

\paragraph{\underline{Update missing values of covariates}}
If a covariate includes missing values for a subject of units, missing value imputation is straightforward and proceeds by drawing missing covariate values conditional on parameters and latent factors from \cref{eq:model_covariates}.

\vspace{20pt}
\paragraph{Note on the importance of study focus and co-occurrence}
We would like to note here that throughout the MCMC, the focus of a study (in whether they are animal- or plant-oriented, denoted by $\bm F$) and the species that occur in the study area (denoted by $\bm O$) only play a role through their product (see for example the update of $L_{ij}$ above). That provides us with some flexibility for the specification of the species occurrence probabilities for species and studies for which the focus is not on these species. Specifically, if $F_{ijs} = 0$ for some pair of species $i,j$ in study $s$, the value of $O_{ijs}$ will play no role.

\subsection{Choice of hyperparameter values}

The choices of hyperparameters used for the simulation and analysis results are included in the table below.

\begin{longtable}{lcc}
\caption{Choice of hyperparameters} \label{app_tab:hyper} \\
\multicolumn{2}{l}{Parameter} & Value \\ \hline
Number of latent factors & $H$ & 5 \\
Parameter in increasing shrinkage prior & $\theta_\infty$ & 0.01 \\
Parameter in increasing shrinkage prior & $\alpha$ & 5 \\
Parameters in inverse Gamma $P_0$ for the increasing shrinkage prior & $\alpha_\theta, \beta_\theta$ &  1, 1 \\
Hyperparameter in inverse Gamma prior on the parameters $\tau$ & $\nu$ & 5 \\
Hyperparameters in Beta prior on $\rho_\latB, \rho_\latP$ & $a_\rho, b_\rho$ & 5, 5 \\
Prior mean and variance coefficients & $\mu_0, \ssq_0$ & 0, 10 \\
Hyperparameters in inverse Gamma prior on the residual variance & $a_Y, b_Y$ & 1, 1 \\
Parameter used in proposal distribution of Metropolis-Hastings steps & $n$ & 100 \\ \hline
\end{longtable}

\section{Variable importance in latent factor network models
}
\label{supp_sec:variable_importance}

Since the latent factors are not identifiable parameters, our approach to investigate variable importance in network models is based on the posterior distribution of the probability of interaction in \cref{eq:model_probinter}. Specifically, assume we want to investigate the importance of the $k^{th}$ covariate for the first set of species, $\covsB_{.k} = (\covB_{1k}, \covB_{2k}, \dots, \covB_{n_\birds k})^T$. Let $l_{ij}^{(r)}$ denote the logit of the fitted probability of interaction between species $i$ and $j$ at the $r^{th}$ iteration of the MCMC and $\bm l_{.j}^{(r)}$ denote the vector of probabilities $l_{ij}^{(r)}$ for all $i$. In what follows we assume that species with missing information on the $k^{th}$ covariate are excluded from both $\covsB_{.k}$ and $\bm l_{.j}^{(r)}$. Then,
\begin{enumerate}
\item For each posterior sample $r$ and species $j$, we calculate the square correlation between  $\bm l_{.j}^{(r)}$ and $\covsB_{.k}$. We denote the value by $T_{jk}^{*(r)}$.
\item We average the values of $T_{jk}^{*(r)}$ across species $j$ and iterations $r$ to acquire $T_k^*$.
\item For a large number of permutations $B$, do
\begin{enumerate}
\item Permute the entries in the vector $\covsB_{.k}$, ``breaking'' any relationship between the probabilities of interaction and the covariate.
\item Perform steps 1-2 using the permuted vector to acquire $T_k^{*(b)}$.
\end{enumerate}
\item Calculate the mean and standard deviation of $T_k^{*(b)}$ across the $B$ permutations.
\item Use $\big[ T_k^* - {\rm mean}(T_k^{*(b)}) \big] \ / \ {\rm sd}(T_k^{*(b)})$ as a measure of variable importance, separately among continuous and binary covariates.
\end{enumerate}
Simulations for this measure of variable importance are shown in Supplement \ref{supp_subsec:simulations_traits}.

We found that this procedure performed worse when species with missing covariate values are {\it included} by using their imputed values (imputed during the MCMC). We believe that this happens because the latent factors are informed by the true interactions in $\bm L$ and they themselves play a role in the imputation of missing covariates. Therefore, a variable importance procedure that uses the imputed values will necessarily bias our understanding about the presence or magnitude of a link between the covariate and interactions. For this reason, our variable importance metric is evaluated only with species for which the corresponding covariate is measured.

\section{Alternative models}
\label{supp_sec:other_methods}

We consider a number of alternative models that vary whether they use latent factors or covariates directly, whether they use the shrinkage prior or not, and whether they accommodate false negatives or not. Any overlap in the notation of coefficients across the models can be ignored. The model that uses latent factors, accommodates false negatives and incorporates the shrinkage prior is the proposed one in \cref{sec:model}. Here, we present the other models.

\subsection{Model that uses covariates directly and accommodates false negatives}
\label{supp_subsec:alt_model_covariates}

\subsubsection{Specification}

The first alternative model we consider includes covariates directly in model components' linear predictors and accommodates false negatives, under the assumption that the independence statements discussed in \cref{supp_sec:likelihood} hold conditional on measured covariates only. This corresponds to the {\sffamily Covariates, bias corrected} model we refer to in the manuscript. The conditional probability of a recorded interaction is specified as in \cref{eq:model_probA_givenL}:
\begin{align*}
& P(A_{ijs} = 1 \mid L_{ij} = l, F_{ijs} = f, O_{ijs} = o, p_i, q_j) =
\begin{cases*}
0,        & if $lfo = 0$, \quad and \\
p_iq_j,   & if $lfo = 1$,
\end{cases*}
\end{align*}
but the submodels are specified using measured covariates only as:
\begin{equation}
\begin{aligned}
& \mathrm{logit} P(L_{ij} = 1 \mid \covsB_i, \covsP_j) = \alpha_0 + \covsB_i^T \bm \alpha_{\covB} + \covsP_j^T \bm \alpha_{\covP} \\
& \logit[p_i] \mid \covsB_i \sim \mathcal{N} (\delta_0 + \covsB_i^T \bm \delta, \ssq_{p,\birds}) \\
& \logit[q_j] \mid \covsP_j \sim \mathcal{N} (\zeta_0 + \covsP_j^T \bm \zeta, \ssq_{q,\plants}).
\end{aligned}
\label{supp_eq:alt_model_covariates}
\end{equation}
The specification in the second and third line of \cref{supp_eq:alt_model_covariates} resembles that of the probability of detection in \cref{eq:model_probobs}, but the latent factors are substituted by the covariates. The latent factors are also substituted by covariates in the linear predictor of the interaction model (first line).
If the covariates include missing values, we extend \cref{supp_eq:alt_model_covariates} to specify $E[\covB_{im}] = \mu_m$ and $E[\covP_{jl}] = \mu_l$. If the covariate is continuous, we assume it is normally distributed with variance $\ssq_m$ and $\ssq_l$ respectively. Doing so allows us to impute missing covariate values. We assume normal and inverse gamma prior distributions on coefficients and variance terms.

\subsubsection{MCMC}

We employ an MCMC scheme that resembles the one for the proposed approach in Supplement \ref{supp_sec:MCMC}: it uses the P\'olya-Gamma data-augmentation of \cite{Polson2013bayesian} to update model parameters of the interaction model, Gibbs updates for the parameters of the probability of observing models, and Metropolis-Hastings steps for updating the actual probabilities of observing a species. Specifically:
\begin{itemize}[label=--, leftmargin=*]

\item The update of the true interaction matrix with entries $L_{ij}$ proceeds exactly as in Supplement \ref{supp_sec:MCMC}, but for $p_{ij}^L = \text{expit} \left\{ \alpha_0 + \covsB_i^T \bm \alpha_{\covB} + \covsP_j^T \bm \alpha_{\covP} \right\} $.

\item We update the parameters of the interaction model $(\alpha_0, \bm \alpha_{\covB}, \bm \alpha_{\covP})$ using P\'olya-Gamma data-augmentation. For each $(i,j)$ pair, we draw latent variables $\omega_{ij}^L \sim \text{PG}(1, \alpha_0 + \covsB_i^T \bm \alpha_{\covB} + \covsP_j^T \bm \alpha_{\covP})$. Conditional on $\omega_{ij}^L$, we sample $(\alpha_0, \bm \alpha_{\covB}, \bm \alpha_{\covP}) \sim \mathcal{N} (\bm \mu_{new}, \bm \Sigma_{new})$ for parameters
\[
\bm \Sigma_{new} = \left[\bm D^T \bm \Omega^L \bm D + \bm \Sigma_0^{-1} \right]^{-1}, \text{ and }
\bm \mu_{new} = \bm \Sigma_{new} \left[\bm D^T \left(\text{vec}(L) - 1 / 2 \right) + \bm \Sigma_0^{-1} \bm \mu_0 \right],
\]
where
\begin{enumerate}
\item $\bm D$ is a matrix with $(n_\birds \times n_\plants)$ rows and $(p_\birds + p_\plants + 1)$ columns, with first column equal to 1, each of the next $p_\birds$ columns equal to $(\underbrace{\covB_{1m}, \covB_{1m}, \dots, \covB_{1m}}_{n_\plants \text{ times}}, \covB_{2m}, \dots, \covB_{2m}, \dots,  \covB_{n_\birds m})$ for $m = 1, 2, \dots, p_\birds$ (the $i^{th}$ entry of the $m^{th}$ covariate is repeated $n_\plants$ number of times), and the next $p_\plants$ columns are $(\covP_{1l}, \covP_{2l}, \dots, \covP_{n_\plants l}, \covP_{1l}, \dots, \covP_{n_\plants l}, \dots \covP_{n_\plants l})$ for $l = 1, 2, \dots, p_\plants$ (the vector of the $l^{th}$ covariate is repeated $n_\birds$ times).
\item $\bm \Omega^L$ is a matrix of dimension $(n_\birds n_\plants \times n_\birds n_\plants)$ with the entries $\text{vec}(\omega_{ij}^L)$ on the diagonal and 0 everywhere else,
\item $\bm \Sigma_0 = \ssq_0 \bm I_{1 + p_\birds + p_\plants}$ is a diagonal matrix with prior variances, and
\item $\bm \mu_0$ is the vector $\bm 0$ of length $1 + p_\birds + p_\plants$ including prior means.
\end{enumerate}

\item We update the parameters of the model for the probability of observing a species interaction by sampling $ (\delta_0, \bm \delta^T)^T \sim \mathcal{N}(\bm \mu_{new}, \bm \Sigma_{new}),$
where
$$ \bm \Sigma_{new} = \left[ \widetilde{\covsB}^T \widetilde{\covsB} / \ssq_{p,\birds} + \bm \Sigma_0^{-1} \right]^{-1} $$
and
$$ \bm \mu_{new} = \bm \Sigma_{new} \left[\widetilde{\covsB}^T [\logit[p]]_{i=1}^{n_\birds} / \ssq_{p,\birds} + \bm \Sigma_0^{-1} \bm \mu_0 \right],$$
where $\widetilde{\covsB} = (\bm 1 \mid \covsB_{.1} \mid \covsB_{.2} \mid \dots \mid \covsB_{.p_\birds})$ is of dimension $n_\birds \times (p_\birds + 1)$, $\bm \Sigma_0 = \ssq_0 \bm I_{p_\birds + 1}$ is the diagonal matrix of prior variances, $\bm \mu_0 = \bm 0$ is the vector of prior means, and $[\logit[p]]_{i=1}^{n_\birds}$ is the vector of length $n_\birds$ including the entries $\logit[p_i]$.

To update the residual variance, we sample $\ssq_{p,\birds}$ from an inverse gamma distribution with parameters $a_0 + n_\birds / 2$ and $b_0 + \sum_{i = 1}^{n_\birds} (\logit[p_i] - \delta_0 - \sum_{m = 1}^{p_\birds} \delta_m \covB_{im})^2 / 2$, where $a_0, b_0$ are the parameters of the inverse gamma prior distribution on $\ssq_{p,\birds}$.

Similarly we update the parameters for the probability of observing an interaction for the second set of species.

\item The updates for the probability of observing an interaction $\logit[p_i], \logit[q_j]$ proceed exactly as in Supplement \ref{supp_sec:MCMC}, for latent factors substituted by the observed covariates.

\item Lastly, if covariates include missing values, models for the covariates are specified which include only an intercept if the covariate is binary, and an intercept and variance term if the covariate is continuous. In the presence of missing data, at each MCMC iteration we update the parameters (intercepts, residual variances for continuous covariates), and impute missing covariate values:
\begin{enumerate}

\item For continuous trait $m$ with $\covB_{im} \sim \mathcal{N}(\mu_m, \ssq_m)$, and priors $\mathcal{N}(\mu_0, \ssq_0)$ and $IG(a_0, b_0)$ for the mean and variance parameters, we update $\mu_m$ from $\mathcal{N}(\mu_{new}, \ssq_{new})$ where $\ssq_{new} = [n_\birds / \ssq_m + (\ssq_0)^{-1} ]^{-1}$, and $\mu_{new} = \ssq_{new}[\sum_{i = 1}^{n_\birds} \covB_{im} / \ssq_m + \mu_0 / \ssq_0]$, and update $\ssq_m$ from and inverse gamma distribution with parameters $a_0 + n_\birds / 2$ and $b_0 + \sum_{i = 1}^{n_\birds} (\covB_{im} - \mu_m)^2 / 2$. For these updates, the full vector $\covsB_{.m}$ is used, including the most current values of the imputed entries.

\item For the continuous trait $m$, we draw new values for $\covB_{im}$ if this entry was missing from $\mathcal{N}(\mu_{i, new}, \ssq_{i, new})$ where
\begin{align*}
\ssq_{i,new} &= \Big(\alpha_{\covB m}^2 \sum_{j = 1}^{n_\plants} \omega_{ij}^L + \delta_m ^2 / \ssq_{p,\birds} + 1 / \ssq_m \Big)^{-1}, \\[10pt]
\mu_{i,new} &= \ssq_{i,new} \Big[
\alpha_{\covB m} \sum_{j = 1} ^ {n_\plants} (L_{ij} - 1/2 - \omega_{ij}^L (\alpha_0 + \covsB_{i(-m)}^T \bm \alpha_{\covB (-m)} + \covsP_j^T \bm \alpha_{\covP}) +\\
& \hspace{50pt}
\delta_m (\logit[p_i] - \delta_0 - \covsB_{i(-m)}^T \bm \delta_{-m}) / \ssq_{p,\birds} + \mu_m / \ssq_m
\Big],
\end{align*}
$\omega_{ij}^L$ are the PG draws discussed above, and the subscript $(-m)$ reflects that the $m^{th}$ covariate (or its coefficient) is excluded.

\item For binary traits $m$ with $\covB_{im} \sim \text{Bern}(\mu_m)$ we assume a normal prior on $\logit[\mu_m]$ with mean $\mu_0$ and variance $\ssq_0$. For the current values of $\mu_m$, we draw $n_\birds$ values from a P\'olya-Gamma$(1, \logit[\mu_m])$ distribution, denoted by $\omega_{im}, i = 1 , 2, \dots, n_\birds$. We sample a new value for $\logit[\mu_m]$ from $\mathcal{N}(\mu_{new}, \ssq_{new})$ where
$\ssq_{new} = [\sum_{i = 1}^{n_\birds} \omega_{im} + (\ssq_0)^{-1}]^{-1}$
and $\mu_{new} = \ssq_{new}[ \sum_{i =1}^{n_\birds} (\covB_{im} - 1 / 2) + (\ssq_0)^{-1}\mu_0]$.

\item 
For binary covariates, if $\covB_{im}$ is missing, we calculate
$p_{imx} = p(\covB_{im} = x \mid \cdot)$ for $x \in \{0, 1\}$ 
(up to a constant)
$$
p_{imx} \propto \left[ \prod_{j = 1}^{n_\plants} (p_{ij}^L)^{L_{ij}} (1 - p_{ij}^L)^{1 - L_{ij}} \right]
p(\logit[p_i] \mid \bm \delta, \covsB_i, \ssq_{p.\birds})
[\mu_m^x(1 - \mu_m)^{1-x}]
,$$
where $p_{ij}^L$ and the likelihood for the $p_i$ model are calculated by setting $\covB_{im} = x$,
and set $\covB_{im}$ equal to $x$ with probability $p_{imx} / (p_{im0} + p_{im1})$.
\end{enumerate}

Updates for the missing covariates of the second set of species are identical, and the updates of all other parameters always use the most recent imputations of the missing covariate values.

\end{itemize}

\subsection{Model that uses covariates directly but does not accommodate false negatives}

\subsubsection{Specification}

An alternative model we consider resembles the one in Supplement \ref{supp_subsec:alt_model_covariates} but assumes that there are no false negatives. We set $L_{ij} = 0$ if $A_{ijs} = 0$ for all $s$, and $L_{ij} = 1$ if $A_{ijs} = 1$ for at least one $s$. Therefore, this model consists solely of the first line in \cref{supp_eq:alt_model_covariates}.

\subsubsection{MCMC}

The MCMC scheme for this model includes solely the updates of the interaction model parameters, which are the ones in Supplement \ref{supp_subsec:alt_model_covariates}. In the presence of missing covariate values, models for these covariates are assumed and updates of these models' parameters are identical to the ones in Supplement \ref{supp_subsec:alt_model_covariates}. However, covariate value imputation is slightly different, since here we do not assume a model for the probability of observation that depends on covariates. Therefore, covariate value imputation is like in Supplement \ref{supp_subsec:alt_model_covariates}, but excluding the term from $\mu_{i,new}$ and $p_{imx}$ corresponding to the $p_i, q_j$-submodel.

\subsection{Model that uses latent factors but does not accommodate false negatives}

\subsubsection{Specification}

We consider a version of our model that ignores the presence of false negatives, but maintains the use of latent factors to link the presence of an interaction and the model for the covariates. Therefore, 
assuming that $L$ is known with entries $L_{ij} = 0$ if $A_{ijs} = 0$ for all $s$, and $L_{ij} = 1$ if $A_{ijs} = 1$ for at least one $s$,
this model specifies
\begin{equation}
\begin{aligned}
& \mathrm{logit} P(L_{ij} = 1 \mid \covsB_i, \latsB_i, \covsP_j, \latsP_j) = \lambda_0 + \sum_{h=1}^H \lambda_h \latB_{ih} \latP_{jh}, \\
& f_m^{-1}(E(\covB_{im} \mid \latsB_i)) = \beta_{m0} +  \latsB_i' \bm \beta_m, \ m = 1, 2, \dots, p_\birds, \text{ and} \\
&
g_l^{-1}(E(\covP_{jl} \mid \latsP_j)) = \gamma_{l0} +  \latsP_j' \bm \gamma_l , \ l = 1, 2, \dots, p_\plants
\end{aligned}
\end{equation}

\subsubsection{MCMC}

The updates for the parameters in the traits models for binary or continuous traits, the parameters $\lambda_0, \bm \lambda$ of the interaction model, the parameters in the covariance matrix of the latent factors $\rho_\latB, \rho_\latP$, and the variance scaling parameters $\tau$ are the same as in Supplement \ref{supp_sec:MCMC}. Therefore, we only have to discuss updates for the latent factors and the increasing shrinkage prior:
\begin{itemize}[label=--,leftmargin=*]
\item To update the latent factors the MCMC proceeds with an update similar to the one in Supplement \ref{supp_sec:MCMC} but accommodating the fact that the latent factors are no longer involved in the model for the probability of observing an interaction from a given species. We describe the update of the latent factors for the first set of units $\latsB_{.h}$ for $h = 1, 2, \dots, H$, and updates for $\latsP_{.h}$ are similar. We use the P\'olya-Gamma draws $\omega_{im}$ for binary traits $m$, and $\omega_{ij}^L = \omega_{ij}^A$ from the interaction model. For each $h = 1,2, \dots, H$, $\latsB_{.h}$ is drawn from $\mathcal{N}_{n_\birds}(\bm \mu_{new}, \bm \Sigma_{new})$ for parameters
\[
\bm \Sigma_{new} = \left( \sum_{\substack{\text{m: } \covB_m \\ \text{continuous}}} \beta_{mh}^2 / \ssq_m \bm I_{n_\birds} + \sum_{\substack{\text{m: }\covB_m \\ \text{binary}}} \beta_{mh}^2 \bm \Omega_m + \sum_{j = 1}^{n_\plants} \lambda_h^2 \latP_{jh}^2 \bm \Omega_j^L + \bm \Sigma_\latB^{-1} \right)^{-1}
\]
and
\begin{align*}
\bm \mu_{new} = \bm \Sigma_{new} \Bigg\{ \sum_{\substack{\text{m: } \covB_m \\ \text{continuous}}} & \beta_{mh} / \ssq_m \ \textbf{part}(m,h) + \\
&
\sum_{\substack{\text{m: } \covB_m \\ \text{binary}}} \beta_{mh} \bm \Omega_m \left[ \left(\frac{\covsB - 1/ 2}{\bm \omega} \right)_m - \left(1 \mid \latsB_{.-h} \right) \bm \beta_{m(-h)} \right] +\\
&
\sum_{j = 1}^{n_\plants} \lambda_h \latP_{jh} \bm \Omega_j \left[ \left(\frac{\bm L - 1/ 2}{\bm \omega} \right)_j - \left(1 \mid \latsB_{.-h}\latsP_{j(-h)} \right) \bm \lambda_{-h} ]\right]
\Bigg\}
\end{align*}
where \(\displaystyle \bm \Omega_j, \bm \Omega_m, \bm \Sigma_{\latB}, \textbf{part}(m,h),
\left(\frac{\covsB - 1/ 2}{\bm \omega} \right)_m, (1 \mid \latsB_{.-h}), \bm \beta_{m(-h)}, \left(\frac{\bm L - 1/ 2}{\bm \omega} \right)_j , \) \(\displaystyle \left(1 \mid \latsB_{.-h}\latsP_{j(-h)} \right) ,\) and  \( \bm\lambda_{-h} \) are defined in Supplement \ref{supp_sec:MCMC}.

\item The updates for the increasing shrinkage prior are as in Supplement \ref{supp_sec:MCMC} with two exceptions:
\begin{itemize}[label=$\cdot$]
\item For the update of $\theta_h$, if $z_h > h$, then $\theta_h$ is drawn from an inverse gamma with parameters $\alpha_\theta + (p_\birds + p_\plants + 1) / 2$ and $\beta_\theta + \left( \sum_m \beta_{mh}^2 / \tau_{mh}^\beta + \sum_l \gamma_{lh}^2 / \tau_{lh}^\gamma + \lambda_h^2 / \tau_h^\lambda \right) / 2$.

\item For the update of $z_h$: $z_h$ are updated from a Multinomial distribution such that
\[
p(z_h = l \mid \cdot, - \bm \theta) \propto \begin{cases}
\omega_l \ \phi(\bm x; \theta_\infty \bm \Sigma) & \text{ for } l = 1, 2, \dots, h \\
\omega_l \ \tau(\bm x; 2\alpha_\theta, \beta_\theta/\alpha_\theta \bm \Sigma) & \text{ for } l = h+1, h+2, \dots, H,
\end{cases}
\]
where $\bm x = (\bm \beta_{.h}^T, \bm \gamma_{.h}^T, \lambda_h)^T$, and $\bm \Sigma$ is a diagonal matrix with entries $\big( (\tau_{.h}^\beta)^T, (\tau_{.h}^\gamma)^T, \tau_h^\lambda \big)$.
\end{itemize}

\end{itemize}

\subsection{Our model without the shrinkage prior}

\subsubsection{Specification}
We investigate how the proposed model performs when $H$ is fixed to a relatively small value, and the shrinkage prior is not used. Specifically, we alter the prior specification on the parameters $\theta_h$ and specify $\theta_h \overset{iid}{\sim} IG(\alpha_\theta, \beta_\theta).$
\subsubsection{MCMC}
The MCMC for this model can be easily implemented using the original scheme in Supplement \ref{supp_sec:MCMC} but setting $\pi_h = 0$ for all $h$.

\subsection{Our model without the parameter-specific variance flexibility}

This model corresponds to the proposed model in \cref{sec:model}, though all $\tau$ parameters are excluded. The MCMC proceeds identically, though all parameters $\tau$ are set to 1 and are not updated.

\subsection{The modularized version of our model}

\subsubsection{Specification}
We consider a modularized version of our model that ``learns'' the latent factors based on the covariate submodel \cref{eq:model_covariates} only. Specifically, even though the latent factors are directly included in the interaction and detectability submodels, the model fit for these submodels is not considered for learning the latent factors, hence ``cutting the feedback'' from the interaction and detectability submodels back into the latent factors. In a sense, this version of the model uses the latent factor representation simply as a dimension reduction technique on the measured covariates, and uses the learnt latent factors in the other models.

\subsubsection{MCMC}
The MCMC for this model is based on the MCMC for the original model described in \cref{supp_sec:MCMC}. The only difference is that updates for the latent factors $\latsB, \latsP$ are performed while dropping {\it any} terms that involve interactions or detectability. Specifically, the update for the $h^{th}$ latent factor of the bird species is:
$\latsB_{.h} \sim \mathcal{N}_{n_\birds}(\bm \mu_{new}, \bm \Sigma_{new})$ for parameters
\[
\bm \Sigma_{new} = \left( \sum_{\substack{\text{m: }\covB_m \\ \text{continuous}}} \beta_{mh}^2 / \ssq_m \bm I_{n_\birds} +
\sum_{\substack{\text{m: }\covB_m \\ \text{binary}}} \beta_{mh}^2 \bm \Omega_m  +
\bm \Sigma_\latB^{-1} \right)^{-1}
\]
and
\begin{align*}
\bm \mu_{new} = \bm \Sigma_{new} \Bigg\{ \sum_{\substack{\text{m: }\covB_m \\ \text{continuous}}} & \beta_{mh} / \ssq_m \ \textbf{part}(m,h) + \\
&
\sum_{\substack{\text{m: } \covB_m \\ \text{binary}}} \beta_{mh} \bm \Omega_m \left[ \left(\frac{\covsB - 1/ 2}{\bm \omega} \right)_m - \left(1 \mid \latsB_{.-h} \right) \bm \beta_{m(-h)} \right] 
\Bigg\}.
\end{align*}
Similarly for $\latsP_{.h}$.

\subsection{Model with covariates and latent factors in the interaction model}

\subsubsection{Specification}
In classic network models (without bias correction for taxonomic or geographical bias) a model that uses both covariates and latent factors is sometimes used. We investigate a related model in our simulations, which uses both covariates and latent factors in the network model, but also performs bias correction for the two sources of bias. Again, we note here that this form of the model (with the bias correction) is of our own construction and to our knowledge does not exist in the literature. Our primary goal for considering this model is to evaluate its performance of the model in identifying covariates that are important for the formation of possible interactions. Specifically, we alter the first line of \cref{supp_eq:alt_model_covariates} to specify
$$
\mathrm{logit} P(L_{ij} = 1 \mid \covsB_i, \covsP_j, \latsB_i, \latsP_j) = \alpha_0 + \covsB_i^T \bm \alpha_{\covB} + \covsP_j^T \bm \alpha_{\covP} + \sum_{h = 1}^H \lambda_h \latB_{ih} \latP_{jh}.
$$
Here, we fix the number of latent factors to $H = 3$, considering that there is already a large number of predictors in the interaction submodel.
We keep the other two lines of \cref{supp_eq:alt_model_covariates} on species detectability as they are. We assume normal and inverse gamma prior distributions on coefficients and variance terms (including the coefficients of the latent factors).

\subsubsection{MCMC}

The MCMC scheme for this model resembles the one in Supplement \ref{supp_subsec:alt_model_covariates}, though we also need to update the latent factors at each MCMC iteration. The differences in the MCMC scheme are summarized below:

\begin{itemize}[label=--, leftmargin=*]

\item The update of the true interaction matrix with entries $L_{ij}$ proceeds exactly as in Supplement \ref{supp_sec:MCMC}, but for $p_{ij}^L = \text{expit} \left\{ \alpha_0 + \covsB_i^T \bm \alpha_{\covB} + \covsP_j^T \bm \alpha_{\covP} + \sum_{h = 1}^H \lambda_h \latB_{ih} \latP_{jh} \right\} $.

\item The parameters of the interaction model $(\alpha_0, \bm \alpha_{\covB}, \bm \alpha_{\covP}, \bm \lambda)$ are updated exactly like in \cref{supp_subsec:alt_model_covariates} with the matrix $\bm D$ updated to include the $H$ vectors $$(\latB_{1h} \latP_{1h}, \latB_{1h} \latP_{2h}, \dots, \latB_{1h} \latP_{n_\plants h}, \latB_{2h}V_{1h}, \dots, \latB_{n_\birds, h} \latP_{n_\plants h})$$
and the dimensions of $\mu_0, \Sigma_0$ are increased by $H$.

\item Lastly, if covariates include missing values, the updates for the parameters of the covariate models are the same as in Supplement \cref{supp_subsec:alt_model_covariates}. What changes is the imputation of the missing values. For binary covariates, the imputation is the same, though using the value of $p_{ij}^L$ defined above which includes both covariates and latent factors.
For the continuous trait $m$, we draw new values for $\covB_{im}$ if this entry was missing from $\mathcal{N}(\mu_{i, new}, \ssq_{i, new})$ where \( \displaystyle \ssq_{i,new} = \Big(\alpha_{\covB m}^2 \sum_{j = 1}^{n_\plants} \omega_{ij}^L + \delta_m ^2 / \ssq_{p,\birds} + 1 / \ssq_m \Big)^{-1} \) and $\mu_{i,new}$ is equal to
\begin{align*}
& \ssq_{i,new} \Big[
\alpha_{\covB m} \sum_{j = 1} ^ {n_\plants} \Big (L_{ij} - 1/2 - \omega_{ij}^L (\alpha_0 + \covsB_{i(-m)}^T \bm \alpha_{\covB (-m)} + \covsP_j^T \bm \alpha_{\covP} + \sum_{h = 1}^H \lambda_h \latB_{ih}\latP_{jh} \Big) +\\
& \hspace{50pt}
\delta_m (\logit[p_i] - \delta_0 - \covsB_{i(-m)}^T \bm \delta_{-m}) / \ssq_{p,\birds} + \mu_m / \ssq_m
\Big],
\end{align*}
for $\omega_{ij}^L$ the PG draws, and the subscript $(-m)$ reflecting that the $m^{th}$ covariate (or its coefficient) is excluded.
Updates for the missing covariates of the second set of species are identical, and the updates of all other parameters always use the most recent imputations of the missing covariate values.

\item The correlation parameters $\rho_\latB, \rho_\latP$ of the latent factors are updated as in Supplement \ref{supp_sec:MCMC}.

\item For each $h$, we draw $\latB_{.h}$ from a multivariate normal distribution with mean $\bm \mu_{new}$ and variance $\bm \Sigma_{new}$ where
\[
\bm \Sigma_{new} = \left( \sum_{j = 1}^{n_\plants} \lambda_h^2 \latP_{jh}^2 \bm \Omega_j^L + \bm \Sigma_\latB^{-1} \right)^{-1}
\]
and
\begin{align*}
\bm \Sigma_{new}^{-1} \ \bm \mu_{new} =
&
\sum_{j = 1}^{n_\plants} \lambda_h \latP_{jh} \bm \Omega_j^L \left[ \left(\frac{\bm L - 1/ 2}{\bm \omega} \right)_j - \left(1 \ \covsB \ \covsP_{j} \ \latsB_{.-h}\latsP_{j(-h)} \right) (\alpha_0, \bm \alpha_{\covsB}^T, \bm \alpha_{\covsP}^T, \bm \lambda_{-h}^T)^T \right]
\end{align*}
for \( \bm \Omega_j^L, \bm \Sigma_{\latB},  \left(\frac{\bm L - 1/ 2}{\bm \omega} \right)_j, \bm\lambda_{-h} \) already defined in the updates of the latent factors in Supplement \ref{supp_sec:MCMC}, and
\(\displaystyle \left(1 \ \covsB \ \covsP_{j} \ \latsB_{.-h}\latsP_{j(-h)} \right) \) is the $n_\birds \times (p_{\birds} + p_\plants + H)$ matrix with first column equal to 1, the next $p_\birds$ including the covariates of the bird species, the next $p_\plants$ columns repeating the covariates of the $j^{th}$ plant $n_\birds$ times across the rows, and the last $H-1$ columns equal to $\latP_{j1} \latsB_{.1} = (\latB_{11} \latP_{j1}, \latB_{21} \latP_{j1}, \dots, \latB_{n_\birds 1} \latP_{j1})^T$, then $\latP_{j2} \latsB_{.2}$, up to the last column which is equal to $\latP_{jH} \latsB_{.H}$, {\it excluding} the $h^{th}$ vector $\latP_{jh}\latsB_{.h}$, and {\it always} using the same unit $j$'s latent factors.

\end{itemize}

% -------- Describing the alternative models we consider -------- %

\section{Simulations: Additional information and results}
\label{supp_sec:simulations}

\subsection{The data generative mechanisms as the choice of important covariates}

There are 10 covariates in $\widetilde{\covsB}$, 2 measured and continuous, 3 measured and binary, 2 unmeasured and continuous, and 3 unmeasured and binary. Similarly there are 24 covariates in $\widetilde{\covsP}$, the first 4 of them measured and continuous, 8 measured and binary, and another set of 4 and 8 of unmeasured continuous and binary covariates. Visually for the bird species:
$$
\widetilde{\covsB}_i = \Big( 
\underbrace{
\underbrace{\widetilde{\covB}_{i1}, \widetilde{\covB}_{i2}}_{\text{continuous}},
\underbrace{\widetilde{\covB}_{i3}, \widetilde{\covB}_{i4}, \widetilde{\covB}_{i5}}_{\text{binary}}}_{\text{measured}},
\underbrace{\underbrace{\widetilde{\covB}_{i6}, \widetilde{\covB}_{i7}}_{\text{continuous}}, \underbrace{\widetilde{\covB}_{i8}, \widetilde{\covB}_{i9}, \widetilde{\covB}_{i10}}_{\text{binary}}}_{\text{unmeasured}} \Big)^T,
$$
and similarly for the plant species.
Intuitively, we would like the covariates $\widetilde{\covsB}, \widetilde{\covsP}$ to include the same amount of information for predicting the true interactions and detectability across DGMs, so that we can attribute differences in performance across DGMs directly to the method rather than the amount of available information to begin with. We do so by equating the number of measured and unmeasured covariates by type and always using the same number of binary and continuous covariates in the different submodels. This led to true interaction models that achieved the same AUROC across all simulation scenarios (approximately equal to 0.78).

Different subsets of these covariates are important in the different DGMs of \cref{tab:sims_important_variables}. In all DGMs, the interaction indicators depend on the covariates in a multiplicative way. For example, in DGM1 we have
$$
\logit[P(L_{ij} = 1)] =  \kappa_0 +
\kappa_1 \covB_{i1}\covP_{j1} +
\kappa_2 \covB_{i3}\covP_{j3} +
\kappa_3 \covB_{i4}\covP_{j4} +
\kappa_4 \covB_{i5}\covP_{j5},
$$
whereas in DGM2 we have
$$
\logit[P(L_{ij} = 1)] =  \kappa_0 +
\kappa_1 \covB_{i1}\covP_{j13} +
\kappa_2 \covB_{i8}\covP_{j3} +
\kappa_3 \covB_{i4}\covP_{j4} +
\kappa_4 \covB_{i5}\covP_{j5}, 
$$
and so on. Therefore, in a subset of the DGMs, the interaction indicators depend on interactions among covariates some or all of which might be unmeasured.
Across DGMs, the detactability scores depend on the covariates linearly. For example, in DGM1, the plant detectability scores depend on covariates 1, 4 and 5, whereas in DGM4 on covariates 2, 4 and 6, even though the generation of the interaction indicators in DGM1 and DGM4 is the same. We used the same coefficients for each submodel across DGMs. These coefficients were generated randomly to avoid picking them ourselves (except for the intercepts that were set to 0).

\paragraph{Same VS different covariates}
The definition of DGMs 1--6 allows us to compare DGM1 to DGM4, DGM2 to DGM5, and DGM3 to DGM6 to investigate model performance when the interaction model is the same but the important covariates are the same (DGMs 1--3) VS different (DGMs 4--6) in the interaction and detection submodels. With these comparisons we investigate whether our approach would suffer from using the {\it same} latent factors across submodels, even if that is not how data are generated. These comparisons will also allow us to compare how the posterior distribution of the variance terms $\theta$ and $\tau$ will perform when the same or different covariates are important.

\paragraph{Measured VS unmeasured covariates}
The definition of DGMs 1--6 also allow us to investigate the importance of having the important covariates measured in the data. We do so by comparing model performance across DGMs 1--3, and DGMs 4--6 separately, since the number of important binary and continuous covariates are the same, and the only difference is whether the important covariates are all measured, mixed, or none measured.

\subsection{Specifying the low and high information scenarios}

Across all scenarios, the coefficients of the covariates in the interaction and the detectability submodels were the same, and we varied {\it which} covariates are multiplied by these coefficients across DGMs 1--6. As described in \cref{sec:simulations}, we also varied whether we were in a low or high information scenario, defined by the amount of sparseness in the species co-occurrence array and the recorded interactions.

In the low information scenario, we set the intercepts of the interaction and detectability submodels to 0. With the interaction intercept equal to 0 for the interaction model, we acquired that on average 53\% of pairs formed a possible interaction within each data set. With the intercept equal to 0 for the detectability models, we acquired an average detectability score for the bird and plant species equal to 0.48 and 0.44, respectively. Clearly, these detectability scores are quite low, indicating that a possible interaction among an average bird and an average plant would be detected with probability of 0.21 (the product of the two) within a study. When using the study focus and co-occurrence from our observed data we acquired recorded interactions for 2.5\% of pairs, on average. Therefore, this is scenario is very sparse, in that we only record 4.7 interactions for each 100 possible ones.

To evaluate model performance in scenarios with less sparsity, we considered an alternative situation where we record a larger proportion of the truly possible interactions. To achieve this, we adjusted the intercepts of the interaction and detectability models and the species co-occurrence array. We did not alter the coefficients of the covariates, as we wanted to maintain the same amount of information in the covariates for predicting the interactions and detectability scores across simulation scenarios. The intercept for the interaction model was set to $-1.1$, leading to an average of 23\% of pairs with a possible interaction, a more reasonable amount of possible interactions. The intercepts for the detectability submodels were set to 1.5 leading to an average detectability score equal to 0.73 and 0.68 for bird and plant species, respectively, and a resulting probability for detecting a possible interaction within a study equal to approximately 0.5 (substantially higher than the probability of 0.21 in the low information scenario). The focus of the studies was kept the same as in the observed data. We adjusted the indicators for species co-occurrence. For the bird and plant species, indicators of occurrence that equal to 1 in the observed data was kept at 1, whereas the indicators of occurrence that were equal to 0 in the observed data were drawn from independent Bernoulli distributions with probability 0.1 for birds and 0.15 for plants. With this species co-occurrence array, we observe on average 12\% of pairs interacting, corresponding to half of the truly possible interactions.

The observational effort for a {\it pair} $(i,j)$ is defined as the total number of studies that could have recorded their interaction: $\sum_s F_{ijs} O_{is}^\birds O_{js}^\plants$. In the low information scenario, the observational effort is the same as in our observed data. Here, about 78\% pairs of species have observational effort equal to 0, meaning that no study could have recorded their interaction, even if it was possible. Also, 53 plants have observational effort equal to 0 in combination with any bird species, and are therefore assumed impossible to observed in interaction with any bird species.
In the high information scenario, 39\% of pairs have observational of effort equal to 0, and 33\% of pairs equal to 1, implying a still relatively sparse scenario, but much less sparse than the low information one. In this scenario, all species have a chance to be observed in an interaction with at least one species of the other type.

The observational effort by pair is an important driver of model performance. That is because an {\it un}recorded interaction among a pair with observational effort equal to 0 is {\it not} informative at all. In the low information (sparse) scenario, the vast majority of pairs are therefore not informative for model estimation and prediction. In \cref{sec:discussion}, we discuss how our data-driven choice of occurrence indicators places us within the lowest-information-possible scenario within the context of our study. This can act as a ``protection'' over setting occurrence indicators to non-zero when the species do not truly occur, which would provide inflated certainty in our predictions.

For each simulated data set, a random sample of 10 bird and 10 plant species were assumed to have corresponding focus, occurrence, and interaction record equal to 0, to represent out-of sample species.

\subsection{Simulation results for all alternative methods considered}

\begin{figure}[p]
\centering
\begin{minipage}{0.035\textwidth}
\includegraphics[width=\textwidth,trim=0 18 522 0, clip]{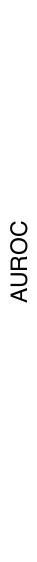}
\end{minipage}
\begin{minipage}{0.88\textwidth}
\includegraphics[width=\textwidth,trim=19 58 0 0, clip]{sims_high_all.pdf} \\[0pt]
\includegraphics[width=\textwidth,trim=19 0 0 22, clip]{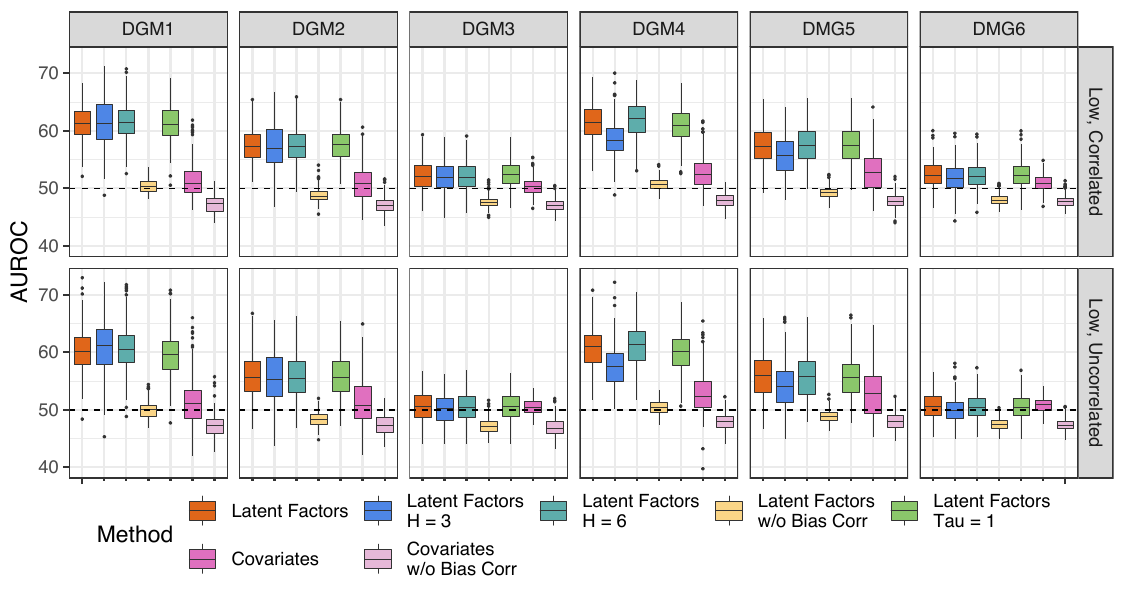}
\end{minipage}
\caption{Predictive Performance of Alternative Approaches in Simulations. AUROC across all pairs of species with an unrecorded interaction by method and simulation scenario. The methods considered are shown using different colors on the x-axis and they are described in Supplement \ref{supp_sec:other_methods}.}
\label{supp_fig:sims_all}
\end{figure}

\cref{supp_fig:sims_all} shows the average AUROC across pairs and simulated data sets for 7 methods: the latent factor model, the latent factor model for fixed $H$ at 3 or 6 without the shrinkage prior, the latent factor model without bias correction, the latent factor model with $\tau$ parameters set to 1, the covariates model, and the covariates model without bias correction. (Results for the model with covariates and latent factors, and for the model that cuts the feedback from the interaction and detectability submodels into the latent factors are discussed below).

First, we note that the models that do not perform bias correction for taxonomic and geographical bias have extremely poor performance in all scenarios. The AUROC for the models without bias correction decreases with observational effort (not shown here). This is expected since the models without bias correction do not incorporate any information on how many times an interaction had the opportunity to be recorded or not. These results illustrate that bias correction is crucial for properly understanding species interactivity.

Second, we see that the model with $H = 3$ and without using the shrinkage prior generally performs worse or as well as the same model with $H = 6$, or as the model which uses the shrinkage prior. The model with $H = 6$ performs comparably to the model that sets $H = 10$ and uses the shrinkage prior. (We found that the model that sets $H = 9$ and does not use the shrinkage prior performs similarly to the model with $H = 6$, though the results are not shown here for brevity.) Therefore, we find that the shrinkage prior effectively reduces the size of the parameter space when possible, while providing the required flexibility when a higher number of latent factors is needed, without having the explicitly specify the effective number of latent factors.

Even though it is not immediately visible from the plot because the differences are small, we found that allowing for the parameters $\tau$ to vary only improved the performance of the proposed model. This reflects that the additional flexibility offered by parameter-specific $\tau$ values is useful in improving our predictions.

\begin{figure}[!t]
\centering
\includegraphics[width=\textwidth]{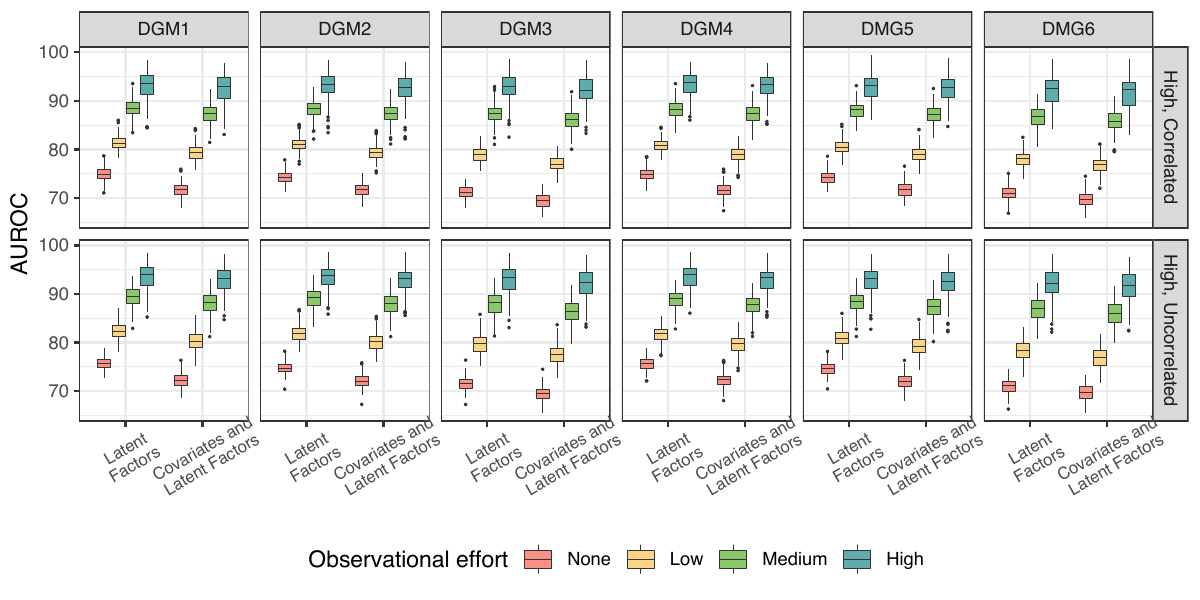}
\caption{Predictive Performance of Model with Covariates and Latent Factors. AUROC across four levels of pair observational effort. The model is introduced in Supplement \ref{supp_sec:other_methods}.}
\label{supp_fig:sims_clf}
\end{figure}

We also investigated the performance of a model that includes both covariates and latent factors in the interaction submodel. Our interest in this model stems mostly from understanding its performance when evaluating the variables' importance for forming interactions, as it is easier to study the coefficients of the covariates in this model, compared to our approach to variable importance introduced in \cref{subsec:variable_importance}. Before we investigate the performance of this model for variable importance, we see that the performance of the model that incorporates both covariates and latent factors for predicting possible interactions, shown in \cref{supp_fig:sims_clf}, is between the performance of the covariates and latent factor models. The AUROC achieved by this model is higher than that of the covariates model, but the latent factor model introduced in \cref{sec:model} still outperforms it. We investigate the performance of this model for variable importance in Supplement \ref{supp_subsec:simulations_traits}.

\begin{figure}[!t]
\centering
\begin{minipage}{0.035\textwidth}
\includegraphics[width=\textwidth,trim=0 20 595 0, clip]{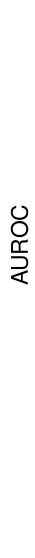} \\[2pt]
\end{minipage}
\begin{minipage}{0.9\textwidth}
\includegraphics[width=\textwidth,trim=19 75 0 0, clip]{sims_high_nofeed.pdf} \\[2pt]
\includegraphics[width=\textwidth,trim=19 0 0 22, clip]{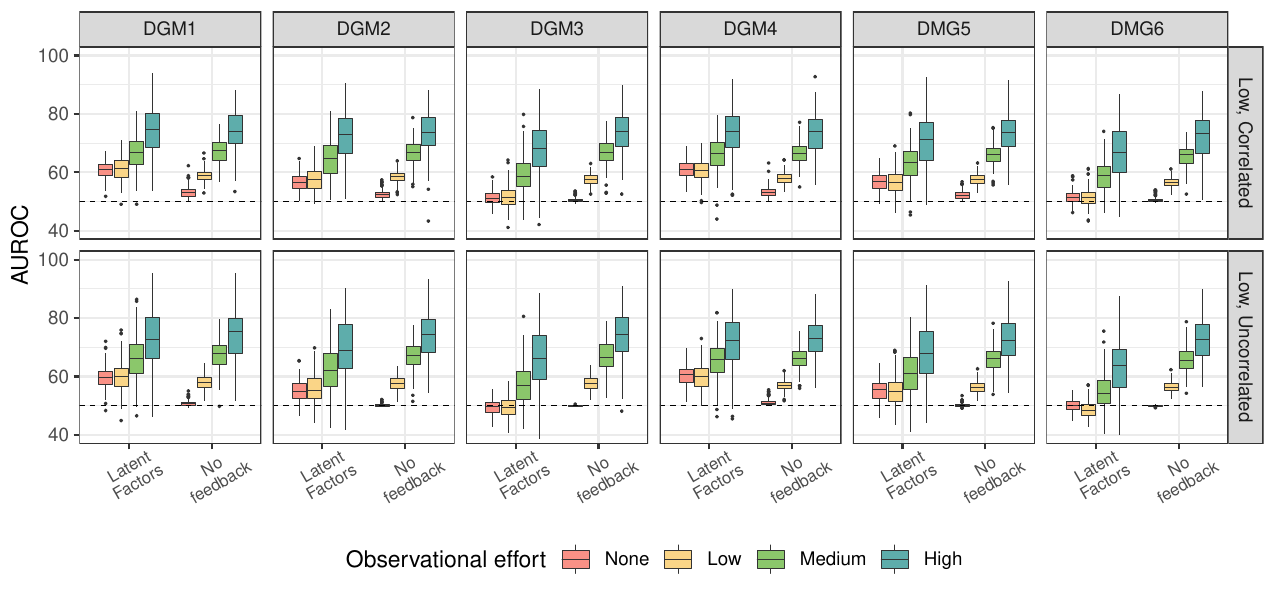}
\end{minipage}
\caption{Predictive Performance in Simulations. The methods considered (shown on the horizontal axis) are the proposed model (Latent Factors) and a modularized version of the same model that cuts the feedback from the interaction and detectability submodels into the latent factors (No Feedback). The columns represent the 6 DGMs in \cref{tab:sims_important_variables}. The rows correspond to combinations of the high and low signal scenarios and the two correlation values. Results are shown by observational effort for pairs of species by color.}
\label{fig:sim_res_nofeed}
\end{figure}

Lastly, we compared our model as proposed, to a modularized version of our model that bases the latent factor updates solely on the covariate model \cref{eq:model_covariates} fit, and cuts the feedback from the interaction and detectability submodels in \cref{eq:model_probinter} and \cref{eq:model_probobs} into the latent factors. By comparing the results of our model to the modularized version of the same model we essentially evaluate the extent to which the full model can harvest information from the measured interactions in informing the latent factors (since the modularized model does not do so). The results from this comparison are shown in \cref{fig:sim_res_nofeed}, across all 24 scenarios we considered and by pair-specific observational effort. In all the high-information scenarios, the full model (that allows for feedback) always performs better than the model that cuts the feedback. This illustrates that, in these scenarios, the latent factors {\it are} indeed informed by the measured interactions in a useful manner in order to predict missing interactions. In the low information setting, the version of the model that cuts the feedback {\it can} perform better than the model that allows for feedback, but only for pairs of species with a lot of observational effort. This illustrates that there can be some efficiency loss in trying to learn the latent factors using the measured interaction profiles, if those measured interactions are very sparse. However, the model that does allow for feedback performs at least as good, and often much better across all scenarios for pairs of species with no observational effort (shown in red in \cref{fig:sim_res_nofeed}). This indicates that using the measured interaction profiles for learning the latent factors can be useful for learning about pairs of species that are rare and not well-represented in the data.

\begin{table}[!b]
    \centering
\caption{Number of simulated data sets with effective dimension of the latent space that is at least three. Results are shown for DGM 1, 2, 4, and 5, for the high information scenario with uncorrelated covariates. Total number of simulated data sets is 200.}
\label{supp_tab:effective_dimension}    
    \begin{tabular}{c|ccc}
    & & \multicolumn{2}{c}{Are the same covariates important in} \\
    & & \multicolumn{2}{c}{interaction and detectability submodels?} \\
are the important &  & & \\
covariates measured? & & same & different\\ 
\hline

 all & & (DGM 1) & (DGM 4) \\
& with feedback: &  11 & 23 \\
& without feedback: & 14 & 11  \\ \hline
 some & & (DGM 2) & (DGM 5) \\
& with feedback: & 8 & 23  \\
& without feedback: & 6 & 2  \\ \hline
    \end{tabular}

\end{table}

We also investigated the effective dimension of the latent space in DGMs 1, 2, 4, and 5 in the high information scenario with uncorrelated covariates. Here, we define the effective number to be the number of $\theta_h$ with posterior mean above 0.05, when $\theta_\infty$ is set to 0.01. In \cref{supp_tab:effective_dimension} we report the number of simulated data sets (out of 200) where the effective dimension was at least three. When the same versus different covariates are important in the interaction and detectability submodels (comparing DGM 1 to DGM 4, and DGM 2 to DGM 5), the modularized approach that cuts the feedback uses at least three latent factors in a similar proportion of simulated data sets. In contrast, the approach that allows for feedback uses a high number of latent factors more frequently when different covariates are important in the two submodels, illustrating that by allowing for feedback the effective dimension of the latent factor space can be larger if needed in order to capture the important information.

% \subsection{The importance of correlation among covariates}

% We investigated whether correlation among the covariates improves the performance of the latent factor model. Specifically, for the different combinations of low and high signal scenarios and the 6 DGMs, we performed a t-test comparing the average AUROC for the latent factor model across 200 simulated data sets when covariates are correlated with correlation 0.3, and when they are uncorrelated.

% \cref{supp_tab:sims_correlation} shows the results.

% \begin{table}[!b]
% \centering
% \caption{P-values from t-test comparing the average AUROC value of the latent factor and covariate models from 200 simulated data sets when correlation among covariates is 0 or 0.3. Asterisk $^*$ is used for entries with p-value below 0.05.}
% \label{supp_tab:sims_correlation}
%     \begin{tabular}{c|c|cccccc}
%   &  & DGM1 & DGM2 & DGM3 & DGM4 & DGM5 & DGM6 \\ \hline
% Latent Factor &  High information  & 0.025* & 0.301 & 0.969 & 0.201 & 0.176 & 0.061 \\
% Model & Low information & 0.006 & 0.000 & 0.000 & 0.114 & 0.000 & 0.000 \\ \hline
% Covariates & High information  & 0.001* & 0.208 & 0.721 & 0.009* & 0.653 & 0.072 \\
% Model & Low information & 
%     \end{tabular}
% \end{table}

\subsection{Simulation results for out of sample species}
\label{supp_subsec:simulations_half_out}

\begin{figure}[p]
\centering
\begin{minipage}{0.035\textwidth}
\includegraphics[width=0.9\textwidth, trim= 0 0 562 0, clip]{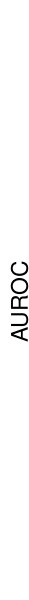}
\end{minipage}
\begin{minipage}{0.9\textwidth}
\includegraphics[width=\textwidth, trim=17 0 0 0, clip]{sims_high_out.pdf}
\includegraphics[width=\textwidth, trim=17 0 0 0, clip]{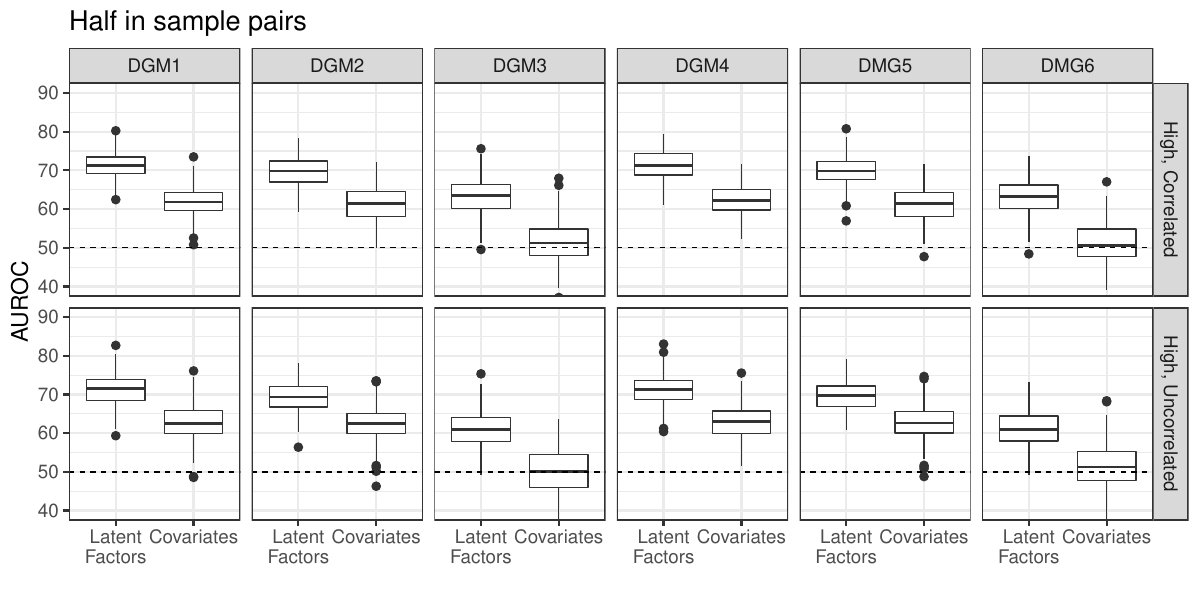}
\end{minipage}
\caption{Predictive Performance of Latent Factor and Covariate Approaches for Out of Sample Species. Results shown for the high information scenario, all DGMs and both correlations. The out of sample species are considered in combination with other out of sample species (rows 1 \& 2) and in combination with in sample species (rows 3 \& 4).}
\label{supp_fig:res_out_sample}
\end{figure}

Out-of-sample species are species with zero observational effort: species that could not have been recorded in an interaction with any other species and in any other study.
For out-of-sample species, we might be interested in predicting their interactions with other out-of-sample species, or with in-sample species. Pairs for which both species are out-of-sample are referred to as ``out-of-sample'' pairs, and pairs of species for which one is in-sample and the other is out-of-sample are referred to as ``half-in-sample''. Results for the high information scenarios for the latent factor and covariates approaches in terms of AUROC achieved are shown in \cref{supp_fig:res_out_sample}.

Again, the latent factor model performs better than the covariates model in all scenarios considered.
We also investigate performance of the latent factor approach for out of sample species.
The pairs of species without a chance of being detected in an interaction are of the following types:
\begin{enumerate*}[label=(\alph*)]
\item in-sample pairs that have zero observation effort between them but non-zero observational effort with other species (illustrated in \cref{fig:sim_res_A0yes} as observation effort `none'),
\item half in sample pairs, and
\item out of sample pairs.
\end{enumerate*}
We see that the AUROC achieved by the latent factor approach is better for in-sample pairs without observational effort, then for half in sample pairs, and then for out of sample pairs. This illustrates that the latent factor approach learns about a species' possible interactions through its interactions with other species.
Conclusions from the low information scenarios are similar though less visible since the approaches have AUROC closer to 50\% even for in-sample pairs of species.

\subsection{Simulation results on variable importance}
\label{supp_subsec:simulations_traits}

\subsubsection{Based on the latent factor model}

We performed simulations to investigate the performance of the variable importance measure of \cref{subsec:variable_importance}. We considered all the simulation scenarios from the main text, though we focus here on the high information scenarios for brevity.

\begin{figure}[p]
\centering
\includegraphics[width=\textwidth]{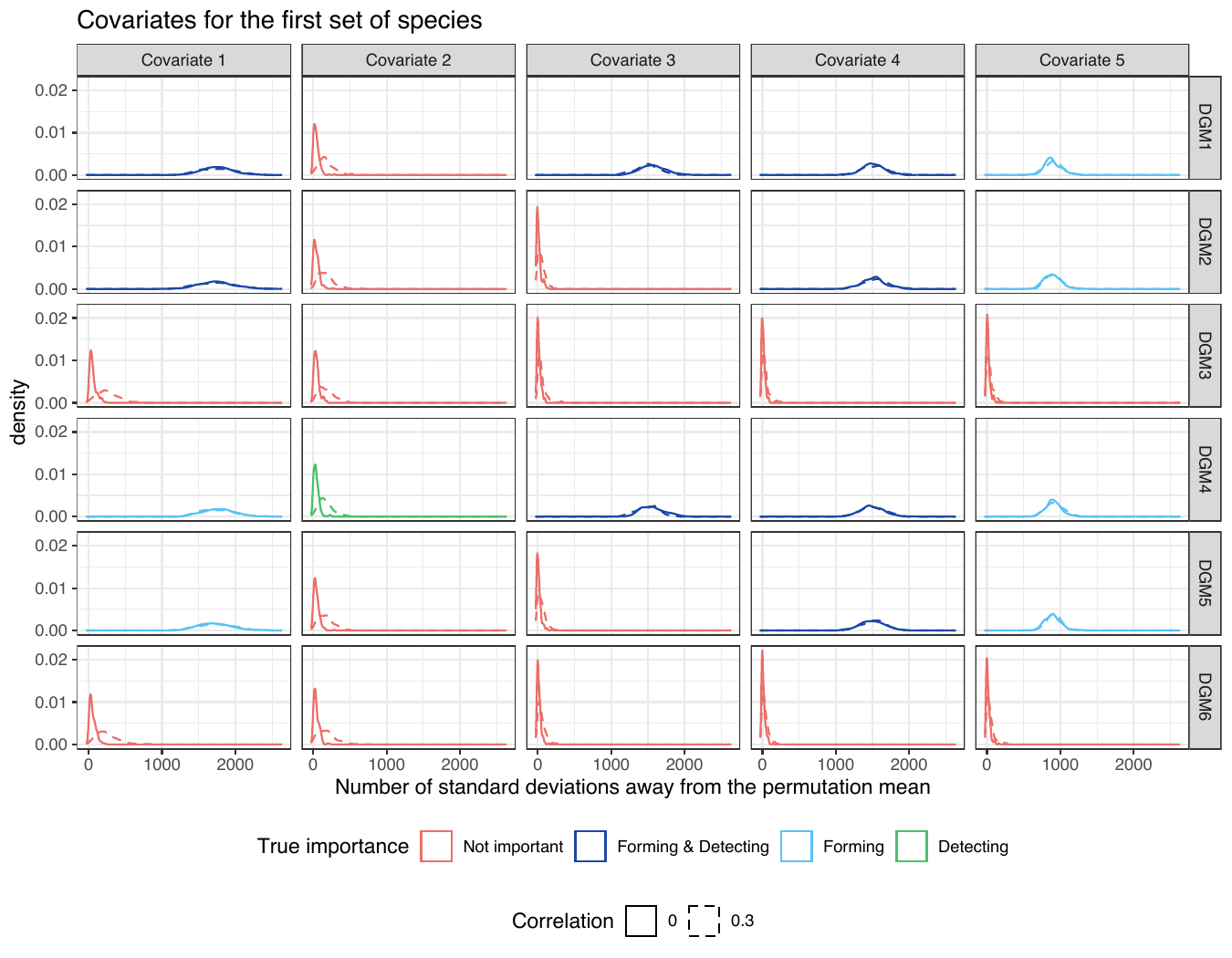}
\caption{Variable importance simulations for the first set of species. Number of standard deviations away from the permutation mean by covariate and data generative model. Different colors are used for variables of different importance (important or not for forming and/or detecting an interaction), in agreement with \cref{tab:sims_important_variables}. Different line type is used for the scenarios with correlated (dashed) and uncorrelated (solid) covariates.}
\label{supp_fig:sims_imp_bird}
\end{figure}

\begin{figure}[p]
\centering
\includegraphics[width=\textwidth]{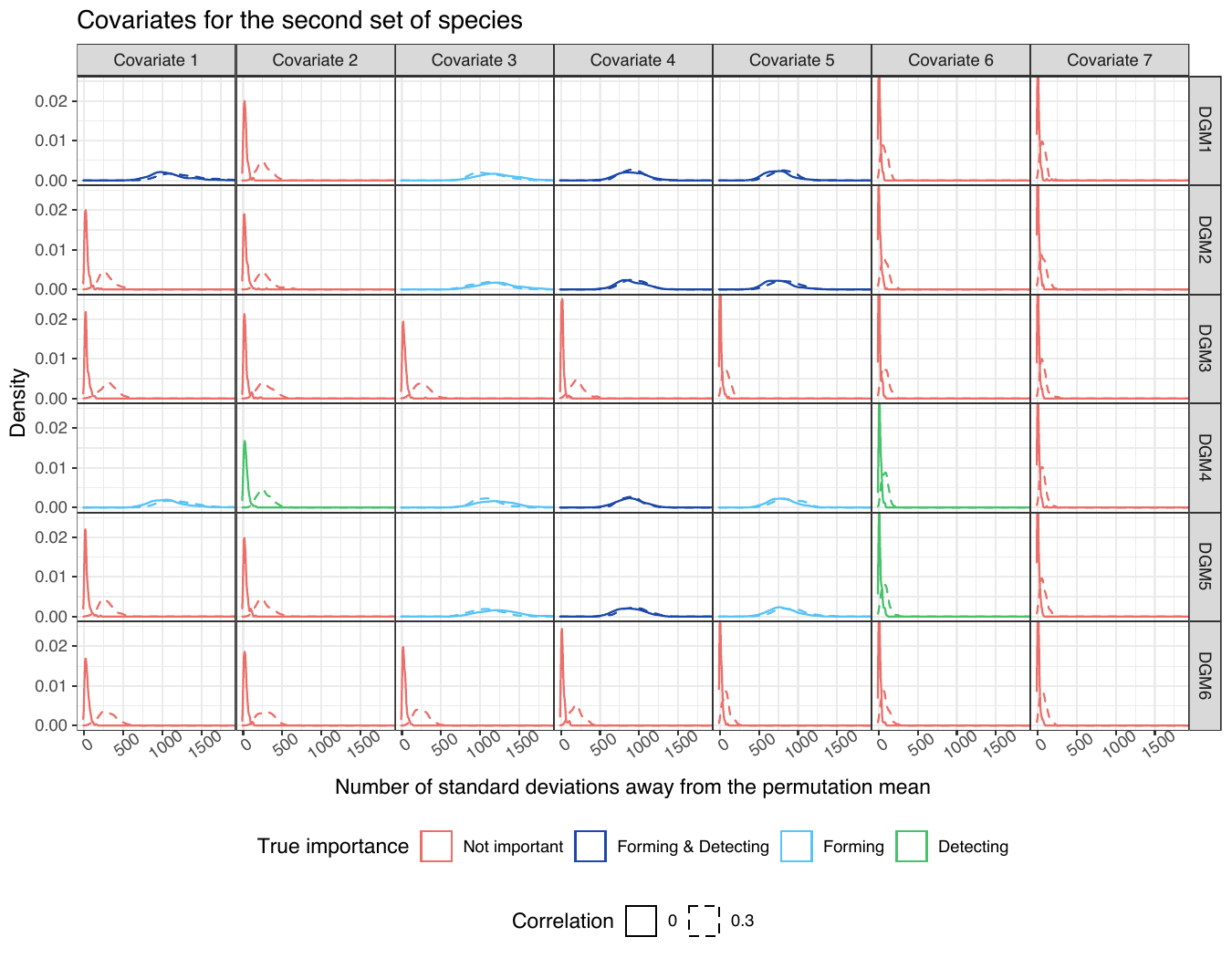}
\caption{Variable importance simulations for the second set of species. Number of standard deviations away from the permutation mean by covariate and data generative model. Different colors are used for variables of different importance (important or not for forming and/or detecting an interaction), in agreement with \cref{tab:sims_important_variables}. Different line type is used for the scenarios with correlated (dashed) and uncorrelated (solid) covariates. Results for plant covariates 8, 9, 11, and 12 are similar to those of covariate 7 and are excluded.}
\label{supp_fig:sims_imp_plant}
\end{figure}

Figures \ref{supp_fig:sims_imp_bird} and \ref{supp_fig:sims_imp_plant} show the results for the first and second sets of species, respectively. Specifically, each panel corresponds to a different combination of covariate and DGM and it shows the distribution across simulated data sets of the number of permutation standard deviations away from the permutation mean that the observed statistic falls. Larger values represent that the observed covariate is more informative of the probability of forming and detecting interactions between species. Dark blue color is used for covariates that are important for both forming or detecting interactions, light blue is used for covariates that are important only for forming interactions, green is used for covariates that are important only for detecting interactions, and red is used for covariates that are important for neither. The importance of each covariate is also shown in \cref{tab:sims_important_variables}. For the first set of species, covariates 1 and 2 are continuous and the rest binary, and for the second set of species covaraites 1 through 4 are continuous and the rest are binary.

Focusing first in the case with uncorrelated covariates, we find that the histograms for covariates that are not important for forming interactions are concentrated near 0 standard deviations, implying that the observed statistic resembles that of the permuted data sets. This means that the approach correctly identifies covariates that are not important for forming interactions, irrespective of whether these covariates are important for detection or not, and covariates that only drive detectability are not identified as important based on this variable importance metric.
In contrast, the histograms for covariates that are important for forming interactions are always well-separated from zero, and correctly identified.
We find that continuous variables tend to have higher variable importance scores than binary variables with the same importance (for example, compare the results for covariates 1 \& 3 of DGM1 of the bird species, and covariates 4 \& 5 of DGM1 for the plant species). Therefore, it might be preferable to focus on variable importance separately for continuous and binary covariates.

Unsurprisingly, when there is correlation among the covariates, the covariates that are not important are correlated with the important ones. Therefore, we see higher variable importance metrics in this case for the unimportant covariates. The variable importance metric for the important covariates remains unaltered.

\subsubsection{Based on the model that includes covariates and latent factors}

To study variable importance based on the model that includes both covariates and latent factors, we investigated the 95\% credible intervals for the coefficients of the covariates. We show here the results for the high information scenarios with uncorrelated covariates. The high information scenarios with uncorrelated covariates can be seen as the ``easiest'' setting for identifying variable importance.

\begin{table}[!t]
\centering
\caption{Variable importance for the model with covariates and latent factors. Percentage of 200 data sets where the 95\% credible interval for the coefficient of each covariate did {\it not} include zero, and the covariate was deemed important. Values close to 1 indicate that the covariate was identified as important in most data sets. Results correspond to the high information scenario with all covariates {\it un}correlated.}
\label{supp_tab:sims_imp_clf}
\begin{tabular}{rrrrrrr}
  \hline
 & DGM1 & DGM2 & DGM3 & DGM4 & DGM5 & DGM6 \\ 
  \hline
$X_1$ & 0.295 & 0.320 & 0.275 & 0.245 & 0.215 & 0.250 \\ 
 $X_2$ & 0.060 & 0.065 & 0.525 & 0.155 & 0.085 & 0.640 \\ 
  $X_3$ & 0.290 & 0.055 & 0.220 & 0.265 & 0.060 & 0.290 \\ 
  $X_4$ & 0.645 & 0.575 & 0.255 & 0.595 & 0.530 & 0.180 \\ 
  $X_5$ & 1.000 & 1.000 & 0.250 & 1.000 & 1.000 & 0.320 \\ 
  $W_1$ & 0.570 & 0.060 & 0.150 & 0.455 & 0.045 & 0.115 \\ 
  $W_2$ & 0.045 & 0.050 & 0.165 & 0.300 & 0.080 & 0.145 \\ 
  $W_3$ & 0.990 & 0.985 & 0.185 & 0.990 & 0.995 & 0.120 \\ 
  $W_4$ & 0.995 & 0.985 & 0.230 & 1.000 & 0.985 & 0.155 \\ 
  $W_5$ & 1.000 & 1.000 & 0.060 & 1.000 & 1.000 & 0.085 \\ 
  $W_6$ & 0.080 & 0.055 & 0.100 & 0.085 & 0.090 & 0.095 \\ 
  $W_7$ & 0.040 & 0.040 & 0.075 & 0.020 & 0.050 & 0.125 \\ 
  $W_8$ & 0.085 & 0.045 & 0.100 & 0.060 & 0.045 & 0.060 \\ 
  $W_9$ & 0.025 & 0.055 & 0.110 & 0.020 & 0.040 & 0.115 \\ 
  $W_{10}$ & 0.040 & 0.050 & 0.140 & 0.050 & 0.055 & 0.090 \\ 
  $W_{11}$ & 0.015 & 0.085 & 0.050 & 0.025 & 0.055 & 0.135 \\ 
  $W_{12}$ & 0.060 & 0.070 & 0.645 & 0.050 & 0.045 & 0.440 \\ 
   \hline
\end{tabular}
\end{table}

Results are shown in \cref{supp_tab:sims_imp_clf}. We see that many covariates that are completely unrelated with species interactivity are identified as important very often. For example, the binary covariate $W_{12}$ that is unrelated to everything, is identified as important in 88 of the 200 data sets from DGM6. If this method for variable importance performed accurately, each null covariate would be identified as important in 5\% of data sets or fewer. However, we see that the proportion of data sets that identify null covariates as important is almost always higher than 5\%. We believe that this is due to collinearity between the phylogenetically-correlated latent factors and phylogenetically-correlated covariates, which leads to issues similar to those of spatial confounding in the spatial literature.

\subsection{Computational time of the proposed approach}

We investigated the computational time for 1,000 posterior samples of our approach when varying the number of bird species, number of plant species, and the number of studies. We considered number of birds equal to 50, 100, 150, and 200, number of plants equal to 50, 200, 350, and 500, and number of studies equal to 10, 40, and 80. For each combination of number of species and studies, we ran a mini-simulation with 10 data sets and 1,000 posterior samples each. The median computational time across the 10 data sets is shown in \cref{supp_tab:comp_time}. The computational time appears to grow linearly in the total number of pairs ($n_\birds \times n_\plants$), whereas the number of measured networks ($S$) does not appear to affect computational time.

\begin{table}[!t]
\centering
\caption{Computational time of 1,000 MCMC iterations when varying the number of species and number of studies. $S$ represents the number of studies, $n_\birds$ the number of species of the first set (birds) and $n_\plants$ the number of species of the second set (plants).}
\label{supp_tab:comp_time}
\begin{tabular}{lcrrrr}
  \hline
  & \backslashbox{$n_\birds$}{$n_\plants$} & 50 & 200 & 350 & 500 \\ 
\hline \hline
$S = 10$ & \hspace{30pt} & & & & \\
  \hline
& 50 & 2.02 & 6.58 & 16.39 & 21.42 \\ 
&  100 & 3.44 & 13.80 & 24.80 & 35.71 \\ 
&  150 & 4.66 & 17.46 & 24.59 & 36.26 \\ 
&  200 & 9.98 & 25.77 & 50.19 & 75.55 \\ 
   \hline \hline
$S = 40$ & & & & & \\
\hline
& 50 & 2.06 & 6.62 & 13.89 & 31.00 \\ 
&  100 & 3.36 & 14.08 & 24.48 & 37.97 \\ 
&  150 & 4.92 & 17.91 & 27.38 & 37.37 \\ 
&  200 & 9.56 & 32.93 & 51.91 & 84.85 \\ 
\hline \hline
$S = 80$ & & & & & \\
\hline
& 50 & 1.85 & 6.84 & 13.67 & 21.18 \\ 
&  100 & 3.50 & 16.20 & 27.08 & 49.02 \\ 
&  150 & 5.26 & 17.33 & 22.85 & 58.47 \\ 
&  200 & 7.19 & 30.05 & 54.23 & 72.50 \\
\hline
\end{tabular}
\end{table}

\section{Out-of-sample species}

\label{subsec:out_of_sample}

Our main focus is in predicting whether a species $i^*$ with covariates $\covsB_{i^*}$ would interact with a species $j^*$ with covariates $\covsP_{j^*}$ if given the opportunity, where the covariates might be available with missingness. If both species are included in the original $n_\birds \times n_\plants$ data, predictions are automatic based on the posterior samples of $\bm L$. However, species might be completely unobserved and have no recorded interactions, and we might be interested in understanding whether these out-of-sample species interact with all other species.

\subsection{The posterior distribution for the presence of interaction for species with unrecorded interactions}

Inference on the probability of interaction between species $i^*$ and $j^*$ when neither of them has any recorded interactions could proceed by extending the observed interaction matrix of $n_\birds$ and $n_\plants$ species to an $(n_\birds + 1) \times (n_\plants + 1)$ interaction matrix with $n_{i^*j} = 0$ and $n_{ij^*} = 0$ for all $i,j$ species in the original data set, and re-fitting the MCMC.

By investigating the posterior probability that these out-of-sample species interact, we better understand how the model borrows information across species. Let $\allpars$ denote all model parameters excluding parameters corresponding to the species $i^*$ and $j^*$, $\alldata$ denote the observed data for all in-sample species, and $\covsB_{i^*}$ and $\covsP_{j^*}$ denote covariates for $i^*$ and $j^*$, respectively, where the covariate vectors can include missing values. Let $\latsB_{i^*}, \latsP_{j^*}$ denote the latent factors corresponding to $i^*, j^*$ respectively.
Then, we wish to predict $L_{i^*j^*}$ based on what we have learnt from the observed data and from the out-of-sample species covariates. That would amount to learning
\( P(L_{i^*j^*} = 1 \mid \alldata, \covsB_{i^*}, \covsP_{j^*}). \), which we rewrite as
{\small
\begin{equation}
\begin{aligned}
P(& L_{i^*j^*} = 1 \mid \alldata, \covsB_{i^*}, \covsP_{j^*}) = \\
&= \int P(L_{i^*j^*} = 1 \mid \allpars, \latsB_{i^*}, \latsP_{j^*} \alldata, \covsB_{i^*}, \covsP_{j^*}) \
p(\allpars, \latsB_{i^*}, \latsP_{j^*} \mid \alldata, \covsB_{i^*}, \covsP_{j^*}) \
\mathrm{d} (\allpars, \latsB_{i^*}, \latsP_{j^*}) \\
&\propto \int P(L_{i^*j^*} = 1 \mid \allpars, \latsB_{i^*}, \latsP_{j^*}) \ 
p(\covsB_{i^*}, \covsP_{j^*} \mid \allpars, \latsB_{i^*}, \latsP_{j^*} \alldata) \
p(\allpars, \latsB_{i^*}, \latsP_{j^*} \mid \alldata) \ 
\mathrm{d} (\allpars, \latsB_{i^*}, \latsP_{j^*})  \\
&\propto \int \overbrace{P(L_{i^*j^*} = 1 \mid \allpars, \latsB_{i^*}, \latsP_{j^*})}^{(A)} \ 
\overbrace{p(\covsB_{i^*} \mid \allpars, \latsB_{i^*}) \
p(\covsP_{j^*} \mid \allpars, \latsP_{j^*})}^{(B)} \\
& \hspace{5cm}
\underbrace{p(\latsB_{i^*}, \latsP_{j^*} \mid \allpars)}_{(C)} \ 
\underbrace{p(\allpars\mid \alldata)}_{(D)} \ 
\mathrm{d} (\allpars, \latsB_{i^*}, \latsP_{j^*}).
\end{aligned}
\label{supp_eq:out_of_sample}
\end{equation}}
In (A) we see that having access to the species' latent factors would allow us to straightforwardly learn their probability of interaction. Therefore, ideally one could draw the species' latent factors from their posterior distribution, and draw their interaction indicator from the corresponding conditional distribution in (A). In that sense, estimation of interaction probabilities for out-of-sample units resembles outcome prediction in mixed models for new units for which the random effect is drawn from the common random effect distribution. The main difference between these parallel settings arise from parts (B) and (C) in \cref{supp_eq:out_of_sample}. Factor (B) tells us that the latent factors for the new species are driven by the latent factors for the in-sample species since their correlation matrix is non-diagonal, and factor (C) tells us that the latent factors for the out-of-sample species needs to also agree with the species' observed covariates. Therefore, learning the interaction probability of out-of-sample species is based largely on the interaction profiles of phylogenetically related species and the species' covariate profile.

\subsection{Computationally efficient algorithm for out-of-sample prediction of species interactions} 

Here, we introduce an alternative approach for acquiring out-of-sample predictions of species interactions which avoids re-fitting the MCMC for every new species. We propose a computationally efficient algorithm for making predictions for out-of-sample species, which combines the MCMC fit to the original data and importance sampling weighting.
Expression \cref{supp_eq:out_of_sample} is the basis of our algorithm. We first describe the algorithm here, and then we provide the explicit steps for its implementation afterwards. We use draws from the posterior distribution of model parameters based on the original data (D) to draw latent factors for the species $i^*,j^*$ (C) where the correlation matrices $\bm C_\latB, \bm C_\latP$ are updated to include the new species. Based on the latent factors and the model parameters of each posterior sample, we draw values for the indicator of whether the species interact (A). Since these draws do not account for the covariates, we up(down)-weigh the draws which use parameters and latent factors that have higher (lower) values of (B). 

To describe the algorithm in full detail, let $\bm C_\latB^\ast$, $\bm C_\latP^\ast$ be the extended correlation matrices with units $i^*$ and $j^*$ included. Then $\bm C_\latB^\ast$, $\bm C_\latP^\ast$ are correlation matrices of dimension $n_\birds + 1$ and $n_\plants + 1$, and the upper left $n_\birds \times n_\birds$ and $n_\plants \times n_\plants$ submatrices are $\bm C_\latB$, $\bm C_\latP$, respectively. In what follows the superscript $(r)$ represents the $r^{th}$ posterior sample from the MCMC out of a total of $R$ samples. The algorithm is as follows:

\begin{enumerate}[label=(\arabic*),leftmargin=*]
\item First, we acquire samples for the species' latent factors based on samples from the posterior distribution of model parameters $\allpars{}^{(r)}$, $r = 1, 2, \dots, R$. We do so as follows:
\begin{itemize}[leftmargin=*]
\item For $r = 1, 2, \dots, R$, we generate latent factors $\latB_{i^*h}$ for species $i^\ast$ from $N(\mu, \ssq)$ where
\[
\ssq = \big[S^{(r)}\big]_{(n_\birds +1),(n_\birds + 1)} - \big[\bm S^{(r)}\big]_{(n_\birds +1),(1 : n_\birds)} \left[ \big[\bm S^{(r)}\big]_{(1 : n_\birds),(1 : n_\birds)} \right] ^{-1} \left[ \big[\bm S^{(r)}\big]_{(n_\birds +1),(1 : n_\birds)} \right]^T
\]
and
\[
\mu = \big[\bm S^{(r)}\big]_{(n_\birds +1),(1 : n_\birds)} \left[ \big[\bm S^{(r)}\big]_{(1 : n_\birds),(1 : n_\birds)} \right] ^{-1} \latsB_{1:n_\birds, h}^{(r)},
\]
where $\bm S^{(r)} = \rho_\latB^{(r)} \bm C_\latB^\ast + (1 - \rho_\latB^{(r)}) \bm{I}_{n_\birds + 1}$, $[\bm S^{(r)}]_{\mathcal{A}, \mathcal{B}}$ represents the submatrix of $\bm S^{(r)}$ with row indices in $\mathcal{A}$ and column indices in $\mathcal{B}$, and $\latsB_{1 : n_\birds, h}^{(r)} = (\latB_{1h}^{(r)}, \latB_{2h}^{(r)},  \dots, \latB_{n_\birds h}^{(r)})$.

\item Generate latent factors for species $j^*$ similarly as for species $i^*$, but substituting $\bm C_\latB$ for $\bm C_\latP$, $\rho_\latB$ for $\rho_\latP$, $\bm \latB$ for $\bm \latP$, and $n_\birds$ for $n_\plants$. 
\end{itemize}
Performing these steps leads to latent factors $\latsB_{i^*}^{(r)} = (\latB_{i^*1}, \latB_{i^*2}, \dots, \latB_{i^*H})^T$ for species $i^*$ and $\latsP_{j^*} = (\latP_{j^*1}, \latP_{j^*2}, \dots, \latP_{j^*H})^T$ for species $j^*$, for all $r$.
We use these latent factors to make an original set of predictions. For each $r$, we generate $\widetilde L_{i^*j^*}^{(r)}$ from a Bernoulli distribution with probability of success equal to $\mathrm{expit}\left( \lambda_0^{(r)} + \sum_h \lambda_{h}^{(r)} \latB_{i^*h}^{(r)} \latP_{j^*h}^{(r)} \right)$.

\item However, the generated latent factors have been sampled taking only the correlation structure of the latent factors across species into consideration, and as a result the latent factors and the predictions do not use the information on the new species' covariates. To account for the covariates we perform importance weighting:
\begin{itemize}[leftmargin=*]
\item For species $i^*$ with generated latent factors $\latsB_{i^\ast}^{(r)}$ and covariates $\covsB_{i^*}$ calculate the importance sampling weight $w_{i^*}^{(r)} = w_{i^*1}^{(r)} w_{i^*2}^{(r)} \dots, w_{i^*p_\birds}^{(r)}$, where $w_{i^*m}^{(r)}$ is 1 if $\covB_{i^*m}$ is missing,
\[
w_{i^*m}^{(r)} = \phi\left(\covB_{i^*m}; l_{i^*m}^{(r)}, (\ssq_m)^{(r)}  \right)
\]
if the $m^{th}$ covariate is continuous, and
\[
w_{i^*m}^{(r)} = u(\covB_{i^*m}; \mathrm{expit}(l_{i^*m}^{(r)}))
\]
if the $m^{th}$ covariate is binary, where
$l_{i^*m}^{(r)} = \beta_{m0}^{(r)} + (\latsB_{i^\ast}^{(r)})^T \bm \beta_m^{(r)}$, $\phi(\cdot; \mu, \ssq)$ is the normal density with mean $\mu$ and variance $\ssq$, and $u(\cdot;p)$ is the mass function for the Bernoulli($p$) random variable.

\item Similarly to the above, we acquire $w_{j^*}$ for species $j^*$.

\item The importance sampling weight for the pair $(i^*, j^*)$ is then defined as $w_{i^*j^*}^{(r)} =  w_{i^*}^{(r)}w_{j^*}^{(r)}$ for $r = 1, 2, \dots, R$.
\end{itemize}

\item
We combine these importance sampling weights with the original predicted interaction values $\widetilde L_{i^*j^*}^{(r)}$. Intuitively, $w_{i^*j^*}^{(r)}$ describes how in-line the generated latent factors $\latsB_{i^*}^{(r)}$ and $\latsP_{j^*}^{(r)}$ are with the species' covariate profiles, and, in a sense, how ``trustworthy'' the $r^{th}$ prediction is. For this reason, we set the posterior probability for an $(i^*, j^*)$ interaction equal to
\[
\left( \sum_{r = 1}^R w_{i^*j^*}^{(r)} \widetilde L_{i^*j^*}^{(r)} \right) \Big/
\left( \sum_{r = 1}^R w_{i^*j^*}^{(r)} \right)
.\]

\end{enumerate}

For making interaction predictions for pairs of species of which one is in the original data set and the other one is not, the algorithm is very similar. For the in-sample species, samples from the posterior distribution for its latent factors are already acquired through the MCMC. Therefore, the procedure above only has to be performed for the species that are out-of-sample, and the importance sampling weights only represent one of the species. For example, if $i^*$ is included in the original data and $j^*$ is out-of-sample, then we use the algorithm described above to acquire $\latsP_{j^*}^{(r)}$, $w_{j^*}{(r)}$, and $\widetilde L_{i^*j^*}^{(r)}$ and set $w_{i^*j^*}^{(r)} = w_{j^*}^{(r)}$.

\subsection{How out-of-sample species might affect predictions for in-sample species}
\label{supp_subsec:in_sample_effect}

Availability of covariate information on the new species requires that we also update our predictions for in-sample species. Therefore, when covariate data on $i^*,j^*$ become available, the posterior for $\allpars$ should be
\begin{equation}
\begin{aligned}
p(\allpars \mid \alldata, \covsB_{i^*}, \covsP_{j^*}) & \propto
p( \alldata \mid \allpars, \covsB_{i^*}, \covsP_{j^*}) \ 
p(\covsB_{i^*}, \covsP_{j^*} \mid \allpars) \ 
p(\allpars) \\
&= p( \alldata \mid \allpars) \ 
p(\covsB_{i^*}, \covsP_{j^*} \mid \allpars) \ 
p(\allpars) \\
&= p(\allpars \mid \alldata) \ 
\frac{p(\allpars \mid \covsB_{i^*}, \covsP_{j^*})}{p(\allpars)} \\
&= p(\allpars \mid \alldata) \ 
\int \frac{p(\allpars \mid \allpars_{i^*,j^*})}{p(\allpars)} p(\allpars_{i^*,j^*} \mid \covsB_{i^*}, \covsP_{j^*}) \ \mathrm{d}\allpars_{i^*, j^*},
\end{aligned}
\label{eq:out_of_sample}
\end{equation}
where we use $\allpars_{i^*,j^*}$ to denote all model parameters for species $i^*, j^*$ (includes latent factors and detection probability).
Intuitively, the covariates $\covsB_{i^*}, \covsP_{j^*}$ drive estimation of the latent factors for $i^*,j^*$ (last term in \cref{eq:out_of_sample}), and correlation of the latent factors across species would imply that the latent factors for $i^*, j^*$ affect the latent factors for in-sample species and $p(\allpars \mid \allpars_{i^*,j^*}) \ / \ p(\allpars) \neq 1$. Therefore, the integral in \cref{eq:out_of_sample} will not be equal to 1, and the posterior distribution that incorporates the new covariate values, $p(\allpars \mid \alldata, \covsB_{i^*}, \covsP_{j^*})$, will not be the same as the original posterior distribution, $p(\allpars \mid \alldata)$.

\section{Additional study results}
\label{supp_sec:app_results}

\subsection{Comparison of results from the two models}

\begin{figure}[p]
\hfill
\begin{minipage}{0.8\textwidth}
\includegraphics[width=\textwidth, trim=1cm 3cm 1cm 1cm, clip]{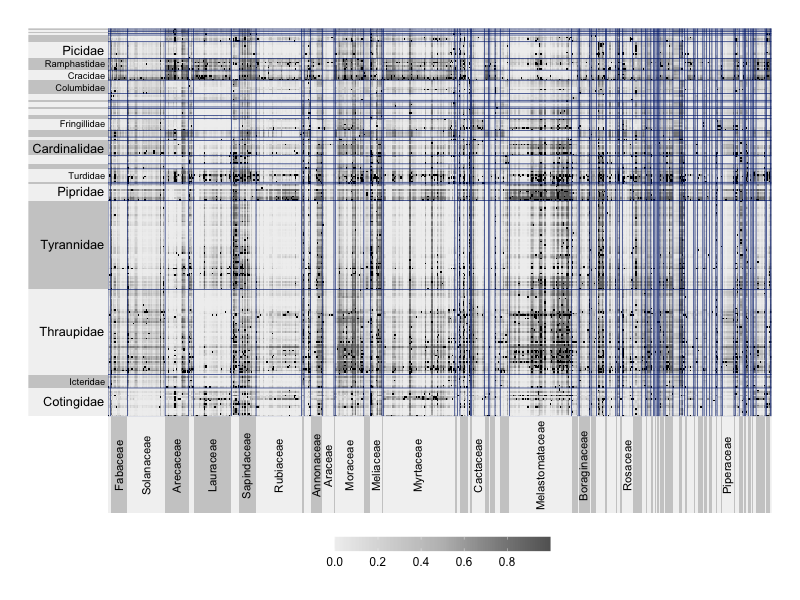} \\
\includegraphics[width=\textwidth, trim=1cm 3cm 1cm 1cm, clip]{app_alt_full.png}
\end{minipage}
\begin{minipage}{0.15\textwidth}
\hspace{20pt}
\includegraphics[width=3\textwidth, angle=90, trim=11.3cm 1cm 7.5cm 19cm, clip]{app_alt_full.png}
\end{minipage}
\caption{Posterior Probability of Possible Interactions for all bird (y-axis) and plant (x-axis) species. Species are organized in taxonomic families separated by blue lines. Black color shows the recorded interactions. \label{fig:results_full}}
\end{figure}

\cref{fig:results_full} shows the posterior probability that an interaction between two species is possible based on our approach and the alternative approach that uses covariates directly. Vertical and horizontal blue lines separate different taxonomic families. A list of all the species included in our analysis in the same ordering as shown in the results is included in Supplement \ref{supp_sec:list_species}. The same conclusions discussed in \cref{sec:application} also hold when showing the full set of species.

\begin{figure}[!t]
\centering
\includegraphics[width = 0.8\textwidth]{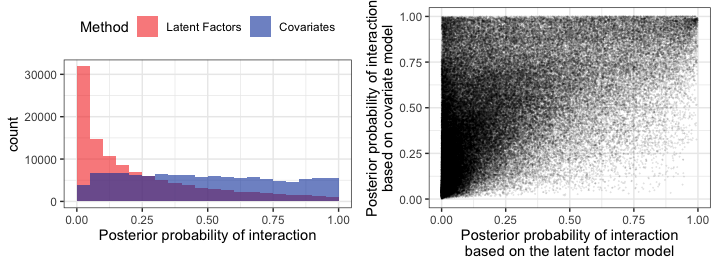} 
\hspace{20pt}
\caption{Comparison of Model Predictions. Values close to 1 indicate that an interaction between the two species is likely to be possible. Observed interactions are excluded from the plot. (Left) Histograms of posterior interaction probabilities for all pairs of species without a recorded interaction based on the two models. (Right) The predictions from the two models are plotted against each other.}
\label{fig:app_compare}
\end{figure}

We further compare results from the two models in \cref{fig:app_compare}. On the left, we see that the model that employs covariates directly returns posterior probabilities of interaction that are above 0.5 for approximately half of the pairs. In contrast, our model predicts that a non-negligible, but more realistic, proportion of pairs are interactive. Furthermore, our model identifies a large number of pairs which are likely {\it impossible} to interact (41\% of pairs have posterior probability of interaction that is below 0.1) in contrast to the covariate model (9\% of pairs).

We study the model predictions against each other in the right side of \cref{fig:app_compare}. We focus on pairs that are likely/not likely to interact, defined as pairs with posterior probability of interaction ranging between 0--0.1 and 0.9--1, respectively.
We find that the two models sometimes return opposite conclusions, though, when that happens, the covariate model has posterior probabilities close to 1 and the latent factor model close to 0. This pattern is more general, in that the latent factor model almost always returns posterior probabilities of interaction that are lower than those based on the covariate model.
The disagreement in models' prediction, in combination with the unrealistically high number of pairs predicted to interact based on the covariates model and the cross-validation results lead us to believe that the predictions of {\it truly interactive} pairs based on the covariates model are not trustworthy. Therefore, we turn our attention to investigating the models' relative performance in identifying pairs that are unlikely to interact. 89\% of the pairs of species which the covariates model identifies as unlikely to interact are also identified as unlikely to interact based on the latent factor model. Moreover, model predictions for these pairs are substantially more precise based on the latent factor model with a median posterior standard deviation equal to 0.13, compared to 0.24 for the covariate model.
Reversely, pairs of species that are unlikely to interact based on our model often also have low probability of interaction according to the covariates model. However, for these pairs, the posterior probabilities of interaction based on the covariate model range more widely over 0--0.5, with 20\% of them being below 0.1 and 76\% below 0.5. Therefore, even though the models seem to partially agree on which pairs of species do not interact, the latent factor model is more confident in these predictions.

The latent factor model also has an overall lower posterior uncertainty in estimated probabilities of interaction than the covariate model, average posterior standard deviation equal to 0.31 and 0.39, respectively.

\begin{figure}[H]
\centering
\includegraphics[width = 0.61 \textwidth, trim = 0 18 0 0, clip]{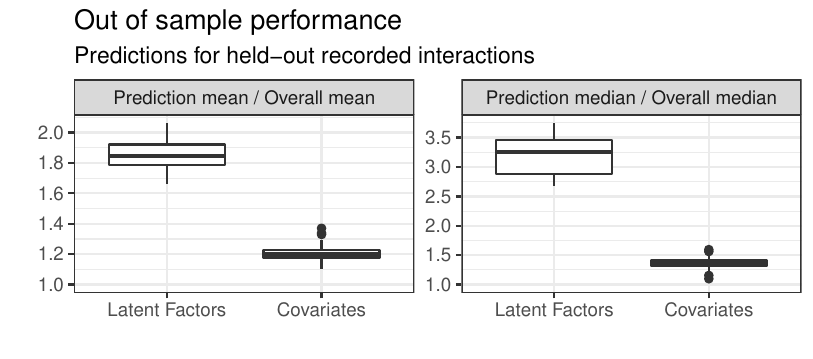}
\caption{Cross Validation Results. For each data set, we calculate the mean and median of the posterior interaction probability in the held-out pairs, and in the overall population. The plot shows boxplots for the ratio of held-out to overall mean (left) and median (right) by model. Higher values indicate that the true interactions were identified more clearly.}
\label{fig:app_cv}
\end{figure}

\cref{fig:app_cv} shows visually the cross-validation results discussed in \cref{subsec:model_comparison}. We see that, for both approaches, the ratio of posterior probability of interaction for the held-out pairs compared to all pairs is above 1, indicating that both of them assign higher posterior probability to the pairs that are known to interact. However, the ratio is much higher for the latent factor approach illustrating that it separates pairs that interact from those that do not more effectively than the approach that uses covariates.

\subsection{Results based on alternative specification of intra-species correlations}

\begin{figure}[t]
\centering
\includegraphics[width=0.55\textwidth]{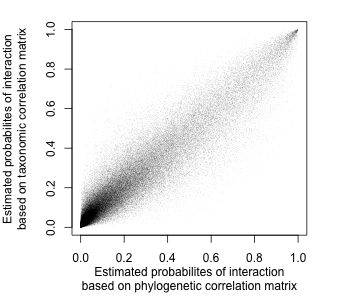} 
\caption{Estimated probabilities of pair interaction using our model with a phylogentic correlation matrix (x-axis) and a taxonomic correlation matrix (y-axis).}
\label{supp_fig:tax_phylo_results}
\end{figure}

To evaluate the robustness of our results to alternative specifications of intra-species dependencies, we also fit our model with taxonomically-structured correlation matrices. 
% Here, we used the taxonomic tree to specify the matrix $\bm C$ in the latent factors' correlation structure as an approximation to the correlation matrix based on the phylogenetic tree \cite[see][Section 6.7]{Ovaskainen2020joint}.
For bird species $i, i'$, we specify $[C_\latB]_{ii'} = 0$ if the species are unrelated, $[C_\latB]_{ii'} = 1$ if $i = i'$, and $[C_\latB]_{ii'}$ equal to 0.75, 0.5, and 0.25 if the species belong to the same genera (very similar), family (similar), or order (somewhat similar), respectively.
% \begin{align*}
% [C_\latB]_{ii'} =
% \begin{cases*}
% 1,     & if $i = i'$, \\
% 0.75,  & if they belong to the same genus (very similar), \\
% 0.5,   & if they belong to the same family (similar), \\
% 0.25   & if they belong to the same order (somewhat similar), and \\
% 0      & if they are unrelated,
% \end{cases*}
% \end{align*}
We use a similar specification for $\bm C_\latP$, where plant species are organized in genera and genera in families. In \cref{supp_fig:tax_phylo_results} we show the estimated probabilities of interaction based when using the taxonomically-based and the the phylogenetically-based correlation matrices. The estimated probabilities are centered and close to the 45 degree line. When regressing the estimated probabilities using the phylogenetic matrix on the estimated probabilities using the taxonomic matrix, the intercept is estimated to be -0.003 (very close to 0), and the slope is estimated to be 0.995 (very close to 1), also illustrating that the models return very similar estimated probabilities of interaction. Therefore, the method is rather robust to the codification of intra-species dependencies. We investigate the out-of-sample predictive accuracy when we use taxonomically-structured correlation matrices in Supplement \ref{supp_sec:trait_phylo_importance} and
\cref{supp_tab:app_trait_importance}.
% Therefore, $\bm C_\latB, \bm C_\latP$ are correlation matrices which are motivated by the species' evolutionary process. In the correlation structure of the latent factors, $\bm \Sigma = \rho \bm C + (1 - \rho) \bm I$,

\subsection{The information content of species traits and phylogenetic information for predicting missing interactions}
\label{supp_sec:trait_phylo_importance}

Our variable importance procedure orders the traits in terms of their importance for predicting interactions. In order to study the information content of species' traits and phylogenetic information for predicting missing interactions, we alter the cross-validation study discussed in \cref{subsec:model_comparison}. In terms of traits, we focus on the top continuous traits for birds and plants: bird body mass and plant fruit diameter.

Our procedure is as follows: We hold out 100 measured interactions by setting the corresponding $A_{ijs} = 0$ for all $s$. We fit our approach when using all the covariates, when using all the covariates excluding bird body mass, when using all the covariates excluding plant fruit diameter, and when using our approach while setting the correlation parameters in the latent factor correlation matrix $\rho_\latB, \rho_\latP$ to zero. For each model fit, we calculate the average and median posterior probability of interaction for the 100 held-out pairs, and the average and median posterior probability of interaction for all pairs with an unrecorded interaction. The latter set necessarily includes pairs that are not able to interact. Therefore, if the model can separate truly interacting pairs from non-interacting pairs the ratio of these two quantities would be large. We repeat this procedure 30 times, each time holding out a different set of 100 interacting pairs. We calculate the average of the ratio of average posterior probabilities and median posterior probabilities.

The results are shown in \cref{supp_tab:app_trait_importance}. We find that when considering the average posterior probability of interaction, all three sets of covariates (a, b, c) perform similarly. However, when studying the ratio of median posterior probabilities for the held-out pairs and for all pairs, the approach's performance when we exclude fruit diameter (c) is relatively lower than the approach's performance when we use all the covariates (a). In contrast, the approach's performance is similar when we use all the covariates and when we exclude body mass. This is true despite the fact that there are substantially more covariates available for plants (12) compared to birds (5), and despite the fact that fruit length is also available in our data (correlation of fruit diameter and length is 0.72). These results suggest that fruit diameter is an important covariate to measure when predicting missing species' interactions. The importance of measuring fruit diameter might be even stronger if the highly-correlated covariate representing fruit length was excluded.

\begin{table}[!t]
    \centering
    \caption{Cross validation results to assess fit improvement attributed to different traits and phylogenetic information. Larger values imply that the approach can better differentiate truly interacting pairs from the set of pairs without a recorded interaction. The approaches considered are the following (as shown in the table): 
    (a) the proposed approach using all covariates and phylogenetic correlation matrix for the latent factors, 
    (b) the proposed approach excluding body mass but using the phylogenetic correlation matrix, 
    (c) same as above but excluding fruit diameter,
    (d) the proposed approach will all covariates but with\textit{out} a phylogenetic correlation matrix ($\rho_\latB, \rho_\latP = 0$), 
    (e) the proposed approach will all covariates but replacing the phylogenetic correlation matrix with a taxonomic correlation matrix, and 
    (f) the model that is based directly on covariates.}
    \label{supp_tab:app_trait_importance}
    \begin{tabular}{ll|cc}
& & Ratio of averages & Ratio of medians \\ \hline
(a) & Latent factor model & 1.85 & 3.19 \\
& (all covariates, phylogeny) & & \\ \hline
(b) & Latent factor model & 1.86 & 3.17  \\
& excluding body mass & & \\
(c) & Latent factor model & 1.84 & 3.06  \\ 
& excluding fruit diameter& &  \\ \hline
(d) & Latent factor model & 1.45 & 1.66 \\
& excluding phylogenetic information & & \\
(e) & Latent factor model & 1.79 & 2.71 \\
& using taxonomic information & & \\ \hline
(f) & Covariate model & 1.21 & 1.36 \\ \hline
    \end{tabular}
\end{table}

Similarly, we notice that the approach's ability in predicting missing interactions is much worse when no aspect of phylogenetic information is used. In \cref{supp_tab:app_trait_importance} we compare the complete model (a), with the model that does not use phylogenetic information by setting $\rho_\latB, \rho_\latP = 0$ (d), and the model that uses taxonomic information instead of phylogenetic information (e). We see that using taxonomic information performs worse than using phylogenetic information, and that using no phylogenetic or taxonomic information at all shows significant loss in predictive accuracy. Therefore, incorporating species' phylogenetic relationships is important for predicting missing interactions. We note here that even when phylogenetic information is completely excluded (d), the latent factor model performs better than the model that uses covariates directly (f).

% ---- DIAGNOSTICS ------- %

\section{MCMC diagnostics}
\label{supp_sec:diagnostics}

\begin{figure}[!p]
\centering
\includegraphics[width = \textwidth]{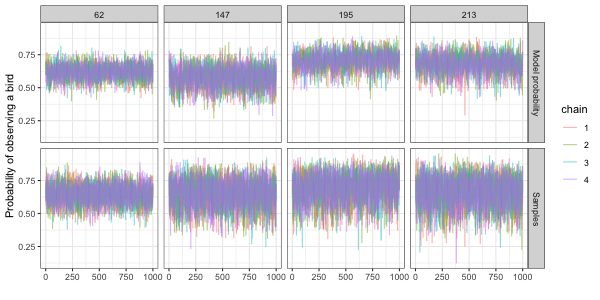} % \\[20pt]
\caption{Detectability scores for four randomly chosen birds. Traceplots for the linear predictor (Row 1) and posterior samples (Row 2). Colors correspond to different MCMC chains.}
\label{supp_fig:diag_prob_obs_B}

\vspace{20pt}

\includegraphics[width = 0.9 \textwidth]{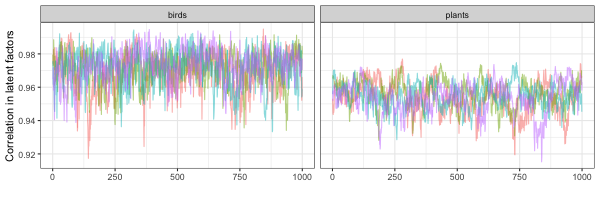}
\caption{Correlation of latent factors. Traceplots showing MCMC samples from the posterior distributions of $\rho_\latB$ (left) and $\rho_\latP$ (right). Colors correspond to different MCMC chains.}
\label{supp_fig:diag_correlations}
\end{figure}

\begin{figure}[!p]
\centering
\includegraphics[width = \textwidth]{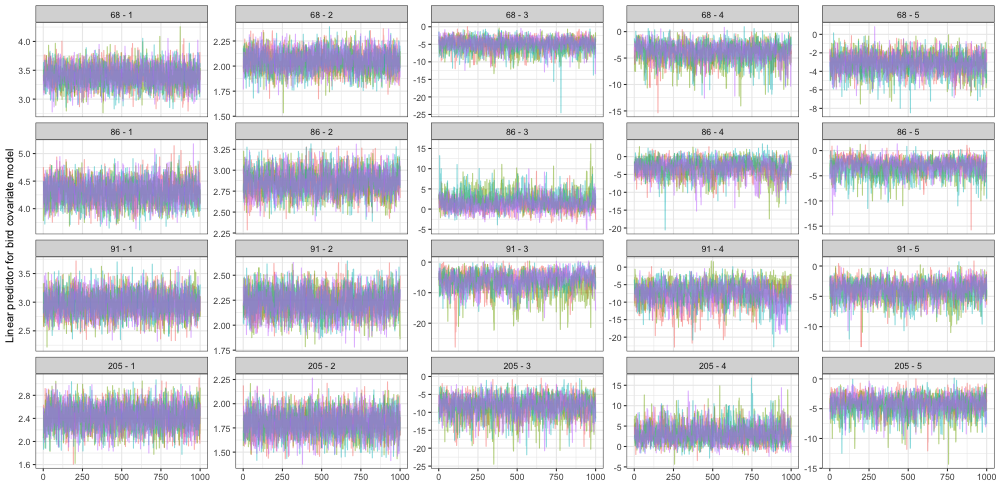}
\caption{Linear predictor of trait models for bird species. Traceplots for the linear predictor of all traits for four randomly chosen species of birds. The rows correspond to different bird species and the columns correspond to observed traits. The first two traits are continuous and the last three traits are binary.}
\label{supp_fig:diag_linpred_X}

\vspace{20pt}

\includegraphics[width = 0.8 \textwidth]{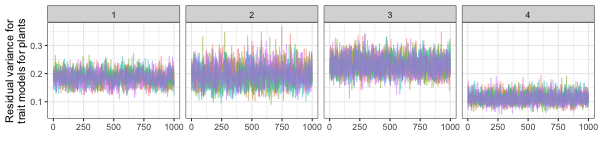}
\caption{Residual variances of continuous traits for plant species.}
\label{supp_fig:diag_sigmasq_l}
\end{figure}

\begin{figure}[!p]
\centering
\includegraphics[width = 0.8 \textwidth]{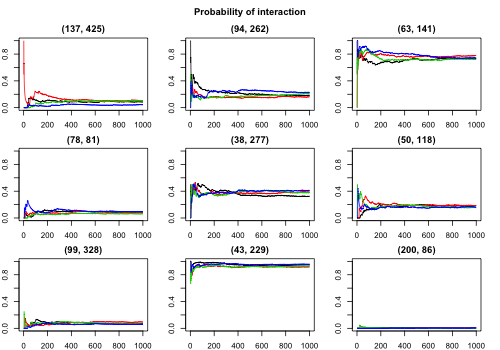}
\caption{Probability of species interaction. Running means of the indicator $L_{ij}$ representing whether species $i, j$ interact. Running means are shown for nine pairs of species without a recorded interaction.}
\label{supp_fig:diag_prob_inter}
\end{figure}

We evaluated convergence of MCMC scheme by studying traceplots of identifiable parameters across chains. Since latent factors and their corresponding coefficients are not identifiable, we focused our attention to: linear predictors and residual variances for the trait models, probabilities of detection and residual variances for both sets of species, and the correlation parameters $\rho_U, \rho_V$ in the latent factors' covariance structure across species in the same set. The traceplots for a subset of parameters are shown in Figures \ref{supp_fig:diag_prob_obs_B}, \ref{supp_fig:diag_correlations}, \ref{supp_fig:diag_linpred_X}, and \ref{supp_fig:diag_sigmasq_l}. Overall convergence seems to be good. We observe some autocorrelation in the posterior samples for the latent factor correlation parameters $\rho$, but mixing seems to be good, and the values over which the MCMC concentrates are similar and very close to 1.
We also investigated running means for the interactions indicators. If the posterior distribution is unimodal and the MCMC has converged sufficiently well, the running means converge to the same point. Running means for nine pairs of species without a recorded interaction are shown in \cref{supp_fig:diag_prob_inter}.

\clearpage

\section{List of all species included in our analysis}
\label{supp_sec:list_species}

Tables \ref{supp_tab:bird_species} and \ref{supp_tab:plant_species} show the bird and plant species, respectively, that are included in our analysis along with their taxonomic information. The column ``Included'' denotes whether the species are shown in the results of the main text or not. The species are listed in the same order shown in the results of \cref{sec:application} and Supplement \ref{supp_sec:app_results}.

{\singlespacing
\footnotesize
\begin{longtable}{rlllc}
\caption{Bird species included in our data and their taxonomic information.} \\
\label{supp_tab:bird_species}
Species & Order & Family & Genus & Included \\ 
  \hline
Pyroderus scutatus & Passeriformes & Cotingidae & Pyroderus & X \\ 
  Pachyramphus validus & Passeriformes & Cotingidae & Pachyramphus & X \\ 
  Pachyramphus castaneus & Passeriformes & Cotingidae & Pachyramphus & X \\ 
  Pachyramphus viridis & Passeriformes & Cotingidae & Pachyramphus & X \\ 
  Pachyramphus polychopterus & Passeriformes & Cotingidae & Pachyramphus & X \\ 
  Lipaugus lanioides & Passeriformes & Cotingidae & Lipaugus & X \\ 
  Lipaugus vociferans & Passeriformes & Cotingidae & Lipaugus & X \\ 
  Tityra cayana & Passeriformes & Cotingidae & Tityra & X \\ 
  Tityra inquisitor & Passeriformes & Cotingidae & Tityra & X \\ 
  Oxyruncus cristatus & Passeriformes & Cotingidae & Oxyruncus & X \\ 
  Carpornis cucullata & Passeriformes & Cotingidae & Carpornis & X \\ 
  Carpornis melanocephala & Passeriformes & Cotingidae & Carpornis & X \\ 
  Schiffornis virescens & Passeriformes & Cotingidae & Schiffornis & X \\ 
  Procnias nudicollis & Passeriformes & Cotingidae & Procnias & X \\ 
  Tijuca atra & Passeriformes & Cotingidae & Tijuca & X \\ 
  Phibalura flavirostris & Passeriformes & Cotingidae & Phibalura & X \\ 
  Laniisoma elegans & Passeriformes & Cotingidae & Laniisoma & X \\ 
  Cacicus haemorrhous & Passeriformes & Icteridae & Cacicus &  \\ 
  Cacicus chrysopterus & Passeriformes & Icteridae & Cacicus &  \\ 
  Chrysomus ruficapillus & Passeriformes & Icteridae & Chrysomus &  \\ 
  Molothrus bonariensis & Passeriformes & Icteridae & Molothrus &  \\ 
  Icterus cayanensis & Passeriformes & Icteridae & Icterus &  \\ 
  Pseudoleistes guirahuro & Passeriformes & Icteridae & Pseudoleistes &  \\ 
  Psarocolius decumanus & Passeriformes & Icteridae & Psarocolius &  \\ 
  Gnorimopsar chopi & Passeriformes & Icteridae & Gnorimopsar &  \\ 
  Sicalis flaveola & Passeriformes & Thraupidae & Sicalis & X \\ 
  Thraupis palmarum & Passeriformes & Thraupidae & Thraupis & X \\ 
  Thraupis episcopus & Passeriformes & Thraupidae & Thraupis & X \\ 
  Thraupis sayaca & Passeriformes & Thraupidae & Thraupis & X \\ 
  Thraupis cyanoptera & Passeriformes & Thraupidae & Thraupis & X \\ 
  Thraupis ornata & Passeriformes & Thraupidae & Thraupis & X \\ 
  Thraupis bonariensis & Passeriformes & Thraupidae & Thraupis & X \\ 
  Tachyphonus coronatus & Passeriformes & Thraupidae & Tachyphonus & X \\ 
  Tachyphonus cristatus & Passeriformes & Thraupidae & Tachyphonus & X \\ 
  Tachyphonus rufus & Passeriformes & Thraupidae & Tachyphonus & X \\ 
  Tangara cayana & Passeriformes & Thraupidae & Tangara & X \\ 
  Tangara seledon & Passeriformes & Thraupidae & Tangara & X \\ 
  Tangara mexicana & Passeriformes & Thraupidae & Tangara & X \\ 
  Tangara desmaresti & Passeriformes & Thraupidae & Tangara & X \\ 
  Tangara cyanocephala & Passeriformes & Thraupidae & Tangara & X \\ 
  Tangara cyanoptera & Passeriformes & Thraupidae & Tangara & X \\ 
  Tangara preciosa & Passeriformes & Thraupidae & Tangara & X \\ 
  Tangara cyanoventris & Passeriformes & Thraupidae & Tangara & X \\ 
  Tangara peruviana & Passeriformes & Thraupidae & Tangara & X \\ 
  Dacnis cayana & Passeriformes & Thraupidae & Dacnis & X \\ 
  Dacnis nigripes & Passeriformes & Thraupidae & Dacnis & X \\ 
  Ramphocelus carbo & Passeriformes & Thraupidae & Ramphocelus & X \\ 
  Ramphocelus bresilius & Passeriformes & Thraupidae & Ramphocelus & X \\ 
  Thlypopsis sordida & Passeriformes & Thraupidae & Thlypopsis & X \\ 
  Conirostrum speciosum & Passeriformes & Thraupidae & Conirostrum & X \\ 
  Hemithraupis guira & Passeriformes & Thraupidae & Hemithraupis & X \\ 
  Hemithraupis ruficapilla & Passeriformes & Thraupidae & Hemithraupis & X \\ 
  Hemithraupis flavicollis & Passeriformes & Thraupidae & Hemithraupis & X \\ 
  Tersina viridis & Passeriformes & Thraupidae & Tersina & X \\ 
  Chlorophanes spiza & Passeriformes & Thraupidae & Chlorophanes & X \\ 
  Pipraeidea melanonota & Passeriformes & Thraupidae & Pipraeidea & X \\ 
  Schistochlamys ruficapillus & Passeriformes & Thraupidae & Schistochlamys & X \\ 
  Schistochlamys melanopis & Passeriformes & Thraupidae & Schistochlamys & X \\ 
  Cissopis leverianus & Passeriformes & Thraupidae & Cissopis & X \\ 
  Orthogonys chloricterus & Passeriformes & Thraupidae & Orthogonys & X \\ 
  Trichothraupis melanops & Passeriformes & Thraupidae & Trichothraupis & X \\ 
  Cyanerpes cyaneus & Passeriformes & Thraupidae & Cyanerpes & X \\ 
  Stephanophorus diadematus & Passeriformes & Thraupidae & Stephanophorus & X \\ 
  Nemosia pileata & Passeriformes & Thraupidae & Nemosia & X \\ 
  Coryphospingus cucullatus & Passeriformes & Thraupidae & Coryphospingus & X \\ 
  Coryphospingus pileatus & Passeriformes & Thraupidae & Coryphospingus & X \\ 
  Volatinia jacarina & Passeriformes & Thraupidae & Volatinia & X \\ 
  Sporophila caerulescens & Passeriformes & Thraupidae & Sporophila & X \\ 
  Sporophila nigricollis & Passeriformes & Thraupidae & Sporophila & X \\ 
  Sporophila leucoptera & Passeriformes & Thraupidae & Sporophila & X \\ 
  Haplospiza unicolor & Passeriformes & Thraupidae & Haplospiza & X \\ 
  Orchesticus abeillei & Passeriformes & Thraupidae & Orchesticus & X \\ 
  Poospiza thoracica & Passeriformes & Thraupidae & Poospiza & X \\ 
  Poospiza lateralis & Passeriformes & Thraupidae & Poospiza & X \\ 
  Eucometis penicillata & Passeriformes & Thraupidae & Eucometis & X \\ 
  Pyrrhocoma ruficeps & Passeriformes & Thraupidae & Pyrrhocoma & X \\ 
  Elaenia flavogaster & Passeriformes & Tyrannidae & Elaenia & X \\ 
  Elaenia spectabilis & Passeriformes & Tyrannidae & Elaenia & X \\ 
  Elaenia chiriquensis & Passeriformes & Tyrannidae & Elaenia & X \\ 
  Elaenia mesoleuca & Passeriformes & Tyrannidae & Elaenia & X \\ 
  Elaenia cristata & Passeriformes & Tyrannidae & Elaenia & X \\ 
  Elaenia obscura & Passeriformes & Tyrannidae & Elaenia & X \\ 
  Elaenia albiceps & Passeriformes & Tyrannidae & Elaenia & X \\ 
  Elaenia parvirostris & Passeriformes & Tyrannidae & Elaenia & X \\ 
  Myiodynastes maculatus & Passeriformes & Tyrannidae & Myiodynastes & X \\ 
  Tyrannus melancholicus & Passeriformes & Tyrannidae & Tyrannus & X \\ 
  Tyrannus savana & Passeriformes & Tyrannidae & Tyrannus & X \\ 
  Tyrannus tyrannus & Passeriformes & Tyrannidae & Tyrannus & X \\ 
  Pitangus sulphuratus & Passeriformes & Tyrannidae & Pitangus & X \\ 
  Myiozetetes similis & Passeriformes & Tyrannidae & Myiozetetes & X \\ 
  Myiozetetes cayanensis & Passeriformes & Tyrannidae & Myiozetetes & X \\ 
  Myiarchus ferox & Passeriformes & Tyrannidae & Myiarchus & X \\ 
  Myiarchus swainsoni & Passeriformes & Tyrannidae & Myiarchus & X \\ 
  Myiarchus tyrannulus & Passeriformes & Tyrannidae & Myiarchus & X \\ 
  Cnemotriccus fuscatus & Passeriformes & Tyrannidae & Cnemotriccus & X \\ 
  Mionectes oleagineus & Passeriformes & Tyrannidae & Mionectes & X \\ 
  Mionectes rufiventris & Passeriformes & Tyrannidae & Mionectes & X \\ 
  Megarynchus pitangua & Passeriformes & Tyrannidae & Megarynchus & X \\ 
  Machetornis rixosa & Passeriformes & Tyrannidae & Machetornis & X \\ 
  Attila rufus & Passeriformes & Tyrannidae & Attila & X \\ 
  Attila phoenicurus & Passeriformes & Tyrannidae & Attila & X \\ 
  Empidonomus varius & Passeriformes & Tyrannidae & Empidonomus & X \\ 
  Colonia colonus & Passeriformes & Tyrannidae & Colonia & X \\ 
  Phyllomyias fasciatus & Passeriformes & Tyrannidae & Phyllomyias & X \\ 
  Phyllomyias griseocapilla & Passeriformes & Tyrannidae & Phyllomyias & X \\ 
  Camptostoma obsoletum & Passeriformes & Tyrannidae & Camptostoma & X \\ 
  Conopias trivirgatus & Passeriformes & Tyrannidae & Conopias & X \\ 
  Leptopogon amaurocephalus & Passeriformes & Tyrannidae & Leptopogon & X \\ 
  Tolmomyias sulphurescens & Passeriformes & Tyrannidae & Tolmomyias & X \\ 
  Tolmomyias flaviventris & Passeriformes & Tyrannidae & Tolmomyias & X \\ 
  Lathrotriccus euleri & Passeriformes & Tyrannidae & Lathrotriccus & X \\ 
  Phylloscartes ventralis & Passeriformes & Tyrannidae & Phylloscartes & X \\ 
  Phylloscartes sylviolus & Passeriformes & Tyrannidae & Phylloscartes & X \\ 
  Phylloscartes oustaleti & Passeriformes & Tyrannidae & Phylloscartes & X \\ 
  Knipolegus nigerrimus & Passeriformes & Tyrannidae & Knipolegus & X \\ 
  Knipolegus cyanirostris & Passeriformes & Tyrannidae & Knipolegus & X \\ 
  Legatus leucophaius & Passeriformes & Tyrannidae & Legatus & X \\ 
  Xolmis cinereus & Passeriformes & Tyrannidae & Xolmis & X \\ 
  Xolmis velatus & Passeriformes & Tyrannidae & Xolmis & X \\ 
  Fluvicola nengeta & Passeriformes & Tyrannidae & Fluvicola & X \\ 
  Serpophaga subcristata & Passeriformes & Tyrannidae & Serpophaga & X \\ 
  Myiophobus fasciatus & Passeriformes & Tyrannidae & Myiophobus & X \\ 
  Satrapa icterophrys & Passeriformes & Tyrannidae & Satrapa & X \\ 
  Capsiempis flaveola & Passeriformes & Tyrannidae & Capsiempis & X \\ 
  Casiornis rufus & Passeriformes & Tyrannidae & Casiornis & X \\ 
  Contopus cinereus & Passeriformes & Tyrannidae & Contopus & X \\ 
  Sirystes sibilator & Passeriformes & Tyrannidae & Sirystes & X \\ 
  Myiopagis caniceps & Passeriformes & Tyrannidae & Myiopagis & X \\ 
  Phaeomyias murina & Passeriformes & Tyrannidae & Phaeomyias & X \\ 
  Chiroxiphia caudata & Passeriformes & Pipridae & Chiroxiphia & X \\ 
  Chiroxiphia pareola & Passeriformes & Pipridae & Chiroxiphia & X \\ 
  Manacus manacus & Passeriformes & Pipridae & Manacus & X \\ 
  Pipra rubrocapilla & Passeriformes & Pipridae & Pipra & X \\ 
  Pipra pipra & Passeriformes & Pipridae & Pipra & X \\ 
  Ilicura militaris & Passeriformes & Pipridae & Ilicura & X \\ 
  Antilophia galeata & Passeriformes & Pipridae & Antilophia & X \\ 
  Neopelma aurifrons & Passeriformes & Pipridae & Neopelma & X \\ 
  Neopelma pallescens & Passeriformes & Pipridae & Neopelma & X \\ 
  Machaeropterus regulus & Passeriformes & Pipridae & Machaeropterus & X \\ 
  Coereba flaveola & Passeriformes & Coerebidae & Coereba &  \\ 
  Turdus amaurochalinus & Passeriformes & Turdidae & Turdus &  \\ 
  Turdus flavipes & Passeriformes & Turdidae & Turdus &  \\ 
  Turdus leucomelas & Passeriformes & Turdidae & Turdus &  \\ 
  Turdus rufiventris & Passeriformes & Turdidae & Turdus &  \\ 
  Turdus albicollis & Passeriformes & Turdidae & Turdus &  \\ 
  Turdus subalaris & Passeriformes & Turdidae & Turdus &  \\ 
  Turdus fumigatus & Passeriformes & Turdidae & Turdus &  \\ 
  Catharus fuscescens & Passeriformes & Turdidae & Catharus &  \\ 
  Zonotrichia capensis & Passeriformes & Emberizidae & Zonotrichia &  \\ 
  Arremon flavirostris & Passeriformes & Emberizidae & Arremon &  \\ 
  Arremon taciturnus & Passeriformes & Emberizidae & Arremon &  \\ 
  Vireo olivaceus & Passeriformes & Vireonidae & Vireo &  \\ 
  Hylophilus amaurocephalus & Passeriformes & Vireonidae & Hylophilus &  \\ 
  Hylophilus thoracicus & Passeriformes & Vireonidae & Hylophilus &  \\ 
  Hylophilus poicilotis & Passeriformes & Vireonidae & Hylophilus &  \\ 
  Cyclarhis gujanensis & Passeriformes & Vireonidae & Cyclarhis &  \\ 
  Saltator maximus & Passeriformes & Cardinalidae & Saltator &  \\ 
  Saltator similis & Passeriformes & Cardinalidae & Saltator &  \\ 
  Saltator fuliginosus & Passeriformes & Cardinalidae & Saltator &  \\ 
  Saltator coerulescens & Passeriformes & Cardinalidae & Saltator &  \\ 
  Saltator maxillosus & Passeriformes & Cardinalidae & Saltator &  \\ 
  Saltator atricollis & Passeriformes & Cardinalidae & Saltator &  \\ 
  Habia rubica & Passeriformes & Cardinalidae & Habia &  \\ 
  Piranga flava & Passeriformes & Cardinalidae & Piranga &  \\ 
  Cyanocompsa brissonii & Passeriformes & Cardinalidae & Cyanocompsa &  \\ 
  Mimus saturninus & Passeriformes & Mimidae & Mimus &  \\ 
  Mimus gilvus & Passeriformes & Mimidae & Mimus &  \\ 
  Cyanocorax cristatellus & Passeriformes & Corvidae & Cyanocorax &  \\ 
  Cyanocorax cyanomelas & Passeriformes & Corvidae & Cyanocorax &  \\ 
  Cyanocorax caeruleus & Passeriformes & Corvidae & Cyanocorax &  \\ 
  Cyanocorax chrysops & Passeriformes & Corvidae & Cyanocorax &  \\ 
  Euphonia violacea & Passeriformes & Fringillidae & Euphonia &  \\ 
  Euphonia pectoralis & Passeriformes & Fringillidae & Euphonia &  \\ 
  Euphonia chlorotica & Passeriformes & Fringillidae & Euphonia &  \\ 
  Euphonia chalybea & Passeriformes & Fringillidae & Euphonia &  \\ 
  Euphonia xanthogaster & Passeriformes & Fringillidae & Euphonia &  \\ 
  Euphonia cyanocephala & Passeriformes & Fringillidae & Euphonia &  \\ 
  Chlorophonia cyanea & Passeriformes & Fringillidae & Chlorophonia &  \\ 
  Thamnophilus caerulescens & Passeriformes & Thamnophilidae & Thamnophilus &  \\ 
  Thamnophilus doliatus & Passeriformes & Thamnophilidae & Thamnophilus &  \\ 
  Myiothlypis flaveola & Passeriformes & Parulidae & Myiothlypis &  \\ 
  Setophaga pitiayumi & Passeriformes & Parulidae & Setophaga &  \\ 
  Basileuterus culicivorus & Passeriformes & Parulidae & Basileuterus &  \\ 
  Geothlypis aequinoctialis & Passeriformes & Parulidae & Geothlypis &  \\ 
  Estrilda astrild & Passeriformes & Estrildidae & Estrilda &  \\ 
  Cranioleuca pallida & Passeriformes & Furnariidae & Cranioleuca &  \\ 
  Synallaxis ruficapilla & Passeriformes & Furnariidae & Synallaxis &  \\ 
  Furnarius rufus & Passeriformes & Furnariidae & Furnarius &  \\ 
  Troglodytes aedon & Passeriformes & Troglodytidae & Troglodytes &  \\ 
  Crotophaga ani & Cuculiformes & Cuculidae & Crotophaga &  \\ 
  Crotophaga major & Cuculiformes & Cuculidae & Crotophaga &  \\ 
  Guira guira & Cuculiformes & Cuculidae & Guira &  \\ 
  Piaya cayana & Cuculiformes & Cuculidae & Piaya &  \\ 
  Patagioenas picazuro & Columbiformes & Columbidae & Patagioenas &  \\ 
  Patagioenas cayennensis & Columbiformes & Columbidae & Patagioenas &  \\ 
  Patagioenas plumbea & Columbiformes & Columbidae & Patagioenas &  \\ 
  Patagioenas speciosa & Columbiformes & Columbidae & Patagioenas &  \\ 
  Leptotila verreauxi & Columbiformes & Columbidae & Leptotila &  \\ 
  Leptotila rufaxilla & Columbiformes & Columbidae & Leptotila &  \\ 
  Zenaida auriculata & Columbiformes & Columbidae & Zenaida &  \\ 
  Columbina talpacoti & Columbiformes & Columbidae & Columbina &  \\ 
  Penelope superciliaris & Craciformes & Cracidae & Penelope &  \\ 
  Penelope obscura & Craciformes & Cracidae & Penelope &  \\ 
  Aburria jacutinga & Craciformes & Cracidae & Aburria &  \\ 
  Ortalis guttata & Craciformes & Cracidae & Ortalis &  \\ 
  Ortalis canicollis & Craciformes & Cracidae & Ortalis &  \\ 
  Crax blumenbachii & Craciformes & Cracidae & Crax &  \\ 
  Ramphastos dicolorus & Piciformes & Ramphastidae & Ramphastos &  \\ 
  Ramphastos toco & Piciformes & Ramphastidae & Ramphastos &  \\ 
  Ramphastos vitellinus & Piciformes & Ramphastidae & Ramphastos &  \\ 
  Baillonius bailloni & Piciformes & Ramphastidae & Baillonius &  \\ 
  Selenidera maculirostris & Piciformes & Ramphastidae & Selenidera &  \\ 
  Pteroglossus aracari & Piciformes & Ramphastidae & Pteroglossus &  \\ 
  Pteroglossus castanotis & Piciformes & Ramphastidae & Pteroglossus &  \\ 
  Picumnus cirratus & Piciformes & Picidae & Picumnus & X \\ 
  Picumnus nebulosus & Piciformes & Picidae & Picumnus & X \\ 
  Celeus flavescens & Piciformes & Picidae & Celeus & X \\ 
  Melanerpes flavifrons & Piciformes & Picidae & Melanerpes & X \\ 
  Melanerpes candidus & Piciformes & Picidae & Melanerpes & X \\ 
  Colaptes campestris & Piciformes & Picidae & Colaptes & X \\ 
  Colaptes melanochloros & Piciformes & Picidae & Colaptes & X \\ 
  Veniliornis spilogaster & Piciformes & Picidae & Veniliornis & X \\ 
  Piculus aurulentus & Piciformes & Picidae & Piculus & X \\ 
  Dryocopus lineatus & Piciformes & Picidae & Dryocopus & X \\ 
  Trogon surrucura & Trogoniformes & Trogonidae & Trogon &  \\ 
  Trogon viridis & Trogoniformes & Trogonidae & Trogon &  \\ 
  Trogon rufus & Trogoniformes & Trogonidae & Trogon &  \\ 
  Trogon curucui & Trogoniformes & Trogonidae & Trogon &  \\ 
  Baryphthengus ruficapillus & Coraciiformes & Momotidae & Baryphthengus &  \\ 
  Coragyps atratus & Accipitriformes & Cathartidae & Coragyps &  \\ 
  Caracara plancus & Falconiformes & Falconidae & Caracara &  \\ 
  Aramides cajanea & Gruiformes & Rallidae & Aramides &  \\ 
   \hline
\end{longtable}}

{\singlespacing
\footnotesize
\begin{longtable}{rllc}
\caption{Plant species included in our data and their taxonomic information.} \\
\label{supp_tab:plant_species}
Species & Family & Genus & Included \\ 
  \hline
Abuta selloana & Menispermaceae & Abuta &  \\ 
  Cissampelos andromorpha & Menispermaceae & Cissampelos &  \\ 
  Acacia auriculiformis & Fabaceae & Acacia &  \\ 
  Andira fraxinifolia & Fabaceae & Andira &  \\ 
  Cajanus cajan & Fabaceae & Cajanus &  \\ 
  Copaifera langsdorffii & Fabaceae & Copaifera &  \\ 
  Copaifera trapezifolia & Fabaceae & Copaifera &  \\ 
  Desmodium incanum & Fabaceae & Desmodium &  \\ 
  Holocalyx balansae & Fabaceae & Holocalyx &  \\ 
  Hymenaea courbaril & Fabaceae & Hymenaea &  \\ 
  Inga edulis & Fabaceae & Inga &  \\ 
  Inga laurina & Fabaceae & Inga &  \\ 
  Inga marginata & Fabaceae & Inga &  \\ 
  Inga sessilis & Fabaceae & Inga &  \\ 
  Samanea tubulosa & Fabaceae & Samanea &  \\ 
  Acnistus arborescens & Solanaceae & Acnistus & X \\ 
  Aureliana fasciculata & Solanaceae & Aureliana & X \\ 
  Cestrum bracteatum & Solanaceae & Cestrum & X \\ 
  Cestrum mariquitense & Solanaceae & Cestrum & X \\ 
  Cestrum schlechtendalii & Solanaceae & Cestrum & X \\ 
  Lycianthes pauciflora & Solanaceae & Lycianthes & X \\ 
  Physalis pubescens & Solanaceae & Physalis & X \\ 
  Solanum aculeatissimum & Solanaceae & Solanum & X \\ 
  Solanum americanum & Solanaceae & Solanum & X \\ 
  Solanum argenteum & Solanaceae & Solanum & X \\ 
  Solanum bullatum & Solanaceae & Solanum & X \\ 
  Solanum corymbiflorum & Solanaceae & Solanum & X \\ 
  Solanum granulosoleprosum & Solanaceae & Solanum & X \\ 
  Solanum inodorum & Solanaceae & Solanum & X \\ 
  Solanum mauritianum & Solanaceae & Solanum & X \\ 
  Solanum megalochiton & Solanaceae & Solanum & X \\ 
  Solanum myrianthum & Solanaceae & Solanum & X \\ 
  Solanum nigrescens & Solanaceae & Solanum & X \\ 
  Solanum paranense & Solanaceae & Solanum & X \\ 
  Solanum pseudoquina & Solanaceae & Solanum & X \\ 
  Solanum rufescens & Solanaceae & Solanum & X \\ 
  Solanum sanctae-catharinae & Solanaceae & Solanum & X \\ 
  Solanum scuticum & Solanaceae & Solanum & X \\ 
  Solanum subsylvestre & Solanaceae & Solanum & X \\ 
  Solanum swartzianum & Solanaceae & Solanum & X \\ 
  Solanum thomasiifolium & Solanaceae & Solanum & X \\ 
  Solanum variabile & Solanaceae & Solanum & X \\ 
  Solanum viscosissimum & Solanaceae & Solanum & X \\ 
  Vassobia breviflora & Solanaceae & Vassobia & X \\ 
  Acrocomia aculeata & Arecaceae & Acrocomia &  \\ 
  Allagoptera arenaria & Arecaceae & Allagoptera &  \\ 
  Archontophoenix cunninghamiana & Arecaceae & Archontophoenix &  \\ 
  Astrocaryum aculeatissimum & Arecaceae & Astrocaryum &  \\ 
  Attalea dubia & Arecaceae & Attalea &  \\ 
  Bactris gasipaes & Arecaceae & Bactris &  \\ 
  Elaeis guineensis & Arecaceae & Elaeis &  \\ 
  Euterpe edulis & Arecaceae & Euterpe &  \\ 
  Euterpe oleracea & Arecaceae & Euterpe &  \\ 
  Geonoma elegans & Arecaceae & Geonoma &  \\ 
  Geonoma gamiova & Arecaceae & Geonoma &  \\ 
  Geonoma pauciflora & Arecaceae & Geonoma &  \\ 
  Livistona  chinensis & Arecaceae & Livistona &  \\ 
  Livistona chinensis & Arecaceae & Livistona &  \\ 
  Phoenix sylvestris & Arecaceae & Phoenix &  \\ 
  Roystonea oleraceae & Arecaceae & Roystonea &  \\ 
  Syagrus pseudococos & Arecaceae & Syagrus &  \\ 
  Syagrus romanzoffiana & Arecaceae & Syagrus &  \\ 
  Aegiphila integrifolia & Lamiaceae & Aegiphila &  \\ 
  Callicarpa reevesii & Lamiaceae & Callicarpa &  \\ 
  Vitex megapotamica & Lamiaceae & Vitex &  \\ 
  Vitex polygama & Lamiaceae & Vitex &  \\ 
  Aiouea saligna & Lauraceae & Aiouea & X \\ 
  Cryptocarya aschersoniana & Lauraceae & Cryptocarya & X \\ 
  Cryptocarya mandioccana & Lauraceae & Cryptocarya & X \\ 
  Cryptocarya moschata & Lauraceae & Cryptocarya & X \\ 
  Endlicheria paniculata & Lauraceae & Endlicheria & X \\ 
  Nectandra cuspidata & Lauraceae & Nectandra & X \\ 
  Nectandra grandiflora & Lauraceae & Nectandra & X \\ 
  Nectandra lanceolata & Lauraceae & Nectandra & X \\ 
  Nectandra megapotamica & Lauraceae & Nectandra & X \\ 
  Nectandra membranacea & Lauraceae & Nectandra & X \\ 
  Nectandra reticulata & Lauraceae & Nectandra & X \\ 
  Ocotea aeciphila & Lauraceae & Ocotea & X \\ 
  Ocotea bicolor & Lauraceae & Ocotea & X \\ 
  Ocotea catharinensis & Lauraceae & Ocotea & X \\ 
  Ocotea corymbosa & Lauraceae & Ocotea & X \\ 
  Ocotea diospyrifolia & Lauraceae & Ocotea & X \\ 
  Ocotea dispersa & Lauraceae & Ocotea & X \\ 
  Ocotea macropoda & Lauraceae & Ocotea & X \\ 
  Ocotea notata & Lauraceae & Ocotea & X \\ 
  Ocotea odorifera & Lauraceae & Ocotea & X \\ 
  Ocotea puberula & Lauraceae & Ocotea & X \\ 
  Ocotea pulchella & Lauraceae & Ocotea & X \\ 
  Ocotea silvestris & Lauraceae & Ocotea & X \\ 
  Ocotea spixiana & Lauraceae & Ocotea & X \\ 
  Ocotea teleiandra & Lauraceae & Ocotea & X \\ 
  Persea alba & Lauraceae & Persea & X \\ 
  Persea major & Lauraceae & Persea & X \\ 
  Persea willdenovii & Lauraceae & Persea & X \\ 
  Phoebe pickelli & Lauraceae & Phoebe & X \\ 
  Alchornea discolor & Euphorbiaceae & Alchornea &  \\ 
  Alchornea glandulosa & Euphorbiaceae & Alchornea &  \\ 
  Alchornea sidifolia & Euphorbiaceae & Alchornea &  \\ 
  Alchornea triplinervia & Euphorbiaceae & Alchornea &  \\ 
  Sapium glandulosum & Euphorbiaceae & Sapium &  \\ 
  Tetrorchidium rubrivenium & Euphorbiaceae & Tetrorchidium &  \\ 
  Allophylus edulis & Sapindaceae & Allophylus &  \\ 
  Cupania emarginata & Sapindaceae & Cupania &  \\ 
  Cupania oblongifolia & Sapindaceae & Cupania &  \\ 
  Cupania riodocensis & Sapindaceae & Cupania &  \\ 
  Cupania vernalis & Sapindaceae & Cupania &  \\ 
  Litchi chinensis & Sapindaceae & Litchi &  \\ 
  Matayba elaeagnoides & Sapindaceae & Matayba &  \\ 
  Matayba guianensis & Sapindaceae & Matayba &  \\ 
  Paullinia carpopoda & Sapindaceae & Paullinia &  \\ 
  Paullinia micrantha & Sapindaceae & Paullinia &  \\ 
  Paullinia rhomboidea & Sapindaceae & Paullinia &  \\ 
  Paullinia uloptera & Sapindaceae & Paullinia &  \\ 
  Sapindus saponaria & Sapindaceae & Sapindus &  \\ 
  Amaioua guianensis & Rubiaceae & Amaioua & X \\ 
  Amaioua intermedia & Rubiaceae & Amaioua & X \\ 
  Chomelia parvifolia & Rubiaceae & Chomelia & X \\ 
  Coccocypselum geophiloides & Rubiaceae & Coccocypselum & X \\ 
  Coccocypselum hasslerianum & Rubiaceae & Coccocypselum & X \\ 
  Coffea arabica & Rubiaceae & Coffea & X \\ 
  Cordiera myrciifolia & Rubiaceae & Cordiera & X \\ 
  Coussarea contracta & Rubiaceae & Coussarea & X \\ 
  Galium hypocarpium & Rubiaceae & Galium & X \\ 
  Genipa americana & Rubiaceae & Genipa & X \\ 
  Geophila macropoda & Rubiaceae & Geophila & X \\ 
  Geophila repens & Rubiaceae & Geophila & X \\ 
  Guettarda viburnoides & Rubiaceae & Guettarda & X \\ 
  Ixora burchelliana & Rubiaceae & Ixora & X \\ 
  Ixora gardneriana & Rubiaceae & Ixora & X \\ 
  Ixora venulosa & Rubiaceae & Ixora & X \\ 
  Margaritopsis astrellantha & Rubiaceae & Margaritopsis & X \\ 
  Margaritopsis chaenotricha & Rubiaceae & Margaritopsis & X \\ 
  Palicourea macrobotrys & Rubiaceae & Palicourea & X \\ 
  Posoqueria latifolia & Rubiaceae & Posoqueria & X \\ 
  Psychotria carthagenensis & Rubiaceae & Psychotria & X \\ 
  Psychotria forsteronioides & Rubiaceae & Psychotria & X \\ 
  Psychotria gracilenta & Rubiaceae & Psychotria & X \\ 
  Psychotria hoffmannseggiana & Rubiaceae & Psychotria & X \\ 
  Psychotria leiocarpa & Rubiaceae & Psychotria & X \\ 
  Psychotria mapourioides & Rubiaceae & Psychotria & X \\ 
  Psychotria nuda & Rubiaceae & Psychotria & X \\ 
  Psychotria racemosa & Rubiaceae & Psychotria & X \\ 
  Psychotria sessilis & Rubiaceae & Psychotria & X \\ 
  Psychotria suterella & Rubiaceae & Psychotria & X \\ 
  Psychotria vellosiana & Rubiaceae & Psychotria & X \\ 
  Rudgea jasminoides & Rubiaceae & Rudgea & X \\ 
  Rudgea recurva & Rubiaceae & Rudgea & X \\ 
  Tocoyena bullata & Rubiaceae & Tocoyena & X \\ 
  Tocoyena formosa & Rubiaceae & Tocoyena & X \\ 
  Amaranthus hybridus & Amaranthaceae & Amaranthus &  \\ 
  Chamissoa altissima & Amaranthaceae & Chamissoa &  \\ 
  Anacardium occidentale & Anacardiaceae & Anacardium &  \\ 
  Lithrea molleoides & Anacardiaceae & Lithrea &  \\ 
  Mangifera indica & Anacardiaceae & Mangifera &  \\ 
  Schinus terebinthifolius & Anacardiaceae & Schinus &  \\ 
  Tapirira guianensis & Anacardiaceae & Tapirira &  \\ 
  Annona cacans & Annonaceae & Annona &  \\ 
  Annona emarginata & Annonaceae & Annona &  \\ 
  Annona neosericea & Annonaceae & Annona &  \\ 
  Guatteria australis & Annonaceae & Guatteria &  \\ 
  Guatteria sellowiana & Annonaceae & Guatteria &  \\ 
  Xylopia aromatica & Annonaceae & Xylopia &  \\ 
  Xylopia brasiliensis & Annonaceae & Xylopia &  \\ 
  Xylopia langsdorfiana & Annonaceae & Xylopia &  \\ 
  Xylopia sericea & Annonaceae & Xylopia &  \\ 
  Anthurium affine & Araceae & Anthurium &  \\ 
  Anthurium scandens & Araceae & Anthurium &  \\ 
  Anthurium sellowianum & Araceae & Anthurium &  \\ 
  Asterostigma lividum & Araceae & Asterostigma &  \\ 
  Heteropsis oblongifolia & Araceae & Heteropsis &  \\ 
  Heteropsis rigidifolia & Araceae & Heteropsis &  \\ 
  Monstera adansonii & Araceae & Monstera &  \\ 
  Philodendron appendiculatum & Araceae & Philodendron &  \\ 
  Philodendron imbe & Araceae & Philodendron &  \\ 
  Araucaria angustifolia & Araucariaceae & Araucaria &  \\ 
  Artocarpus heterophyllus & Moraceae & Artocarpus & X \\ 
  Ficus  carica & Moraceae & Ficus & X \\ 
  Ficus benjami & Moraceae & Ficus & X \\ 
  Ficus benjamina & Moraceae & Ficus & X \\ 
  Ficus carica & Moraceae & Ficus & X \\ 
  Ficus cestrifolia & Moraceae & Ficus & X \\ 
  Ficus citrifolia & Moraceae & Ficus & X \\ 
  Ficus enormis & Moraceae & Ficus & X \\ 
  Ficus eximia & Moraceae & Ficus & X \\ 
  Ficus guaranitica & Moraceae & Ficus & X \\ 
  Ficus hirsuta & Moraceae & Ficus & X \\ 
  Ficus insipida & Moraceae & Ficus & X \\ 
  Ficus luschnathiana & Moraceae & Ficus & X \\ 
  Ficus luschthia & Moraceae & Ficus & X \\ 
  Ficus microcarpa & Moraceae & Ficus & X \\ 
  Ficus organensis & Moraceae & Ficus & X \\ 
  Ficus pertusa & Moraceae & Ficus & X \\ 
  Ficus trigona & Moraceae & Ficus & X \\ 
  Maclura tinctoria & Moraceae & Maclura & X \\ 
  Morus alba & Moraceae & Morus & X \\ 
  Morus nigra & Moraceae & Morus & X \\ 
  Sorocea bonplandii & Moraceae & Sorocea & X \\ 
  Byrsonima cydoniifolia & Malpighiaceae & Byrsonima &  \\ 
  Byrsonima ligustrifolia & Malpighiaceae & Byrsonima &  \\ 
  Byrsonima sericea & Malpighiaceae & Byrsonima &  \\ 
  Byrsonima variabilis & Malpighiaceae & Byrsonima &  \\ 
  Malpighia glabra & Malpighiaceae & Malpighia &  \\ 
  Cabralea canjerana & Meliaceae & Cabralea &  \\ 
  Guarea guidonia & Meliaceae & Guarea &  \\ 
  Guarea kunthiana & Meliaceae & Guarea &  \\ 
  Guarea macrophylla & Meliaceae & Guarea &  \\ 
  Melia azedarach & Meliaceae & Melia &  \\ 
  Trichilia catigua & Meliaceae & Trichilia &  \\ 
  Trichilia clausseni & Meliaceae & Trichilia &  \\ 
  Trichilia elegans & Meliaceae & Trichilia &  \\ 
  Trichilia pallida & Meliaceae & Trichilia &  \\ 
  Calophyllum brasiliense & Calophyllaceae & Calophyllum &  \\ 
  Calyptranthes clusiifolia & Myrtaceae & Calyptranthes & X \\ 
  Calyptranthes concinna & Myrtaceae & Calyptranthes & X \\ 
  Campomanesia guaviroba & Myrtaceae & Campomanesia & X \\ 
  Campomanesia guazumifolia & Myrtaceae & Campomanesia & X \\ 
  Campomanesia neriiflora & Myrtaceae & Campomanesia & X \\ 
  Campomanesia phaea & Myrtaceae & Campomanesia & X \\ 
  Campomanesia xanthocarpa & Myrtaceae & Campomanesia & X \\ 
  Eugenia astringens & Myrtaceae & Eugenia & X \\ 
  Eugenia brasiliensis & Myrtaceae & Eugenia & X \\ 
  Eugenia cerasiflora & Myrtaceae & Eugenia & X \\ 
  Eugenia cuprea & Myrtaceae & Eugenia & X \\ 
  Eugenia florida & Myrtaceae & Eugenia & X \\ 
  Eugenia handroi & Myrtaceae & Eugenia & X \\ 
  Eugenia hiemalis & Myrtaceae & Eugenia & X \\ 
  Eugenia involucrata & Myrtaceae & Eugenia & X \\ 
  Eugenia melanogyna & Myrtaceae & Eugenia & X \\ 
  Eugenia mosenii & Myrtaceae & Eugenia & X \\ 
  Eugenia neoglomerata & Myrtaceae & Eugenia & X \\ 
  Eugenia oblongata & Myrtaceae & Eugenia & X \\ 
  Eugenia pyriformis & Myrtaceae & Eugenia & X \\ 
  Eugenia umbelliflora & Myrtaceae & Eugenia & X \\ 
  Eugenia uniflora & Myrtaceae & Eugenia & X \\ 
  Eugenia uruguayensis & Myrtaceae & Eugenia & X \\ 
  Eugenia verticillata & Myrtaceae & Eugenia & X \\ 
  Marlierea neuwiediana & Myrtaceae & Marlierea & X \\ 
  Marlierea obscura & Myrtaceae & Marlierea & X \\ 
  Marlierea reitzii & Myrtaceae & Marlierea & X \\ 
  Marlierea suaveolens & Myrtaceae & Marlierea & X \\ 
  Marlierea tomentosa & Myrtaceae & Marlierea & X \\ 
  Myrceugenia myrcioides & Myrtaceae & Myrceugenia & X \\ 
  Myrcia anacardiifolia & Myrtaceae & Myrcia & X \\ 
  Myrcia brasiliensis & Myrtaceae & Myrcia & X \\ 
  Myrcia ferruginea & Myrtaceae & Myrcia & X \\ 
  Myrcia hartwegiana & Myrtaceae & Myrcia & X \\ 
  Myrcia hebepetala & Myrtaceae & Myrcia & X \\ 
  Myrcia ilheosensis & Myrtaceae & Myrcia & X \\ 
  Myrcia oblongata & Myrtaceae & Myrcia & X \\ 
  Myrcia palustris & Myrtaceae & Myrcia & X \\ 
  Myrcia pubipetala & Myrtaceae & Myrcia & X \\ 
  Myrcia pulchra & Myrtaceae & Myrcia & X \\ 
  Myrcia spectabilis & Myrtaceae & Myrcia & X \\ 
  Myrcia splendens & Myrtaceae & Myrcia & X \\ 
  Myrcia tomentosa & Myrtaceae & Myrcia & X \\ 
  Myrciaria  glomerata & Myrtaceae & Myrciaria & X \\ 
  Myrciaria cuspidata & Myrtaceae & Myrciaria & X \\ 
  Myrciaria floribunda & Myrtaceae & Myrciaria & X \\ 
  Myrciaria trunciflora & Myrtaceae & Myrciaria & X \\ 
  Myrrhinium atropurpureum & Myrtaceae & Myrrhinium & X \\ 
  Neomitranthes glomerata & Myrtaceae & Neomitranthes & X \\ 
  Neomitranthes obscura & Myrtaceae & Neomitranthes & X \\ 
  Plinia cauliflora & Myrtaceae & Plinia & X \\ 
  Psidium cattleianum & Myrtaceae & Psidium & X \\ 
  Psidium guajava & Myrtaceae & Psidium & X \\ 
  Siphoneugena densiflora & Myrtaceae & Siphoneugena & X \\ 
  Syzygium cumini & Myrtaceae & Syzygium & X \\ 
  Carica papaya & Caricaceae & Carica &  \\ 
  Jacaratia spinosa & Caricaceae & Jacaratia &  \\ 
  Casearia decandra & Salicaceae & Casearia &  \\ 
  Casearia sylvestris & Salicaceae & Casearia &  \\ 
  Cecropia glaziovii & Urticaceae & Cecropia &  \\ 
  Cecropia hololeuca & Urticaceae & Cecropia &  \\ 
  Cecropia pachystachya & Urticaceae & Cecropia &  \\ 
  Coussapoa microcarpa & Urticaceae & Coussapoa &  \\ 
  Pourouma guianensis & Urticaceae & Pourouma &  \\ 
  Urera baccifera & Urticaceae & Urera &  \\ 
  Celtis iguanaea & Cannabaceae & Celtis &  \\ 
  Trema micrantha & Cannabaceae & Trema &  \\ 
  Cerastium glomeratum & Caryophyllaceae & Cerastium &  \\ 
  Cereus fernambucensis & Cactaceae & Cereus &  \\ 
  Cereus hildmannianus & Cactaceae & Cereus &  \\ 
  Opuntia monacantha & Cactaceae & Opuntia &  \\ 
  Pereskia aculeata & Cactaceae & Pereskia &  \\ 
  Pilosocereus arrabidae & Cactaceae & Pilosocereus &  \\ 
  Rhipsalis campos-portoana & Cactaceae & Rhipsalis &  \\ 
  Rhipsalis elliptica & Cactaceae & Rhipsalis &  \\ 
  Rhipsalis paradoxa & Cactaceae & Rhipsalis &  \\ 
  Rhipsalis teres & Cactaceae & Rhipsalis &  \\ 
  Stephanocereus luetzelburgii & Cactaceae & Stephanocereus &  \\ 
  Chrysophyllum flexuosum & Sapotaceae & Chrysophyllum &  \\ 
  Chrysophyllum gonocarpum & Sapotaceae & Chrysophyllum &  \\ 
  Chrysophyllum viride & Sapotaceae & Chrysophyllum &  \\ 
  Cinnamodendron dinisii & Canellaceae & Cinnamodendron &  \\ 
  Cissus paulliniifolia & Vitaceae & Cissus &  \\ 
  Cissus selloana & Vitaceae & Cissus &  \\ 
  Cissus striata & Vitaceae & Cissus &  \\ 
  Cissus verticillata & Vitaceae & Cissus &  \\ 
  Citharexylum myrianthum & Verbenaceae & Citharexylum &  \\ 
  Duranta erecta & Verbenaceae & Duranta &  \\ 
  Lantana camara & Verbenaceae & Lantana &  \\ 
  Lantana pohliana & Verbenaceae & Lantana &  \\ 
  Citrus reticulata & Rutaceae & Citrus &  \\ 
  Citrus x aurantium & Rutaceae & Citrus &  \\ 
  Clausena excavata & Rutaceae & Clausena &  \\ 
  Murraya paniculata & Rutaceae & Murraya &  \\ 
  Zanthoxylum hyemale & Rutaceae & Zanthoxylum &  \\ 
  Zanthoxylum rhoifolium & Rutaceae & Zanthoxylum &  \\ 
  Zanthoxylum riedelianum & Rutaceae & Zanthoxylum &  \\ 
  Clidemia hirta & Melastomataceae & Clidemia & X \\ 
  Clidemia urceolata & Melastomataceae & Clidemia & X \\ 
  Henriettea saldanhaei & Melastomataceae & Henriettea & X \\ 
  Leandra acutiflora & Melastomataceae & Leandra & X \\ 
  Leandra aurea & Melastomataceae & Leandra & X \\ 
  Leandra australis & Melastomataceae & Leandra & X \\ 
  Leandra barbinervis & Melastomataceae & Leandra & X \\ 
  Leandra carassana & Melastomataceae & Leandra & X \\ 
  Leandra laevigata & Melastomataceae & Leandra & X \\ 
  Leandra melastomoides & Melastomataceae & Leandra & X \\ 
  Leandra pilonensis & Melastomataceae & Leandra & X \\ 
  Leandra refracta & Melastomataceae & Leandra & X \\ 
  Leandra regnellii & Melastomataceae & Leandra & X \\ 
  Leandra sabiaensis & Melastomataceae & Leandra & X \\ 
  Leandra variabilis & Melastomataceae & Leandra & X \\ 
  Leandra xanthocoma & Melastomataceae & Leandra & X \\ 
  Miconia affinis & Melastomataceae & Miconia & X \\ 
  Miconia albicans & Melastomataceae & Miconia & X \\ 
  Miconia alborufescens & Melastomataceae & Miconia & X \\ 
  Miconia brasiliensis & Melastomataceae & Miconia & X \\ 
  Miconia budlejoides & Melastomataceae & Miconia & X \\ 
  Miconia cabucu & Melastomataceae & Miconia & X \\ 
  Miconia chartacea & Melastomataceae & Miconia & X \\ 
  Miconia cinerascens & Melastomataceae & Miconia & X \\ 
  Miconia cinnamomifolia & Melastomataceae & Miconia & X \\ 
  Miconia collatata & Melastomataceae & Miconia & X \\ 
  Miconia cubatanensis & Melastomataceae & Miconia & X \\ 
  Miconia cuspidata & Melastomataceae & Miconia & X \\ 
  Miconia discolor & Melastomataceae & Miconia & X \\ 
  Miconia elegans & Melastomataceae & Miconia & X \\ 
  Miconia inaequidens & Melastomataceae & Miconia & X \\ 
  Miconia inconspicua & Melastomataceae & Miconia & X \\ 
  Miconia latecrenata & Melastomataceae & Miconia & X \\ 
  Miconia ligustroides & Melastomataceae & Miconia & X \\ 
  Miconia minutiflora & Melastomataceae & Miconia & X \\ 
  Miconia paniculata & Melastomataceae & Miconia & X \\ 
  Miconia pepericarpa & Melastomataceae & Miconia & X \\ 
  Miconia prasina & Melastomataceae & Miconia & X \\ 
  Miconia pusilliflora & Melastomataceae & Miconia & X \\ 
  Miconia racemifera & Melastomataceae & Miconia & X \\ 
  Miconia rubiginosa & Melastomataceae & Miconia & X \\ 
  Miconia sellowiana & Melastomataceae & Miconia & X \\ 
  Miconia tentaculifera & Melastomataceae & Miconia & X \\ 
  Miconia theizans & Melastomataceae & Miconia & X \\ 
  Miconia tristis & Melastomataceae & Miconia & X \\ 
  Miconia urophylla & Melastomataceae & Miconia & X \\ 
  Miconia valtheri & Melastomataceae & Miconia & X \\ 
  Ossaea amygdaloides & Melastomataceae & Ossaea & X \\ 
  Clusia criuva & Clusiaceae & Clusia &  \\ 
  Clusia hilariana & Clusiaceae & Clusia &  \\ 
  Clusia lanceolata & Clusiaceae & Clusia &  \\ 
  Clusia organensis & Clusiaceae & Clusia &  \\ 
  Garcinia gardneriana & Clusiaceae & Garcinia &  \\ 
  Codonanthe cordifolia & Gesneriaceae & Codonanthe &  \\ 
  Cordia abyssinica & Boraginaceae & Cordia &  \\ 
  Cordia axillaris & Boraginaceae & Cordia &  \\ 
  Cordia corymbosa & Boraginaceae & Cordia &  \\ 
  Cordia ecalyculata & Boraginaceae & Cordia &  \\ 
  Cordia sellowiana & Boraginaceae & Cordia &  \\ 
  Cordia silvestris & Boraginaceae & Cordia &  \\ 
  Myriopus paniculatus & Boraginaceae & Myriopus &  \\ 
  Varronia curassavica & Boraginaceae & Varronia &  \\ 
  Costus spiralis & Costaceae & Costus &  \\ 
  Curatella americana & Dilleniaceae & Curatella &  \\ 
  Davilla elliptica & Dilleniaceae & Davilla &  \\ 
  Davilla rugosa & Dilleniaceae & Davilla &  \\ 
  Doliocarpus dentatus & Dilleniaceae & Doliocarpus &  \\ 
  Cybianthus peruvianus & Primulaceae & Cybianthus &  \\ 
  Myrsine coriacea & Primulaceae & Myrsine &  \\ 
  Myrsine ferruginea & Primulaceae & Myrsine &  \\ 
  Myrsine gardneriana & Primulaceae & Myrsine &  \\ 
  Myrsine lancifolia & Primulaceae & Myrsine &  \\ 
  Myrsine umbellata & Primulaceae & Myrsine &  \\ 
  Myrsine venosa & Primulaceae & Myrsine &  \\ 
  Daphnopsis brasiliensis & Thymelaeaceae & Daphnopsis &  \\ 
  Dendropanax cuneatus & Araliaceae & Dendropanax &  \\ 
  Hedera nepalensis & Araliaceae & Hedera &  \\ 
  Schefflera actinophylla & Araliaceae & Schefflera &  \\ 
  Schefflera angustissima & Araliaceae & Schefflera &  \\ 
  Schefflera arboricola & Araliaceae & Schefflera &  \\ 
  Schefflera macrocarpa & Araliaceae & Schefflera &  \\ 
  Schefflera morototoni & Araliaceae & Schefflera &  \\ 
  Dichorisandra thyrsiflora & Commelinaceae & Dichorisandra &  \\ 
  Diospyros inconstans & Ebenaceae & Diospyros &  \\ 
  Diospyros kaki & Ebenaceae & Diospyros &  \\ 
  Drimys brasiliensis & Winteraceae & Drimys &  \\ 
  Drimys winteri & Winteraceae & Drimys &  \\ 
  Eriobotrya japonica & Rosaceae & Eriobotrya &  \\ 
  Prunus myrtifolia & Rosaceae & Prunus &  \\ 
  Prunus persica & Rosaceae & Prunus &  \\ 
  Pyracantha coccinea & Rosaceae & Pyracantha &  \\ 
  Rubus brasiliensis & Rosaceae & Rubus &  \\ 
  Rubus erythroclados & Rosaceae & Rubus &  \\ 
  Rubus rosifolius & Rosaceae & Rubus &  \\ 
  Rubus urticifolius & Rosaceae & Rubus &  \\ 
  Erythroxylum ambiguum & Erythroxylaceae & Erythroxylum &  \\ 
  Erythroxylum argentinum & Erythroxylaceae & Erythroxylum &  \\ 
  Erythroxylum deciduum & Erythroxylaceae & Erythroxylum &  \\ 
  Erythroxylum gonocladum & Erythroxylaceae & Erythroxylum &  \\ 
  Erythroxylum pauferrense & Erythroxylaceae & Erythroxylum &  \\ 
  Erythroxylum pulchrum & Erythroxylaceae & Erythroxylum &  \\ 
  Erythroxylum simonis & Erythroxylaceae & Erythroxylum &  \\ 
  Frangula purshiana & Rhamnaceae & Frangula &  \\ 
  Hovenia dulcis & Rhamnaceae & Hovenia &  \\ 
  Scutia buxifolia & Rhamnaceae & Scutia &  \\ 
  Fuchsia regia & Onagraceae & Fuchsia &  \\ 
  Gaylussacia brasiliensis & Ericaceae & Gaylussacia &  \\ 
  Gaylussacia pulchra & Ericaceae & Gaylussacia &  \\ 
  Gaylussacia virgata & Ericaceae & Gaylussacia &  \\ 
  Guapira opposita & Nyctaginaceae & Guapira &  \\ 
  Guapira pernambucensis & Nyctaginaceae & Guapira &  \\ 
  Hedychium coronarium & Zingiberaceae & Hedychium &  \\ 
  Hedyosmum brasiliense & Chloranthaceae & Hedyosmum &  \\ 
  Heisteria silvianii & Olacaceae & Heisteria &  \\ 
  Hohenbergia ramageana & Bromeliaceae & Hohenbergia &  \\ 
  Humiria balsamifera & Humiriaceae & Humiria &  \\ 
  Hyeronima alchorneoides & Phyllanthaceae & Hyeronima &  \\ 
  Margaritaria nobilis & Phyllanthaceae & Margaritaria &  \\ 
  Richeria grandis & Phyllanthaceae & Richeria &  \\ 
  Hypochaeris brasiliensis & Asteraceae & Hypochaeris &  \\ 
  Ilex affinis & Aquifoliaceae & Ilex &  \\ 
  Ilex brevicuspis & Aquifoliaceae & Ilex &  \\ 
  Ilex microdonta & Aquifoliaceae & Ilex &  \\ 
  Ilex paraguariensis & Aquifoliaceae & Ilex &  \\ 
  Ilex pseudobuxus & Aquifoliaceae & Ilex &  \\ 
  Ilex theezans & Aquifoliaceae & Ilex &  \\ 
  Lasiacis sorghoidea & Poaceae & Lasiacis &  \\ 
  Megathyrsus maximus & Poaceae & Megathyrsus &  \\ 
  Triticum aestivum & Poaceae & Triticum &  \\ 
  Urochloa decumbens & Poaceae & Urochloa &  \\ 
  Urochloa plantaginea & Poaceae & Urochloa &  \\ 
  Ligustrum japonicum & Oleaceae & Ligustrum &  \\ 
  Ligustrum lucidum & Oleaceae & Ligustrum &  \\ 
  Magnolia champaca & Magnoliaceae & Magnolia &  \\ 
  Magnolia ovata & Magnoliaceae & Magnolia &  \\ 
  Marcgravia polyantha & Marcgraviaceae & Marcgravia &  \\ 
  Schwartzia brasiliensis & Marcgraviaceae & Schwartzia &  \\ 
  Maytenus aquifolia & Celastraceae & Maytenus &  \\ 
  Maytenus brasiliensis & Celastraceae & Maytenus &  \\ 
  Maytenus gonoclada & Celastraceae & Maytenus &  \\ 
  Maytenus littoralis & Celastraceae & Maytenus &  \\ 
  Schaefferia argentinensis & Celastraceae & Schaefferia &  \\ 
  Meliosma sellowii & Sabiaceae & Meliosma &  \\ 
  Melothria cucumis & Cucurbitaceae & Melothria &  \\ 
  Momordica charantia & Cucurbitaceae & Momordica &  \\ 
  Mollinedia boracensis & Monimiaceae & Mollinedia &  \\ 
  Mollinedia schottiana & Monimiaceae & Mollinedia &  \\ 
  Mollinedia triflora & Monimiaceae & Mollinedia &  \\ 
  Mollinedia uleana & Monimiaceae & Mollinedia &  \\ 
  Muntingia calabura & Muntingiaceae & Muntingia &  \\ 
  Musa paradisiaca & Musaceae & Musa &  \\ 
  Musa rosacea & Musaceae & Musa &  \\ 
  Ouratea polygyna & Ochnaceae & Ouratea &  \\ 
  Ouratea vaccinioides & Ochnaceae & Ouratea &  \\ 
  Passiflora actinia & Passifloraceae & Passiflora &  \\ 
  Passiflora edulis & Passifloraceae & Passiflora &  \\ 
  Peplonia organensis & Apocynaceae & Peplonia &  \\ 
  Peschiera catharinensis & Apocynaceae & Peschiera &  \\ 
  Tabernaemontana hystrix & Apocynaceae & Tabernaemontana &  \\ 
  Pera glabrata & Peraceae & Pera &  \\ 
  Phoradendron crassifolium & Santalaceae & Phoradendron &  \\ 
  Phoradendron piperoides & Santalaceae & Phoradendron &  \\ 
  Phoradendron quadrangulare & Santalaceae & Phoradendron &  \\ 
  Phytolacca dioica & Phytolaccaceae & Phytolacca &  \\ 
  Piper aduncum & Piperaceae & Piper &  \\ 
  Piper amalago & Piperaceae & Piper &  \\ 
  Piper corintoanum & Piperaceae & Piper &  \\ 
  Piper dilatatum & Piperaceae & Piper &  \\ 
  Piper gaudichaudianum & Piperaceae & Piper &  \\ 
  Piper hispidinervum & Piperaceae & Piper &  \\ 
  Piper miquelianum & Piperaceae & Piper &  \\ 
  Piper mollicomum & Piperaceae & Piper &  \\ 
  Piper tectoniifolium & Piperaceae & Piper &  \\ 
  Podocarpus sellowii & Podocarpaceae & Podocarpus &  \\ 
  Protium heptaphyllum & Burseraceae & Protium &  \\ 
  Protium spruceanum & Burseraceae & Protium &  \\ 
  Protium widgrenii & Burseraceae & Protium &  \\ 
  Psittacanthus robustus & Loranthaceae & Psittacanthus &  \\ 
  Struthanthus concinnus & Loranthaceae & Struthanthus &  \\ 
  Struthanthus vulgaris & Loranthaceae & Struthanthus &  \\ 
  Quiina glazovii & Quiinaceae & Quiina &  \\ 
  Scaevola plumieri & Goodeniaceae & Scaevola &  \\ 
  Sloanea guianensis & Elaeocarpaceae & Sloanea &  \\ 
  Sloanea hirsuta & Elaeocarpaceae & Sloanea &  \\ 
  Smilax elastica & Smilacaceae & Smilax &  \\ 
  Smilax rufescens & Smilacaceae & Smilax &  \\ 
  Stromanthe tonckat & Marantaceae & Stromanthe &  \\ 
  Strychnos brasiliensis & Loganiaceae & Strychnos &  \\ 
  Styrax leprosus & Styracaceae & Styrax &  \\ 
  Styrax pohlii & Styracaceae & Styrax &  \\ 
  Symplocos estrellensis & Symplocaceae & Symplocos &  \\ 
  Symplocos glandulosomarginata & Symplocaceae & Symplocos &  \\ 
  Symplocos laxiflora & Symplocaceae & Symplocos &  \\ 
  Symplocos pubescens & Symplocaceae & Symplocos &  \\ 
  Symplocos revoluta & Symplocaceae & Symplocos &  \\ 
  Symplocos tetrandra & Symplocaceae & Symplocos &  \\ 
  Symplocos uniflora & Symplocaceae & Symplocos &  \\ 
  Turnera ulmifolia & Turneraceae & Turnera &  \\ 
  Virola bicuhyba & Myristicaceae & Virola &  \\ 
  Virola gardneri & Myristicaceae & Virola &  \\ 
  Virola sebifera & Myristicaceae & Virola &  \\ 
  Vismia brasiliensis & Hypericaceae & Vismia &  \\ 
   \hline
\end{longtable}}

\end{document}